\documentclass[journal]{IEEEtran}
\ifCLASSINFOpdf
  \usepackage[pdftex]{graphicx}
\else
\fi

\usepackage{comment}

\usepackage{array}
\usepackage{cite}
\usepackage{multirow}
\usepackage{color,soul}

\usepackage{subfigure}
\usepackage{upgreek}

\usepackage{amssymb,booktabs}
\usepackage{graphicx}
\usepackage[table,xcdraw]{xcolor}
\usepackage{bbding}

\usepackage[acronyms,nonumberlist,nopostdot,nomain,nogroupskip]{glossaries}
\newacronym{iont}{IoNT}{Internet of Nano-Things}
\newacronym{spp}{SPP}{Surface Plasmon-Polariton}
\newacronym{pec}{PEC}{Perfect Electric Conductor}
\newacronym{oam}{OAM}{Orbital Angular Momentum}
\newacronym{pic}{PIC}{Photonic Integrated Circuits}
\newacronym{2d}{2D}{two-dimensional}
\newacronym{gnr}{GNR}{Graphene NanoRibbon}
\newacronym{cnt}{CNT}{Carbon NanoTubes}
\newacronym{gaas}{GaAs}{Gallium Arsenide}
\newacronym{ingaas}{InGaAs}{Indium Gallium Arsenide}
\newacronym{gan}{GaN}{Gallium Nitride}
\newacronym{hbn}{h-BN}{hexagonal Boron Nitride}
\newacronym{mos2}{MoS2}{Molybdenum disulfide}
\newacronym{hemt}{HEMT}{High-Electron Mobility Transistors}
\newacronym{2deg}{2DEG}{Two-Dimensional Electron Gas}
\newacronym{nems}{NEMS}{Nano-ElectroMechanical System}
\newacronym{mems}{MEMS}{Micro-ElectroMechanical System}
\newacronym{wnoc}{WNoC}{Wireless Network-on-Chip}
\newacronym{ieee}{IEEE}{Institute of Electrical and Electronics Engineers}
\newacronym{wpan}{WPAN}{Wireless Personal Area Network}
\newacronym{wlan}{WLAN}{Wireless Local Area Network}
\newacronym{nr}{NR}{New Radio}
\newacronym{3gpp}{3GPP}{3rd Generation Partnership Project}
\newacronym{vpn}{VPN}{Virtual Private Network}
\newacronym{dsp}{DSP}{Digital Signal Processing}
\newacronym{iobnt}{IoBNT}{Internet of Bio-Nano Things}
\newacronym{vna}{VNA}{Vector Network Analyzer}
\newacronym{mac}{MAC}{Medium access control}
\newacronym{ris}{RIS}{Reconfigurable Intelligent Surface}
\newacronym{fpga}{FPGA}{Field-Programmable Gate Array}

\begin{document}
%
\title{Electromagnetic Nanonetworks Beyond 6G:\\From Wearable and Implantable Networks to On-chip and Quantum Communication}
%
%
%

\author{Sergi~Abadal,~\IEEEmembership{Member,~IEEE,}
        Chong~Han,~\IEEEmembership{Senior Member,~IEEE,}
		Vitaly~Petrov,~\IEEEmembership{Member,~IEEE,}
        Laura~Galluccio,~\IEEEmembership{Senior Member,~IEEE,}
        Ian~F.~Akyildiz,~\IEEEmembership{Life Fellow,~IEEE,}
        and~Josep~M.~Jornet,~\IEEEmembership{Fellow,~IEEE}%
\thanks{S. Abadal is with the NaNoNetworking Center in Catalonia (N3Cat), Universitat Polit\`{e}cnica de Catalunya, 08034 Barcelona, Spain (e-mail: abadal@ac.upc.edu).}
\thanks{C. Han is with the Terahertz Wireless Communications (TWC) Laboratory, Department of Electronic Engineering and Cooperative Medianet Innovation Center (CMIC), Shanghai Jiao Tong University, Shanghai, China (Email:~chong.han@sjtu.edu.cn).}
\thanks{V. Petrov is with the Division of Communication Systems, KTH~Royal Institute of Technology, Stockholm, Sweden, and was with the Institute for the Wireless Internet of Things, Northeastern University, Boston, USA (e-mail: v.petrov@northeastern.edu).}
\thanks{L. Galluccio is with the Dipartimento di Ingegneria Elettrica Elettronica e Informatica (DIEEI), University of Catania, Italy and Consorzio Nazionale Interuniversitario per le Telecomunicazioni (CNIT)  (laura.galluccio@unict.it).}
\thanks{I. F. Akyildiz is with the Department of Electrical and Computer Engineering, University of Iceland, Reykjavik, Iceland (e-mail: ianaky@hi.is).}
\thanks{J. M. Jornet is with the Institute for the Wireless Internet of Things, Northeastern University, Boston, USA (e-mail:j.jornet@northeastern.edu).}
\thanks{Manuscript received Month DD, YYYY; revised Month DD, YYYY.}
}

%
%

\markboth{IEEE Journal on Selected Areas in Communications, 2024}
{Abadal \MakeLowercase{\textit{et al.}}: Electromagnetic Nanonetworks}
%



\maketitle

\begin{abstract}
Emerging from the symbiotic combination of nanotechnology and communications, the field of nanonetworking has come a long way since its inception more than fifteen years ago. Significant progress has been achieved in several key communication technologies as enablers of the paradigm, as well as in the multiple application areas that it opens. In this paper, the focus is placed on the electromagnetic nanonetworking paradigm, providing an overview of the advances made in wireless nanocommunication technology from microwave through terahertz to optical bands. The characteristics and potential of the compared technologies are then confronted with the requirements and challenges of the broad set of nanonetworking applications in the Internet of NanoThings (IoNT) and on-chip networks paradigms, including quantum computing applications for the first time. Finally, a selection of cross-cutting issues and possible directions for future work are given, aiming to guide researchers and practitioners towards the next generation of electromagnetic nanonetworks. 
\end{abstract}


%
\IEEEpeerreviewmaketitle

\section{Introduction}
\label{sec:intro}
\IEEEPARstart{N}{anonetworks} emerged as a forward-looking vision around fifteen years ago, supported by the advancements in nanotechnology at the time~\cite{akyildiz2008nanonetworks}. Not only had transistors scaled below the 50~nm barrier by that time, leading to the integration of hundreds of them in less than a square micrometer~\cite{kuhn2008managing}, but also significant progress was made in the fabrication of ultra-precise nanosensors~\cite{anker2008biosensing}, in the functionalization of materials for energy harvesting at the nanoscale~\cite{wang2008energy}, and in the development of nano-antennas~\cite{jornet2010graphene}. All these ingredients allowed to envisage nanonetworks in the form of swarms of nanosensor motes communicating to reach unprecedented locations and sense events with unprecedented accuracy, mostly for biomedical applications~\cite{yang2020comprehensive}. 

One of the key elements behind the inception of the nanonetworking paradigm was the pioneering work towards developing two key communication technologies. On the one hand, molecular communications, consisting in the use of physical particles and/or their density for the encoding of data and of different means of transport to reach the receiver, was proposed due to its energy efficiency and biocompatibility. The latter aspects are of fundamental relevance for a nanonetworking paradigm where biomedical applications were paramount~\cite{akyildiz2015internet}. On the other hand, electromagnetic communications in the nanoscale were studied given their higher capacity and wider applicability, also spurred in part by the advancement in miniaturized plasmonic antennas in the terahertz and optical ranges~\cite{akyildiz2010electromagnetic}. Hence, the concept of electromagnetic nanonetworks (the main focus of this manuscript) was born.

In the last decade, nanotechnology has continued its relentless progress leading to transistor scaling towards the sub-nm regime~\cite{cao2023future}, outstanding achievements in CMOS-RF circuits towards true terahertz support~\cite{yu2020mmwave}, silicon photonics and other fully integrable photonic technologies~\cite{pasricha2020survey}, and energy harvesting and power transfer means enabling perpetual operation of devices~\cite{ntontin2024perpetual}, to name a few. Transversal to them all, the last decade has also seen tremendous advances in the synthesis and application of graphene, which in turn heralded the arrival of a new breed of two-dimensional nanomaterials with outstanding opto-electromechanical properties~\cite{ferrari2015science, briggs2019roadmap}. Such progress also helped broaden the application scope of electromagnetic nanonetworks, leading to a body of work aiming to employ them to create intelligent and reconfigurable materials~\cite{dash2022active}, to accelerate existing computing systems~\cite{abadal2022graphene}, or to scale quantum computers~\cite{alarcon2023scalable}.

In parallel to the technological evolution and the new application areas envisaged for electromagnetic nanonetworks, the research community has also devoted considerable effort in developing protocol stacks uniquely tailored to the extremely stringent requirements of this paradigm. Novel and often opportunistic techniques for modulation, coding, medium access control, routing, addressing, localization, energy management, just to name a few, have been developed with the goal of becoming part of the communications backbone of the first generations of electromagnetic nanonetworks. 

As the nanonetworking vision evolves and enters a stage closer to implementation, the research community finds new challenges stemming not only from the complexity of developing appropriate communication means, but also from the maturing of technology (such as the need to find ways to integrate, prototype, and test nanonetworking devices) and from the new application domains that emerge (such as operation in cryogenic temperatures within quantum computers).

The present paper, as well as the special issue to which it belongs, aims to provide an updated overview of the field of electromagnetic nanonetworks from the technological and application perspectives. As shown in Table~\ref{table_surveyall}, previous survey papers in this area have either considered only body-centric communications or focused on multiple application scenarios but covering only communications aspects, mostly investigated by following a layered perspective and listing relevant design aspects going from the physical layer up to the higher layer issues.

\begin{table*}
\centering
\label{table_surveyall}
\caption{Literature surveys on nanonetworks.}
\resizebox{\textwidth}{!}{%
\begin{tabular}{@{}p{4.7cm}<{\centering}cccp{2cm}<{\centering}p{2cm}<{\centering}p{3cm}<{\centering}p{2cm}<{\centering}@{}}
\toprule
 &
   &
  \multicolumn{2}{c}{\textbf{Technologies}} &
  \multicolumn{4}{c}{\textbf{Application Scenarios}} \\ \cmidrule(l){3-8} 
\multirow{-2}{*}{\textbf{Title and Reference}} &
  \multirow{-2}{*}{\textbf{Year}} &
  EM &
  Non-EM &
  Internet of Nano-Things &
  Body-Area Networks &
  Wireless Networks within Computing Packages &
  Quantum Computing \\ \midrule
Nanonetworks: A new communication paradigm~\cite{akyildiz2008nanonetworks} &
  2008 &
  $\checkmark$ &
  $\checkmark$ &
   &
  $\checkmark$ &
   &
   \\
\rowcolor[HTML]{EFEFEF} 
Electromagnetic wireless nanosensor networks~\cite{akyildiz2010electromagnetic} &
  2010 &
  $\checkmark$ &
   &
  $\checkmark$ &
  $\checkmark$ &
   &
   \\
The internet of nano-things~\cite{akyildiz2010internet} &
  2010 &
  $\checkmark$ &
  $\checkmark$ &
  $\checkmark$ &
  $\checkmark$ &
   &
   \\
\rowcolor[HTML]{EFEFEF} 
The internet of multimedia nano-things~\cite{jornet2012internet} &
  2012 &
  $\checkmark$ &
   &
  $\checkmark$ &
  $\checkmark$ &
   &
   \\
Realizing the internet of nano things: challenges, solutions, and applications~\cite{balasubramaniam2012realizing} &
  2012 &
  $\checkmark$ &
  $\checkmark$ &
  $\checkmark$ &
  $\checkmark$ &
   &
   \\
\rowcolor[HTML]{EFEFEF} 
A brief survey on molecular and electromagnetic communications in nano-networks~\cite{rikhtegar2013brief} &
  2013 &
  $\checkmark$ &
  $\checkmark$ &
   &
  $\checkmark$ &
   &
   \\
The internet of bio-nano things~\cite{akyildiz2015internet} &
  2015 &
   &
  $\checkmark$ &
  $\checkmark$ &
  $\checkmark$ &
   &
   \\
\rowcolor[HTML]{EFEFEF} 
Connecting in-body nano communication with body area networks: Challenges and opportunities of the Internet of Nano Things~\cite{dressler2015connecting} &
  2015 &
  $\checkmark$ &
  $\checkmark$ &
  $\checkmark$ &
  $\checkmark$ &
   &
   \\
Molecular communication and nanonetwork for targeted drug delivery: A survey~\cite{chude2017molecular} &
  2017 &
   &
  $\checkmark$ &
   &
  $\checkmark$ &
   &
   \\
\rowcolor[HTML]{EFEFEF} 
A review on the role of nano-communication in future healthcare systems: A big data analytics perspective~\cite{rizwan2018review} &
  2018 &
  $\checkmark$ &
  $\checkmark$ &
   &
  $\checkmark$ &
   &
   \\
Moving forward with molecular communication: From theory to human health applications [point of view]~\cite{akyildiz2019moving} &
  2019 &
   &
  $\checkmark$ &
   &
  $\checkmark$ &
   &
   \\
\rowcolor[HTML]{EFEFEF} 
Nanonetworks in biomedical applications~\cite{marzo2019nanonetworks} &
  2019 &
  $\checkmark$ &
  $\checkmark$ &
   &
  $\checkmark$ &
   &
   \\
A comprehensive survey on hybrid communication in context of molecular communication and terahertz communication for body-centric nanonetworks~\cite{yang2020comprehensive} &
  2020 &
  $\checkmark$ &
  $\checkmark$ &
   &
  $\checkmark$ &
   &
   \\
\rowcolor[HTML]{EFEFEF} 
Electromagnetic nanocommunication networks: Principles, applications, and challenges~\cite{kabir2021electromagnetic} &
  2021 &
  $\checkmark$ &
  $\checkmark$ &
   &
  $\checkmark$ &
  $\checkmark$ &
   \\
A survey of molecular communication in cell biology: Establishing a new hierarchy for interdisciplinary applications~\cite{bi2021survey} &
  2021 &
   &
  $\checkmark$ &
   &
  $\checkmark$ &
   &
   \\
\rowcolor[HTML]{EFEFEF} 
Survey on terahertz nanocommunication and networking: A top-down perspective~\cite{lemic2021survey} &
  2021 &
  $\checkmark$ &
   &
   &
  $\checkmark$ &
  $\checkmark$ &
   \\
New insights on molecular communication in nano communication networks and their applications~\cite{koshy2022new} &
  2022 &
   &
  $\checkmark$ &
  $\checkmark$ &
  $\checkmark$ &
  $\checkmark$ &
   \\
\rowcolor[HTML]{EFEFEF} 
A biomedical perspective in terahertz nano-communications: A review~\cite{yin2022biomedical} &
  2022 &
  $\checkmark$ &
  $\checkmark$ &
  $\checkmark$ &
  $\checkmark$ &
   &
   \\
Nanotechnology applications in biomedical systems~\cite{buniyamin2022nanotechnology} &
  2022 &
   &
   &
  $\checkmark$ &
  $\checkmark$ &
   &
   \\
\rowcolor[HTML]{EFEFEF} 
Nanonetworking in the Terahertz Band and Beyond~\cite{jornet2023nanonetworking} &
  2023 &
  $\checkmark$ &
  &
  $\checkmark$ &
  $\checkmark$ &
   &
  $\checkmark$ \\
Internet of nano and bio-nano things: A review~\cite{csenturk2023internet} &
  2023 &
  $\checkmark$ &
  $\checkmark$ &
  $\checkmark$ &
  $\checkmark$ &
   &
   \\
\rowcolor[HTML]{EFEFEF}
Internet of nano-things (IoNT): A comprehensive review from architecture to security and privacy challenges~\cite{alabdulatif2023internet} &
  2023 &
   &
   &
  $\checkmark$ &
  $\checkmark$ &
   &
  $\checkmark$ \\
\rowcolor[HTML]{C0C0C0} 
This work &
  - &
  $\checkmark$ &
  $\checkmark$ &
  $\checkmark$ &
  $\checkmark$ &
  $\checkmark$ &
  $\checkmark$ \\ \bottomrule
\end{tabular}%
}
\end{table*}

In this survey paper, we consider a more general perspective where nanonetworks can be employed in different application scenarios ranging from environmental monitoring, smart manufacturing, and healthcare, to wireless networks within computing packages and quantum information processing. Then, we broaden the focus to outline the cross-cutting issues of the field, placing special attention to aspects that might hinder the actual implementation and deployment of electromagnetic nanonetworks in the years to come. 

\begin{figure}[!t]
\centering
\includegraphics[width=\columnwidth]{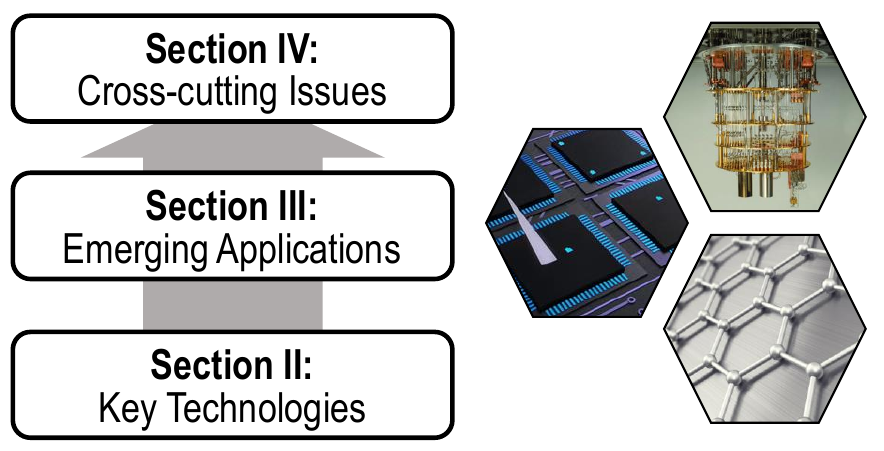}
\caption{Organization of this paper, from key communication technologies to emerging applications and cross-cutting issues.}
\label{fig:abstract}
\end{figure}

The remainder of this paper is organized as depicted in Figure~\ref{fig:abstract}. In Section~\ref{sec:tech}, we review the evolution and current state of the art of the key wireless technologies that enable electromagnetic nanonetworks, including plasmonic and magnetoelectric antennas, to then briefly describe non-radiative communication techniques also used in the nanonetworking paradigm. In Section~\ref{sec:apps}, we provide a summary of the main application areas for electromagnetic nanonetworks, from the \gls{iont} paradigm to new areas such as wireless networks within computing packages and quantum computing. Finally, in Section~\ref{sec:discussion}, we analyze issues common to all technologies and application areas of electromagnetic nanonetworks, and conclude the paper in Section~\ref{sec:conclusion}. 

\section{Key Communication Technologies for Electromagnetic Nanonetworks}
\label{sec:tech}
\begin{figure}
\centering
\includegraphics[width=\columnwidth]{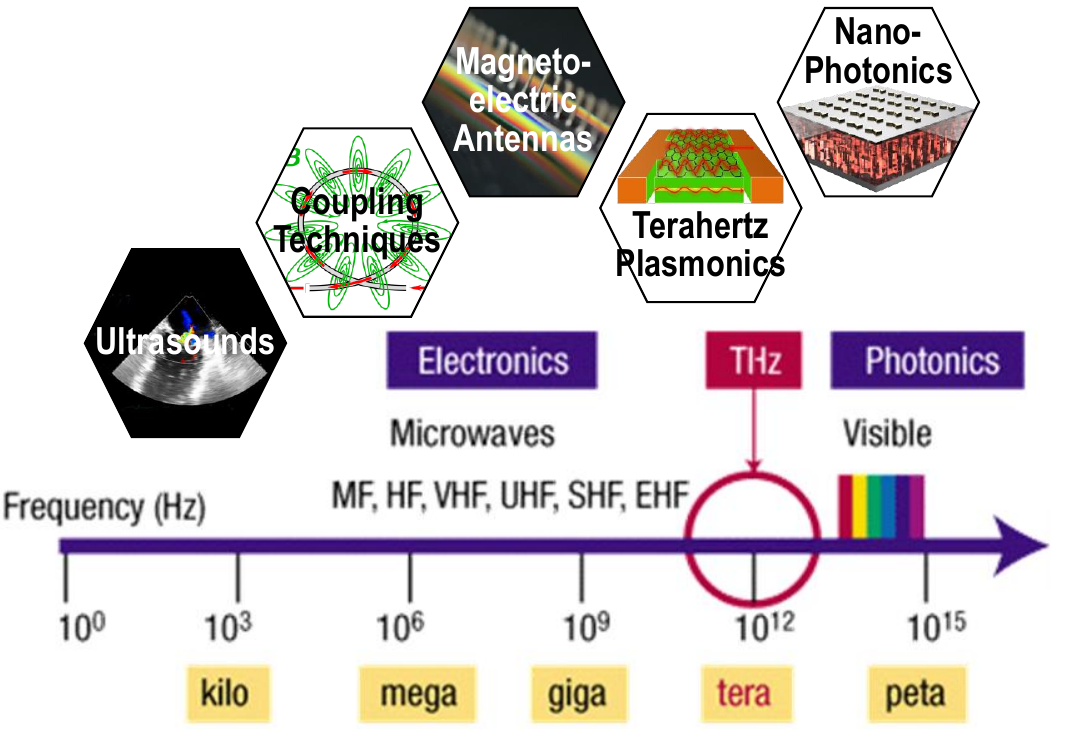}
\caption{Summary of wave-based key communication technologies for nanonetworks discussed in this survey.}
\label{fig:tech}
\end{figure}

In this section, we discuss the architectural and design features that emerge in electromagnetic nanonetworks and present their various enabling communication technologies, as illustrated in Figure~\ref{fig:tech}.


\subsection{Nano-Radio Architecture and Target Specifications}

The fundamental building block of electromagnetic nanonetworks is the nano-radio or miniaturized communication unit that can fit within an embedded nano-system or nanomachine. The nano-radio consists of two elements: 
\begin{itemize}
    \item The \textbf{transceiver} is responsible for generating, modulating, power-amplifying, and filtering signals in transmission and detecting, filtering, low-noise-amplifying, and demodulating the signals in reception. It traditionally consists of two parts: the \gls{dsp} back-end and the analog front-end. In nanonetworks, where low complexity is the key, the signal processing is mainly expected to take place in the analog domain. The key parameters include operating frequency range, modulation bandwidth, transmission power, and receiver noise figure and sensitivity. 
    \item The \textbf{antenna} is in charge of radiating the signals generated by the transmitter and reciprocally coupling incoming electromagnetic signals to the receiver. The key specifications of an antenna include the resonant frequency(s), bandwidth(s), three-dimensional radiation diagram (including directivity), and effective area.
\end{itemize}

Although recent emerging applications of nanonetworks do not necessarily impose micro-level size restrictions, the volume of a nanomachine has been traditionally defined as being of a few cubic micrometers at most~\cite{akyildiz2011nanonetworks}. Therefore, the communication unit will generally be smaller than that. In conventional communication systems, the antenna is usually the largest element and imposes or limits the frequency range of operation and the communication range, as discussed in Sec.~\ref{subsec:technology_options}.

Besides a very small footprint, nano-radios have other desirable properties. Some are true for all application scenarios, whereas others are application-dependent. In particular,
\begin{itemize}
    \item \textbf{Energy efficiency} is critical for all applications. An increased energy efficiency leads to lower heat generation, longer lifespan of the nano-devices, increased conversation of resources, and, ultimately, reduced greenhouse gas emissions. This becomes even more important when nanomachines are powered by nano-batteries or energy-harvesting nano-systems~\cite{jornet2012joint}, such as in the \gls{iont}  (Sec.~\ref{subsec:nanothings}).
    \item \textbf{Low latency and high data rate} are expected in several applications. For example, in computing applications of nanonetworks (Sec.~\ref{subsec:computing}), both low latency and high data rates are needed to meet the expectations of high-performance computing. However, in most applications of the \gls{iont}, latency is commonly more important than high data rate (for example, in the detection of a biological event, there might not be much information to transfer, but timeliness becomes critical). Still, high data rates can contribute to lowering multi-user interference and simplifying medium access control.
    \item \textbf{Mechanical flexibility} and \textbf{bio-compatibility} are similar properties that are highly relevant for some applications, such as in wearable and implantable nanomachines for biomedical \gls{iont} applications, but not a priority for others. 
\end{itemize}

Next, we describe the existing nano-radio technologies, including key specifications, evolution, maturity, and open challenges. First, we discuss the radiative technologies, and then, for completeness, we discuss additional non-radiative technologies often presented in the context of nanonetworks.



\subsection{Radiative Technologies across the Electromagnetic Spectrum}
\label{subsec:technology_options}

\subsubsection{Optical Frequencies: Infrared, Visible and Beyond}
The miniaturization of a conventional metallic antenna to meet the size requirements of nanomachines would result in operating frequencies in the optical spectrum. For example, a one-micrometer-long antenna built with a \gls{pec} material would resonate at 150~THz, i.e., in the infrared region of the electromagnetic spectrum. Under the same assumptions, an antenna for blue light (with a wavelength of 480~nm or, equivalently, a frequency of 625~THz), would be just 240~nm in its longest dimension. Traditionally, such very small antennas were not realizable, but today's optical nano-antennas are a reality thanks to nanotechnology~\cite{dorfmuller2010plasmonic,agio2013optical}. Alternatively, or simultaneously, nano-lenses can be used to further control the propagation of optical signals~\cite{petrov2017polymer}.

However, some observations are needed. First, conventional metals, such as gold, silver, or copper, cannot be modeled as \gls{pec} materials at optical frequencies. Their conductivity is not infinite but finite and complex-valued. As a result, electromagnetic fields can exist within the penetration depth of the antenna. Such surface waves, also known as \gls{spp} waves, propagate at a lower speed than electromagnetic waves in free space. As a result, the \gls{spp} wavelength is smaller than the free-space wavelength and the optical antenna resonant frequency is lower than that of an ideal \gls{pec} antenna. The ratio between the plasmonic and free-space wavelength is known as the plasmonic confinement factor and is usually lower than 10 for conventional metals. The higher the confinement factor, the smaller the antenna, but the lower its efficiency~\cite{nafari2017modeling}. Therefore, adopting plasmonic structures does not help traditional optical wireless communication systems, but it is advantageous for nanonetworks.

In addition to the optical nano-antenna, an optical nano-transceiver is needed. Miniature lasers are needed to generate the photo carriers at the target resonant frequency in transmission. Among others, micro-ring lasers with a footprint in the order of a few micrometers have been experimentally demonstrated~\cite{feng2014single,wong2021epitaxially}. Such lasers can emit microWatt-level powers and can be dynamically tuned to support high-speed modulation in amplitude, frequency/wavelength, and in \gls{oam}~\cite{zhang2020tunable}. Other ways to achieve on-chip optical modulation includes the adoption of plasmonic Mach-Zender modulators at the nanoscale~\cite{haffner2015all,thomaschewski2022plasmonic} or tunable plasmonic cavities for optical phase modulators~\cite{liu2011graphene,li2017single}. In reception, miniature on-chip photodetectors using different technologies have been demonstrated~\cite{chen2012infrared,nozaki2016photonic}.

The development of on-chip and nanoscale optical components has been tremendously benefited by the development of silicon photonic technologies~\cite{thomson2016roadmap}. The goal of silicon photonics is to develop \glspl{pic}, i.e., basically miniature circuits that use light instead of electricity to transmit signals, leveraging well-established silicon manufacturing techniques (similar to those used for electronic chips) to fabricate them at scale and low-cost. 
Leveraging these technologies, arrays of optical nano-transceivers and nano-antennas can be engineered to increase the communication range of nanomachines~\cite{sangwan2021beamforming,qiao2021higher}. 
Nevertheless, the very small wavelength of optical signals makes their propagation challenging in many of the application scenarios discussed in Sec.~\ref{sec:apps}. This motivates the exploration of technologies that support operating with longer wavelengths (i.e., lower frequencies) but still with nanoscale antennas.


\noindent
\subsubsection{Terahertz Band: Graphene and 2D Nanomaterials}
Graphene enables the development of nano-antennas that can operate at terahertz-band frequencies (broadly, between 100~GHz and 10~THz). Graphene is a \gls{2d} material, which had been studied since the XIX century but experimentally obtained and characterized in 2004. This one-atom-thick layer of carbon atoms arranged in a honeycomb crystal lattice has many unique mechanical, electrical and optical properties. It is extremely light and bendable, it offers very high electron mobility, and, despite being one atom thick, it offers interesting non-linear optical interactions. When it comes to nanonetworking, the main property of graphene is that it supports \gls{spp} waves~\cite{vakil2011transformation}, but at terahertz-band frequencies and with a much higher plasmonic confinement factor than that of metals at optical frequencies (it can exceed 100). As a result, it can be utilized to develop plasmonic antennas with sub-micrometric footprints that operate at terahertz-band frequencies. 

Graphene can be utilized as a \gls{2d} material, cut in narrow \glspl{gnr}, or rolled into~\glspl{cnt}. This has opened the door to different types of plasmonic nano-antennas. In 2006, Burke \emph{et al.} proposed for the first time the utilization of \glspl{cnt} as plasmonic nano-dipole antennas~\cite{burke2006quantitative}. In 2010, Jornet and Akyildiz proposed the utilization of \glspl{gnr} as nano-patch antennas and compared its performance to nano-dipoles~\cite{jornet2010graphene}. In 2012, Llatser \emph{et al.} studied the properties of graphene-based nano-antennas utilizing the conductivity model for infinitely large graphene sheets~\cite{llatser2012graphene}. In 2013, Jornet and Akyildiz developed a new conductivity model able to cover both \glspl{gnr} and graphene sheets and derived analytically the performance of nano-antennas in the terahertz band. These works are at the basis of electromagnetic nanonetworks.

Besides small footprints, graphene-based nano-antennas offer additional interesting properties. They are highly tunable. In particular, electrostatic bias or chemical doping can modify the propagation speed of \gls{spp} waves, effectively changing the plasmonic wavelength. As a result, the antenna resonant frequency can be changed~\cite{perruisseau2013graphene} or, at a fixed frequency, the antenna's phase and radiation diagram can be modified in near real-time~\cite{tamagnone2012reconfigurable}. Moreover, the mutual coupling between plasmonic nano-antennas is determined again by the plasmonic wavelength~\cite{zakrajsek2017design}, and this being much smaller than the free space wavelength, opens the door to very high-density antenna arrays~\cite{singh2020design}. These antenna arrays can be leveraged to increase the communication range of nanomachines or directly as part of macroscale applications of the terahertz band~\cite{akyildiz2016realizing}.

In addition to the antennas, graphene can also be leveraged to build nano-transceivers, many times in conjunction with silicon, III-V semiconductor materials, such as~\gls{ingaas}, \gls{gaas} and \gls{gan}, and other \gls{2d} nanomaterials, such as~\gls{hbn} and \gls{mos2}. For example, graphene can be utilized as the \gls{2deg} channel in a \gls{hemt}. When creating asymmetric boundary conditions between the source and drain of the \gls{hemt}, on-chip terahertz \gls{spp} waves can be generated through the Dyakonov-Shur instability~\cite{crabb2021hydrodynamic, crabb2022plasma}. The same structure can be used as a direct on-chip terahertz amplitude and frequency modulator~\cite{crabb2022amplitude}. Similarly, by leveraging the tunability of graphene, a fixed-length plasmonic waveguide can be utilized as an on-chip phase modulator~\cite{singh2016graphene}. The source, modulator, and antenna can all be integrated and co-designed in a single device~\cite{crabb2023chip}.

In nanonetworking applications where size constraints are important but not the main limiting factor, non-plasmonic terahertz systems can be adopted. Several groups \cite{Park2012, Yu2014, Thyagarajan2015, Fritsche2017, Tokgoz2018, Byeon2020, Yi2021, callender2022fully, agrawal2023128} have pushed the envelope in terms of frequency (towards 300 GHz) and speed (over 100 Gb/s), also seeking to improve their bandwidth density (over 100 Gb/s/mm\textsuperscript{2}) and energy per bit (below 1 pJ/bit) to fit ultra-short range applications with high speed requirements. In this context, graphene has also been explored to realise RF circuits such as amplifiers, power detectors, rectifiers, frequency multipliers, oscillators, mixers, and receivers with improved characteristics by virtue of the graphene's high charge carrier mobility and saturation velocity \cite{saeed2018graphene, Saeed2021, hamed2020graphene}. 



While significant progress has been made in relation to graphene plasmonics, the experimental development of such devices is still in its early stages. Different techniques are being explored for the large-scale and low-cost fabrication of graphene and graphene-based heterostructures, ranging from photolithography-based methods in a top-down approach to self-assembly methods in a bottom-up approach~\cite{you2020laser, rudrapati2020graphene, wittmann2023assessment}. 

Besides the device technology, the adoption of the terahertz band for nanonetworking applications is also impacted by the propagation of terahertz waves in different media~\cite{akyildiz2022terahertz, lemic2021survey, kokkoniemi2016wideband}. For example, terahertz signals propagate very well through silicon and, thus, can be adopted in computing applications (\cite{abadal2019wave}, see Sec.~\ref{subsec:computing}). However, terahertz radiation is absorbed by most liquids, starting with water, which can compromise its use in biomedical applications of nanonetworks (\cite{elayan2017terahertz, elayan2017multi}, see Sec.~\ref{subsec:nanothings}). While terahertz signals propagate better than optical signals~\cite{750kokkoniemi2016frequency}, adopting even lower frequencies is desirable when, for example, much longer transmission distances or propagation through biological tissues, including the human body, is pursued.

\noindent
\subsubsection{Radio Frequencies: Magnetoelectric Antennas}
Magnetoelectric antennas are becoming increasingly popular as an alternative to metallic and plasmonic nano-antennas. They are based on multiferroic materials that exhibit piezoelectricity and exploit the magnetoelectric effect. Their fundamental working principle is as follows. In transmission, a modulated voltage applied across the piezoelectric component of the magnetoelectric antenna induces a strain or mechanical deformation within the piezoelectric material. Due to the magnetoelectric effect, the strain in the piezoelectric material creates a corresponding magnetic field in the vicinity of the antenna. The time-varying magnetic field, following the principles of electromagnetism, generates a radiating electromagnetic wave. This electromagnetic wave carries the information encoded in the original electrical signal. Reciprocally, in reception, an incoming electromagnetic wave induces a magnetic field in the magnetoelectric antenna. This field creates a corresponding strain, ultimately generating a voltage output from the piezoelectric component.

Magnetoelectric antennas can be dramatically smaller than the traditional antennas operating at the same frequency. This is because the relationship between antenna size and electromagnetic wavelength does not limit them. For example, micrometric antennas can be designed to resonate at tens to hundreds of MHz~\cite{nan2017acoustically, zaeimbashi2019nanoneurorfid}. A traditional metallic resonant antenna at the same frequency would be nearly 3~m. The size advantage makes them attractive for applications where space is limited, such as in wearable technology, implantable bioelectronics and, overall, many of the relevant application scenarios of nanonetworks.

\begin{figure*}
\centering
\includegraphics[width=0.7\textwidth]{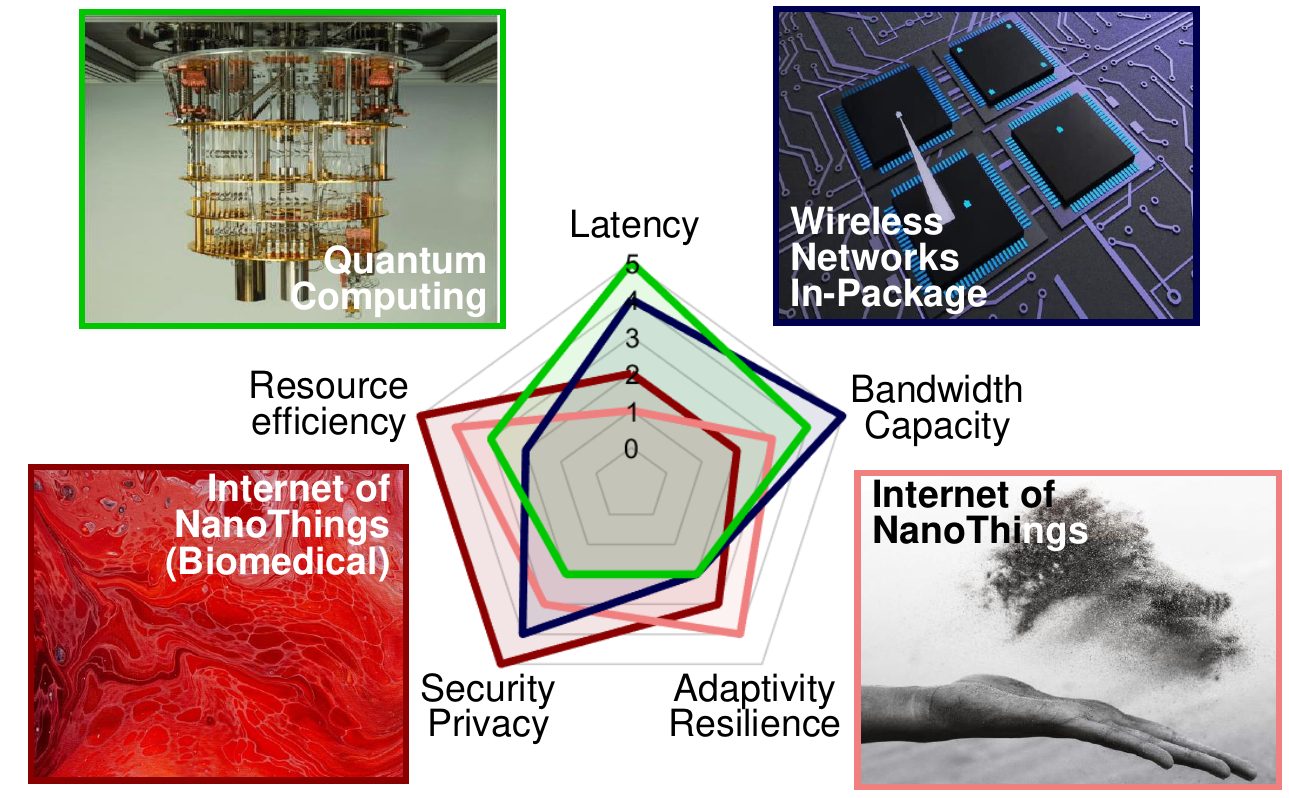}
\caption{Coarse assessment of the communication requirements of the four nanonetworking application areas analyzed in this survey paper.}
\label{fig:apps}
\end{figure*}

\Gls{nems} technology provides a platform for miniaturized and efficient magnetoelectric antennas~\cite{will2022tutorial}. The field of \gls{nems} deals with the design and fabrication of mechanical devices on the nanoscale (1-100~nm) with precise control over their dimensions and properties. As a result, \gls{nems} technology enables the fabrication of magnetoelectric antennas with significantly small dimensions compared to typical resonant antennas, particularly at low frequencies. Moreover, the miniaturization achieved through \gls{nems} can also improve efficiency in some cases, as smaller structures can experience less energy loss due to factors like ohmic losses and current crowding.

The field of magnetoelectric antennas is still in its infancy. Two main challenges to overcome include finding suitable materials with strong magnetoelectric coupling at desired frequencies and developing large-scale manufacturing techniques.  



\subsection{Non-Radiative Technologies} 


In addition to the electromagnetic technologies discussed until this point, there are other non-radiative technologies that are often presented as enablers of electromagnetic nanonetworks, namely, coupling techniques and ultrasounds.

\textbf{Coupling technologies} is an umbrella term under which we include different techniques suitable for nanonetworks, including inductive or magnetic coupling, capacity coupling, and galvanic coupling~\cite{Vizziello1, Vizziello2}. In\textit{ inductive coupling}, a magnetic field between two coils is utilized to carry information \cite{GC6}. 
For example, for a micrometric coil system, the expected resonant frequency is in the order of hundreds of MHz \cite{MDPI2020}. 
In \textit{capacitive coupling,} the electric field between two conductors or electrodes is utilized for the exchange of information. Similarly, for micrometric electrodes, the expected operation frequency is in the hundreds of MHz \cite{WCMC}. 
In \textit{galvanic coupling,} modulated electrical currents are transferred between electrodes, utilizing a conductive medium (e.g., the human body in the context of body area networks). The operation frequency of this technique is similar to that of other coupling techniques.

\textbf{Ultrasounds} or acoustic waves at ultrasonic frequencies (i.e., above the audible limit of 20~kHz), have also been discussed in the context of nanonetworks~\cite{Galluccio1, Galluccio2}. \gls{nems} can be utilized to both generate and detect pressure waves at the nano and micro scales. When modulated, such waves can be utilized to exchange information between nanomachines, specially when in a liquid or solid medium, such as the human body. 

Finally, it is relevant to note that both coupling techniques and ultrasounds are often presented in the context of nanonetworks not only as a potential communication technique but as a wireless power transfer or energy harvesting technology~\cite{jornet2012joint,harvest1,harvest2,harvest3,GC6,15}.
Harvesting energy from the environment serves as a crucial enabler for the establishment of long-lasting nanonetworks. In particular, developing wireless devices, especially suitable for implanted medical applications at sub-millimeter scale, capable of operating at depths extending centimeters into tissues, presents significant challenges in both power delivery and data transmission.




\section{Emerging Novel Applications of Electromagnetic Nanonetworks}
\label{sec:apps}

In this section, we review and describe the key cutting-edge applications enabled by nanocommunications and nanonetworks, as illustrated in Figure~\ref{fig:apps} and summarized in Table~\ref{tab:apps}. Selected nanonetwork proposals for each application are listed in Table~\ref{tab:appsAndTechs} at the end of the section.

\begin{figure*}[!ht]
\centering
\includegraphics[width=\textwidth]{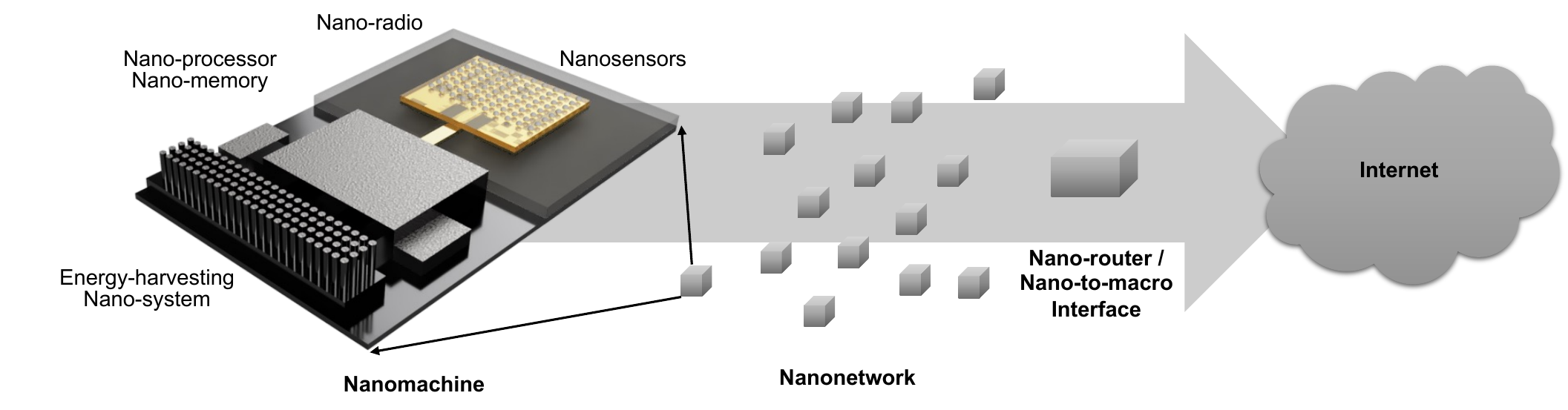}
\caption{The Internet of Nano-Things. From left to right: individual nanomachines communicate with each other creating nanonetworks; through a nano-router that serves as a nano-to-macro interface, nanonetworks can be connected to the macro-scale networks and eventually the Internet.}
\label{fig:iont_architecture}
\end{figure*}

\begin{table*}[ht]
\centering
\caption{Summary of the key aspects and challenges of the application areas described in Section \ref{sec:apps}.}
\label{tab:apps}
\begin{tabular}{|p{3.2cm}||p{3.8cm}|p{4.5cm}|p{4.5cm}|}
\hline
\textbf{Application Areas} & \textbf{Enabling Technologies} & \textbf{Key Communication Requirements} & \textbf{Salient Challenges} \\ \hline\hline
\textbf{Internet of NanoThings (biomedical)} & Energy harvesting, nano-sensors, bio-compatible electronics & Privacy, ultra-high resource efficiency & Biocompatibility, intermittence, propagation through biological media \\ \hline
\textbf{Internet of NanoThings (other)} & Energy harvesting, nano-sensors & Ultra-high resource efficiency, scalability & Intermittence, mobility, changing propagation media, swarm-like coordination \\ \hline
\textbf{Wireless Networks within Computing Packages} & High-speed and fully-integrated mixed-signal circuits & Low latency, high speed, wire-like reliability & Security, enclosed propagation environment, co-integration with processors \\ \hline
\textbf{Quantum Computing} & Cryo-electronics (digital and RF) & Ultra-low latency, energy consumption & Operation $<$4K, enclosed environment, interference with RF-operated qubits \\ \hline
\end{tabular}
\end{table*}


\subsection{The Internet of Nano-Things}
\label{subsec:nanothings}

The \gls{iont} is a cyber-physical system that enables the wireless exchange of information between nanomachines and macroscale networks, processes, and users~\cite{akyildiz2010internet}. Nanomachines are embedded nano-systems with limited sensing and actuation, computing and data storing, and communication capabilities, often powered by an energy harvesting system\footnote{The \gls{iobnt}~\cite{akyildiz2015internet}, the biological counterpart of the \gls{iont}, in which nanomachines are built using living cells and communicate through molecular communications, is out of the scope of this manuscript.}. The volume of the complete nanomachine is expected to be in the order of a few cubic micrometers. As shown in Figure \ref{fig:iont_architecture}, nanomachines can talk to each other directly or through nano-routers, which can serve as nano-to-macro interfaces and gateways to connect to macroscale networks. While individual nanomachines cannot do much, large networks with thousands or even millions of miniature devices can accomplish complex tasks in a distributed manner.

The applications of the \gls{iont} are diverse and posed to revolutionize many fields. Environmental monitoring \cite{environmentsurvey, environmentsurvey1} stands to benefit greatly from the~\gls{iont}, with nanodevices capable of detecting pollutants, monitoring ecological systems, and mitigating environmental risks with unparalleled precision. For example, the utilization of nanomachines to perform over-the-air spectroscopy and so detect the presence of greenhouse gases that accelerate climate change is proposed in~\cite{wedage2023climate}. In manufacturing and logistics \cite{manufacturing}, the \gls{iont} facilitates the optimization of supply chains, asset tracking, and quality control through the deployment of nanoscale sensors for real-time monitoring and process optimization. Agriculture~\cite{agrisurvey, agrisurvey1} stands poised for a technological revolution, with the \gls{iont} enabling precision agriculture through soil monitoring, crop health assessment, and optimized resource utilization. Furthermore, the \gls{iont} holds promise for enhancing safety and security in smart cities \cite{smartcities}, with nanoscale sensors monitoring infrastructure, detecting hazards, and optimizing urban systems for efficiency and resilience.

\begin{figure*}[!ht]
\centering
\includegraphics[width=0.8\textwidth]{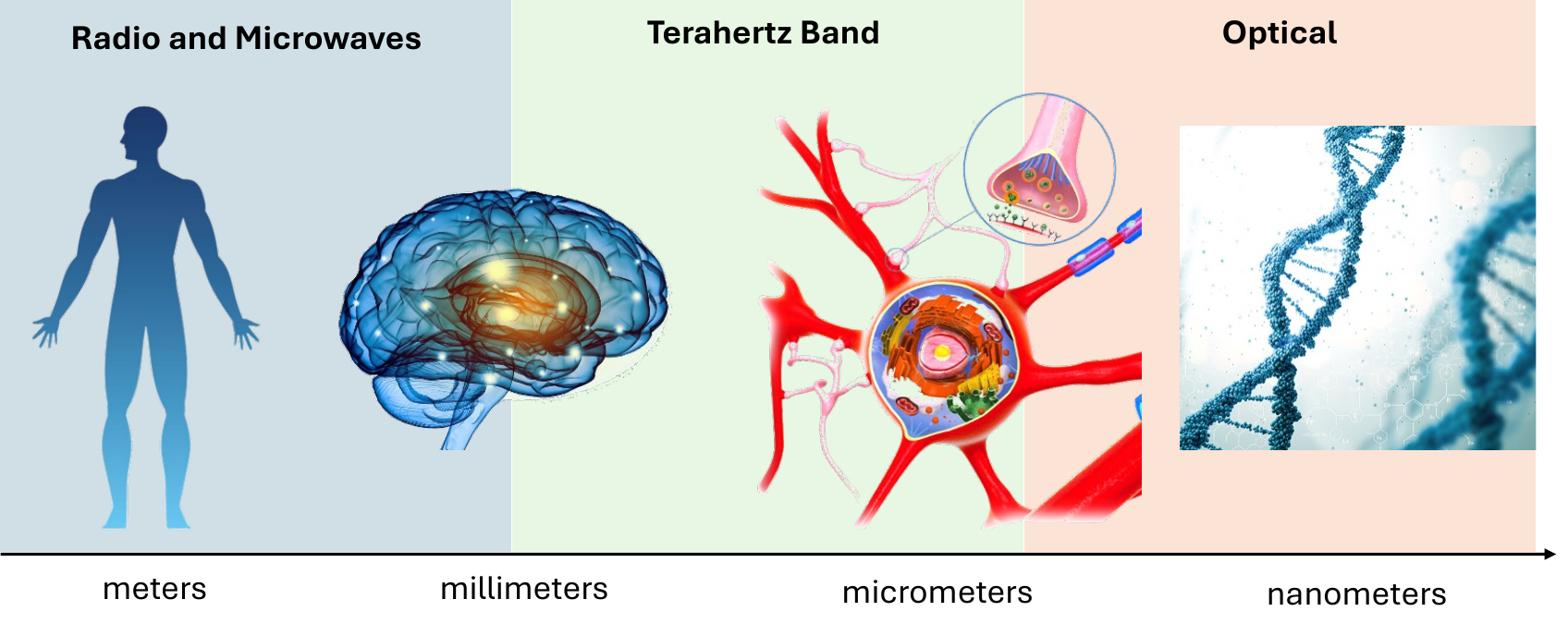}
\vspace{-0.2cm}
\caption{Interaction of electromagnetic radiation with biological systems, as a function of frequency. From left to right: radiowaves and microwaves perceive the human body as a single entity; terahertz-band radiation interacts with the human body at the organ and tissue levels; optical radiation can interact with cells and their building blocks.}
\label{fig:bio_interactions}
\end{figure*}

One critical application domain of the \gls{iont} is healthcare~\cite{yang2020comprehensive}. The \gls{iont} offers unprecedented opportunities for personalized medicine, with nanoscale sensors and actuators enabling real-time monitoring of vital signs, early detection of diseases including cancer, brain-machine interfaces to restore or repair neuronal damage, and targeted drug delivery within the human body, among many others. 
First, it is relevant to note that the field of body area networks has been one of the fastest-growing paradigms in the last decade~\cite{vizziello2023intra}. Still, today, existing demonstrated solutions rely on millimetric devices, i.e., orders of magnitude larger than the \gls{iont} vision. This directly impacts their invasiveness, limiting their applicability to primarily wearable devices or sub-cutaneous implantation. 

Moving from body area networks to the true \gls{iont} opens the door to applications that are not possible at a larger scale, not only because of the size but also because of how higher frequency (smaller wavelength) electromagnetic radiation interacts with the human body (see Figure~\ref{fig:bio_interactions}). For example, nanomachines can be deployed in the human brain to control cellular activity through optogenetic and optogenomic techniques. In optogenetics, blue light is utilized to control the flow of action potential signals between neurons that have been transfected with channel rhodopsins~\cite{deisseroth2011optogenetics}. While classical studies generally rely on the use of millimetric lasers to illuminate neuronal tissues~\cite{93,Park2015}, often in conjunction with optical fibers to guide the light towards the target, nanomachines equipped with optical nano-radios can be deployed in the brain (for example, through nasal injection) and utilized to target individual neurons or small group of neurons~\cite{balasubramaniam2018wireless,wirdatmadja2017wireless}. In optogenomics, two different red wavelengths are used to actuate a molecular toggle switch that can induce or stop the expression of specific genes~\cite{jornet2019optogenomic}. The genome is the set of genetic instructions written in the DNA of every cell in the human body. Different parts of the genome describe different operations that cells might perform. The ability to control the genome results in the possibility of reprogramming cells, including restoring functionalities that have been lost due to neurodegenerative diseases, for example, or due to cancer agents.

Besides the technology-related aspects discussed in Sec.~\ref{sec:tech}, there are many challenges that need to be addressed by the communication and networking communities. First, when adopting optical signals for actuation (e.g., optogenetics and optogenomics), sensing (e.g., through fluorescent proteins), or direct communications between nanomachines, it is essential to note that the human body is a challenging propagation environment. Indeed, the very small wavelength of optical signals (hundreds of nanometers) makes cells and their components appear as obstacles. Nevertheless, it has been experimentally demonstrated that different types of cells can act as focusing lenses (e.g., red blood cells~\cite{miccio2015red,johari2017nanoscale}) or guiding waveguides~\cite{wirdatmadja2019analysis}. Second, when utilized for actuation, the type of optical modulations are set by biology-set protocols (e.g., nano-lasers are switched on and off at fixed intervals for a given time). In the case of sensing, the use of chirp-based modulations as the basis of innovative joint intra-body communication and sensing systems has been recently proposed~\cite{sangwan2022joint}.

Some of these challenges can be addressed by moving to the terahertz band. The larger wavelength of terahertz signals (tens to hundreds of micrometers) results in lower spreading and scattering losses. However, the photon energy of terahertz radiation can induce resonant modes in different types of molecules, including proteins, through absorption~\cite{elayan2017terahertz, elayan2023terahertz}. Molecular absorption in free space, where gaseous molecules can freely vibrate, results in a propagation loss, as electromagnetic energy is converted into kinetic energies within the molecules. However, in the human body, where molecules cannot freely vibrate, the friction between the molecules and the solid or liquid media results in heat through photothermal effects~\cite{elayan2017photothermal}. Such photothermal effects ultimately constrain the total radiated power and the spatial illumination and temporal modulation patterns to meet safety limits. Moving from optical to terahertz signals allows the decoupling of the sensing and actuation from the communication processes. Finally, one could adopt magnetoelectric \cite{yu2022magnetoelectric, gong2024} and coupling techniques to operate at significantly lower frequencies. In this case, the much-expected better propagation of the signals will bring interesting medium access control and networking challenges.

Independently of the frequency of operation, an additional aspect frequently present in the \gls{iont} has to do with mobility. While in many scenarios the nanomachines are expectedly fixed in place (e.g., implanted in the body, weaved in the fibers of clothing, or embedded in the paint of a room), there are scenarios in which nanomachines might be uncontrollably moving and whose turbulent and swarm dynamics might require a holistic multiphysics approach to guide the design of systems. For instance, nanomachines in environmental applications might be transported by the air. Another example, this time in the context of intra-body networks, relates to the active transport of nanomachines within the circulatory system~\cite{canovas2020understanding, garcia2023dynamic, gomez2023optimizing}. Ultimately, the new mobility models resulting from the physics that transport nanomoachines inside or outside the human body require the study and optimization of multi-hop communication and routing for nanonetworks.

\subsection{Wireless Networks within Computing Packages}
\label{subsec:computing}
In the last decade, the field of computer architecture has transitioned from monolithic, general-purpose, single-chip processors towards disintegrated, specialized, multi-chip architectures. In this direction, recent years have seen the emergence of the concept of chiplet, small chips that are interconnected through a Printed Circuit Board (PCB) or an interposer chip. Traditionally, Network-on-Chip (NoC) has been the \emph{de facto} interconnect fabric within chips to satisfy the internal communication needs of computer processors \cite{Nychis2012}. Yet with the arrival of multi-chiplet architectures, the concept of NoC had to be extended to incorporate off-chip links and form the Network-in-Package (NiP) paradigm \cite{Shao2019}. 

NoCs are packet-switched networks of integrated routers and wires typically arranged in a grid topology; an approach with important drawbacks when scaled beyond a handful processors. For example, the latency and energy consumption of chip-wide and broadcast communications increases significantly due to the number of hops needed to reach the destination(s) through the grid \cite{Nychis2012}. 
When shifting to multi-chiplet systems, NiPs inherit the problems of NoCs and add new ones such as a further increase of the latency when going from chip to chip, or the necessity to stay with grid topologies because the fan-out of chiplets is limited by the amount of connection pins used for chip-to-chip links.

\begin{figure}
  \centering
  \includegraphics[width=0.9\columnwidth]{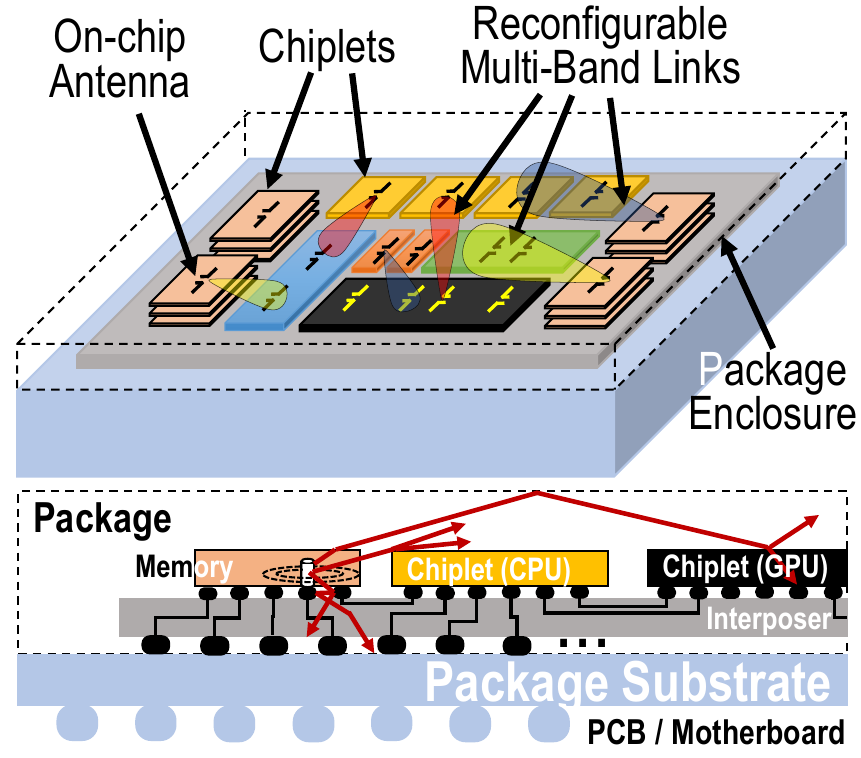}
    \vspace{-0.2cm}
\caption{Schematic diagram of a System-in-Package (SiP) composed of a heterogeneous set of chiplets interconnected via a silicon interposer Network-in-Package (NiP) augmented with a multi-band and reconfigurable electromagnetic nanonetwork within a computing package. Bottom panel shows a cross-section of the architecture, together with the different propagation mechanisms within the package.}
  \label{fig:vision}
\end{figure}

In this context, several research lines have proposed to use wireless links from millimeter-wave to optical frequencies to form connections within and across the chips of a computing system \cite{Shamim2017, Yi2021, abadal2022graphene, nafari2017chip, Petrov2017}. Figure~\ref{fig:vision} illustrates such an electromagnetic nanonetworking vision, wherein tiny antennas and transceivers are co-integrated with the computing elements and use the system package as the wireless propagation medium. The wireless links form a network within and across chiplets that becomes a natural complement to the rigid yet effective wired NiPs.

With these in mind, wireless nanonetworks offer multiple benefits. By not needing to lay down extra wires between the chips, wireless NiPs can bypass pin limitations or wire routing constraints. The newly available bandwidth can be shared dynamically to adapt to the needs of each computing system. The latency, which is critical in this application, can be reduced by an order of magnitude, especially in transfers that would otherwise require the traversal of long multihop paths within the dense maze of wired interconnects. Broadcast communications are a very relevant example of this: they require flooding the network in wired chip-scale networks, while they become scalable in wireless nanonetworks due to the inherent broadcast capabilities of wireless communications.

The use of graphene in this application scenario turns a wireless interconnect into an actual electromagnetic nanonetwork. As hinted in Section \ref{sec:tech}, graphene allows to develop compact, beam-steerable, and frequency-tunable antennas as well as faster and more efficient transceivers. This could lead to significant gains in capacity and network-level flexibility, which can be exploited in computing systems through a judicious orchestration of space, time, and frequency channels. 

The chip-scale scenario exhibits a unique combination of unexplored aspects, stringent constraints and high performance requirements. Indeed, nanonetworks are typically associated to very stringent resource constraints (i.e. area and power), but not as much to high performance in latency and capacity. Yet in the case of electromagnetic nanonetworks within computing systems, performance targets are around 1--10 ns and 10--100 Gb/s with extremely low error rate, far away from the aims of IoNT applications. Further, the chip-scale THz channel remains largely unexplored, as very few works have studied wave propagation from millimeter-wave to optics in enclosed packages like those of modern computing systems \cite{Timoneda2018ADAPT, abadal2019wave}. Clearly, this prevents the direct application of channel models and protocols used in other scenarios and calls for novel, opportunistic, and highly optimized solutions instead. 

In the envisaged scenario, physical layer protocols are challenged by three main constraints. First, coherent schemes are relatively costly due to the need for bulky and power-hungry components such as Phase-Locked Loops (PLLs). Second, the channel is prone to (static) multipath \cite{Timoneda2018ADAPT}, which could be exploited with techniques such as channel shaping \cite{imani2021smart} or time reversal \cite{rodriguez2023collective}. Third, the bit error rate required in this scenario is $\sim$10\textsuperscript{-12}. At upper layers of design, the protocols must manage the reconfigurability properties of the graphene antennas, if available. This represents an open issue as none of the existing wireless chip-scale networks proposals, e.g. \cite{Shamim2017}, support dynamic beam-steering or frequency tuning simply because conventional on-chip antennas cannot have such capabilities. In other words, there are no protocols in the literature that can orchestrate the space-time-frequency channels offered by graphene antennas with the simplicity needed at the chip scale, while adapting to the traffic requirements. 

Finally, it is worth noting that, in the wireless NiP context, the communications serve a compute purpose and the end goal is to minimize the execution time of a given application. Hence, a computer science perspective is crucial to fully exploit the benefits of wireless communication. In particular, a cross-disciplinary approach where the architecture guides the MAC protocol operation, instead of the protocol blindly trying to adapt to the traffic, is promising and unique to this monolithic scenario. This is the strategy followed in \cite{jog2021one}, where a MAC protocol was designed to give priority to latency-sensitive messages from the computer architecture; or like in \cite{fernando2019replica}, where protocols were simplified (and made faster) exploiting the fact that certain compute applications are tolerant to communication errors.

\begin{figure*}
     \centering
     \includegraphics[width=0.83\textwidth]{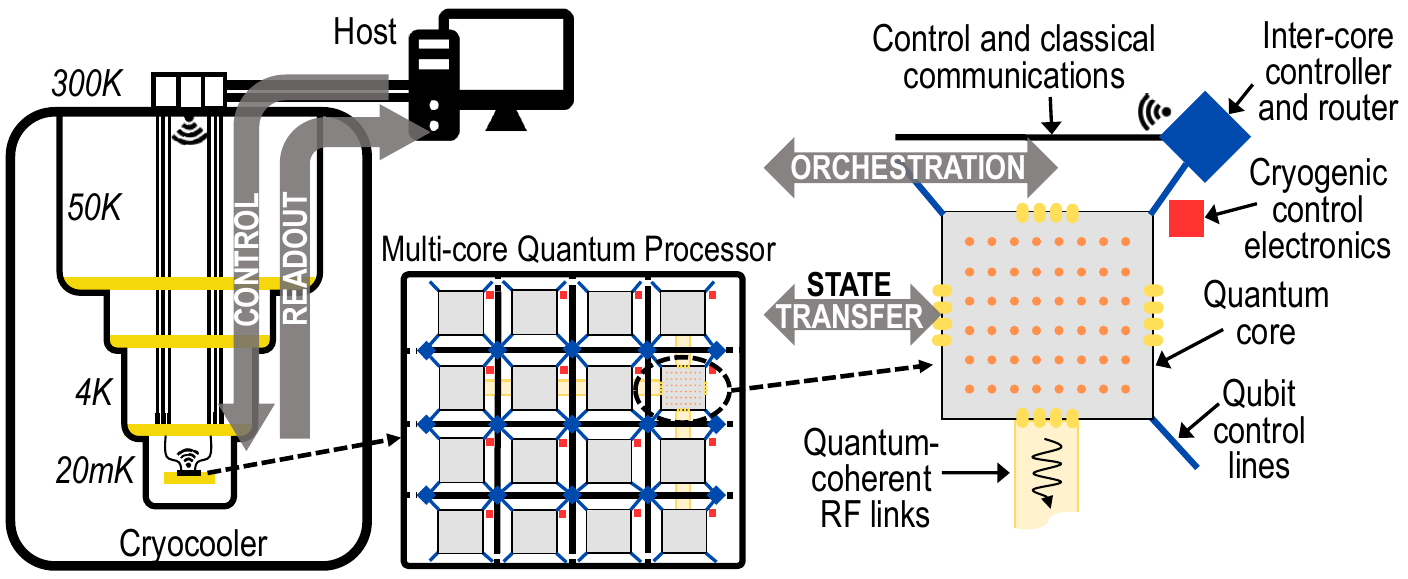}
     \vspace{-0.2cm}
     \caption{Overview of a possible system architecture and communication flows of a modular quantum computer with cores residing in different chips. Electromagnetic nanonetworks might be used vertically, for the wireless control and readout of qubits \cite{Wang2023}, or horizontally for the state transfer of qubits via waveguided quantum-coherent links and for their orchestration via wireless links between quantum cores \cite{alarcon2023scalable}.} 
     \vspace{-0.1cm}
     \label{fig:comms}
\end{figure*}

\subsection{Quantum Computing}
\label{subsec:quantum}
Quantum computing has heralded a revolution in computer science thanks to its promise to solve computational problems that are classically intractable, which has huge implications in fields such as physics \cite{Low_2019}, chemistry \cite{Peruzzo_2014}, finance \cite{Woerner_2019}, and healthcare \cite{fi15030094}. Nevertheless, for this potential to be realized, quantum computers need to scale to millions of qubits, where a qubit is the fundamental unit of information in quantum computing. Unfortunately, current state-of-the-art computers only have a thousand qubits \cite{choi2023ibm} and scaling three orders of magnitude towards the million barrier is extremely complex due to multiple technological issues. 

Among the strategies proposed to scale quantum computers, there is the quantum multi-core approach \cite{rodrigo2021double, smith2022scaling, jnane2022multicore}, which consists in interconnecting multiple quantum processing units or \emph{quantum cores} with links that can handle both classic information and quantum states \cite{gold2021entanglement}. Hence, the interconnect fabric within such quantum architectures emerges as a critical sub-system as the number of quantum cores increases.

The general system architecture of a multi-core quantum computer is shown in Figure~\ref{fig:comms} \cite{escofet2023interconnect}. It consists of a host computer containing the circuits to control the execution of a quantum circuit, and a set of quantum cores containing the qubits where the quantum circuit is actually executed. On the one hand, the host computer is in charge of running the compiled circuit, which implies (i) sending the required signals to the quantum computer to apply quantum gates or to read out a particular set of qubits at the required instants, and (ii) receiving the result of the readout operations. On the other hand, the quantum cores essentially consist of an array of qubits and the necessary circuits to route the control and readout signals to each specific qubit. Being qubits possibly placed in multiple chips, extra circuitry is required to transfer the quantum state of the qubits across chips and to orchestrate their operation, i.e. to avoid conflicting transfers. Hence, we can conclude that multi-core quantum computers implement two types of communication, namely, communications between the host computer and the quantum cores, and across the quantum cores. We refer the reader to \cite{escofet2023interconnect, rached2024spatio, rodrigo2022characterizing} for more details on the different communication flows in a quantum computer.


Multiple technologies have been proposed to implement qubits \cite{nakamura_1999_coherent, pieter_2007_linear, PhysRevLettQDS, PhysRevLettTI}, each one imposing its own means for control and readout, as well as for quantum state transfer \cite{kaushal2019shuttlingbased, llewellyn2020chip, c877b567b53e44cda1a0346b58a90d0f, marinelli2023dynamically}. Several of them, however, have in common that Radio-Frequency (RF) signals can be used for the aforementioned functions. For instance, superconducting qubits are manipulated via precisely shaped RF pulses at the qubit microwave frequency, which is typically 4-8 GHz \cite{Krantz_2019}, thus requiring a maze of coaxial cables (possibly driven by a network of RF switches \cite{potovcnik2021millikelvin}) to connect the host computer with the quantum cores, although recent works advocate for optical signals through fibers for control and readout \cite{lecocq2021control}. For the state transfer of qubits across quantum cores, one might use cryogenic waveguides connecting the different chips \cite{magnard2020microwave}, possibly at frequencies closer to the THz band to miniaturize the form factor of the waveguides. Therefore, a quantum computer can be seen as an electromagnetic nanonetwork, albeit most likely a wired one.


The role of wireless technology for the creation of electromagnetic nanonetworks is yet unclear in this context, but it has been proposed in two fronts. First, as a means for the readout of complete quantum core from the host computer without the need for a bulky, expensive, and non-scalable set of coaxial cables. Precisely, Wang \emph{et al.} recently proposed an ultra-low power terahertz backscattering system to this end \cite{Wang2023}. Second, wireless communications have also been proposed to implement the classical communications backbone necessary to orchestrate the different quantum state transfers of a multi-core quantum computer. In particular, Alarc\'on \emph{et al.} \cite{alarcon2023scalable} argue that the ultra-low latency requirements of these computers, together with the opportunity arising from the existence of RF circuits on-chip to drive the qubits, pose wireless technology as a suitable candidate.


\begin{table*}[ht]
\centering
\caption{Summary of references for each combination of application area and communication technology and frequency band.}
\vspace{-0.2cm}
\label{tab:appsAndTechs}
\begin{tabular}{l|c|c|c|}
\cline{2-4}
                                         & \textbf{Internet of NanoThings} &  \textbf{Wireless Networks within Package} & \textbf{Quantum Computing}\\ \hline

\multicolumn{1}{|l|}{\textbf{Optical}}   
& \cite{jornet2019optogenomic, 93,Park2015}                
&  \cite{nafari2017chip, calo2022reconfigurable}               
&  \cite{lecocq2021control}                
\\ \hline
\multicolumn{1}{|l|}{\textbf{THz}}       
&  \cite{elayan2017terahertz, elayan2017photothermal}              
&  \cite{Park2012, Yi2021, abadal2022graphene, Petrov2017, DiTomaso2015}             
&  \cite{Wang2023}             
\\ \hline
\multicolumn{1}{|l|}{\textbf{mmWave}}    
&  \cite{pellegrini2013antennas}           
& \cite{Shamim2017, Yu2014, DiTomaso2015, extra_chang2008, extra_Mineo2016, Baniya2019, Duraisamy2016, gade2019}  
& \cite{alarcon2023scalable}            
\\ \hline
\multicolumn{1}{|l|}{\textbf{Microwave}} 
& \cite{yu2022magnetoelectric, gong2024, 37, 38}              
& N/A             
& \cite{Krantz_2019, potovcnik2021millikelvin, magnard2020microwave}              
\\ \hline \hline
\multicolumn{1}{|l|}{\textbf{Non-Radiative}}    
& \cite{Galluccio1, Galluccio2}               
& N/A              
& N/A              
\\ \hline
\end{tabular}
\end{table*}

The creation of electromagnetic nanonetworks within quantum computers is obviously not exempt of significant challenges requiring a multi-disciplinary physics-driven approach \cite{ganguly2022interconnects, escofet2023interconnect}. Next, we revisit some of them:
\begin{itemize}
    \item \textbf{Cryogenic operation:} Most quantum computers nowadays rely on cryogenic operation, placing the qubits inside a cryocooler at temperatures below 4K. This requires a careful revisit of not only the channel models, as thermal noise is reduced dramatically, but also of the RF circuit design, now that most RF device non-idealities are largely attenuated.
    \item \textbf{Enclosed environment:} Communications happen, similarly to in chip-scale networks, within a heavily enclosed environment and inside a metallic cryocooler. This implies that RF fields will be spread across the entire cryocooler, bouncing back and forth, which has a doublefold effect in quantum computers. On the one hand, it increases the delay spread significantly, limiting the speed of wireless links; on the other hand, the RF signals used to communicate might leak inside the qubit cavities and quantum-coherent link waveguides, interfering them in ways that are not yet fully understood.
    \item \textbf{High performance:} Communications are very latency sensitive, since qubits decohere and lose information as time passes and, hence, any delay can lead to erroneous computations. Moreover, the bandwidth requirement generally increases with the number of qubits, reaching the Tb/s barrier as we scale towards millions of qubits \cite{ganguly2022interconnects}.
    \item \textbf{Ultra-low power:} Cryocoolers have a limited power budget associated to their limited heat dissipation capacity. Since very high communication speeds are envisioned, ultra-low energy consumption towards the fJ/bit level is also expected.
    \item \textbf{Variability:} Qubits and circuits may behave differently in different chips due to common fabrication process variations \cite{smith2022scaling}. This leads to a diverse range of error profiles that interconnect designers must consider when designing the communications stack.
\end{itemize}

\section{Cross-Cutting Issues}
\label{sec:discussion}

\begin{figure*}[!ht]
\centering
\includegraphics[width=0.6\textwidth]{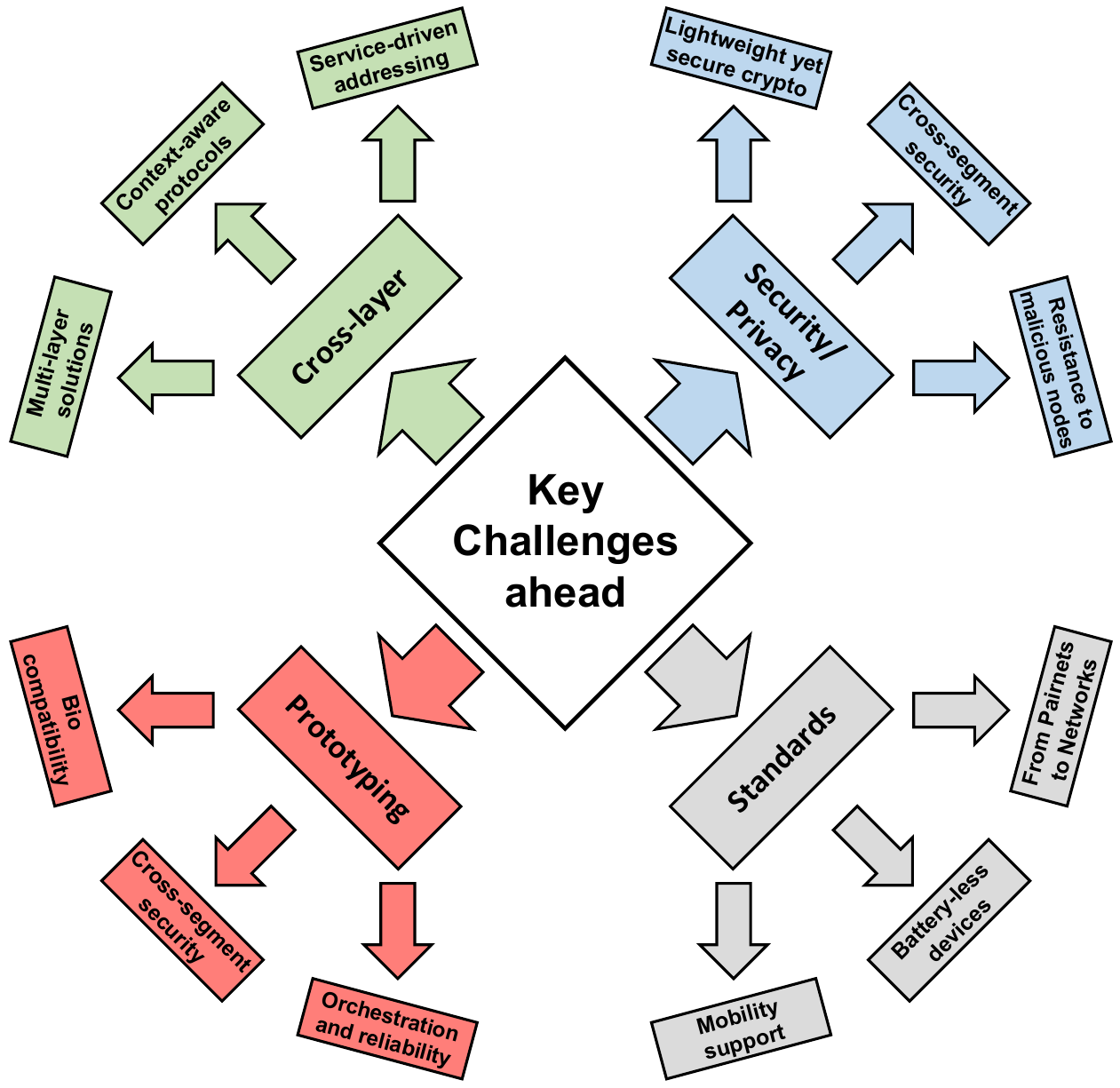}
\caption{Key challenges from the current proof-of-concept stage toward operational and commercially viable electromagnetic nanonetworks as a part of the \gls{iont}. 
}
\label{fig:challenges}
\end{figure*}

In this section, we review some challenges and implementation aspects that pierce through the protocol stack of electromagnetic nanonetworks. An overview and key highlights are also illustrated in Figure~\ref{fig:challenges}.

\subsection{Cross-layer Design}

In the realm of nanonetworks, the traditional protocol stack model faces significant challenges due to the unique characteristics of nanodevices and the environment in which they operate. Therefore, in connection with the cross-disciplinary nature of nanonetworks, a crucial approach to address these challenges is the adoption of cross-layer design principles. By breaking the barriers between conventional protocol layers and allowing for dynamic interactions and information exchange among different layers, cross-layer design enables the optimization of network performance and efficiency in nanonetworks. Cross-layer design facilitates adaptive and context-aware communication protocols tailored to the requirements and constraints of nanodevices through the integration of physical, data link, network, and application layers. This holistic approach fosters enhanced reliability, energy efficiency, and scalability, paving the way for achieving robust and sustainable applications in the IoNT.

Specifically, the proposed solutions for the physical, data link, and network layers of electromagnetic nanonetworks are interrelated and heavily tailored to the nano-device capabilities and the channel peculiarities. For example, the use of hundred-femtosecond-long pulses by following an On-Off Keying modulation and spread in time (TS-OOK)~\cite{jornet2014femtosecond} was proposed as a way to minimize the nano-transceiver complexity while leveraging the very large bandwidth of the terahertz channel. Building on this, an innovative low-sampling-rate (LSR) synchronization algorithm is proposed in~\cite{han2017sync} for terahertz pulse communications, by extending the idea of sampling signals with finite rate of innovation from compressive sampling in signal processing to the communication context. 
Specifically, to use the features of the annihilating filters, it is 
possible to reliably decode a received signal by sampling it at or above the rate of innovation, rather than at the sampling rate dictated by the bandwidth of the transmit signal.

Interestingly, when utilizing TS-OOK, the channel can no longer be modeled as a binary symmetric channel: the probability of decoding a logical ``1'' (transmitted as a THz pulse) is \emph{not} equal to the probability of decoding a logical ``0'' (transmitted as silence). Further, when several nodes use TS-OOK, this leads to a non-symmetric interference picture, as logical ``1''s create interference, while logical ``0''s do not~\cite{hossain2019stochastic}. This motivated the development of new low-weight coding schemes~\cite{jornet2011low} that prioritize the transmission of logical ``0"s over ``1"s to reduce noise and interference and, ultimately, prevent errors from happening. Ultimately, the spreading rate of the symbols in TS-OOK and the coding weight of error-preventing codes are parameters that the \gls{mac} protocol at the link layer can jointly optimize~\cite{jornet2012phlame}. 

Similar cross-layer aspects appear when going up in the protocol stack. For example, in energy harvesting nanonetworks~\cite{harv1, harv2, harv3, harv4}, the selection of routes needs to capture not only the delay or throughput associated to each possible link, but also the energy status of each node along the route~\cite{pierobon2014routing}. As another example, the works by Tsioliaridou \emph{et al.}~\cite{tsioliaridou2017packet, tsioliaridou2015corona} propose a solution for joint localization and routing using specific fixed nanomachines as anchors, so that the address sort of encodes the physical position of a nanomachine in the network's coordinates. Hence, routing is greatly simplified as it becomes a problem of finding a linear path from the source position to destination.

Besides routing, addressing is another aspect that drastically changes within electromagnetic nanonetworks. For example, in many use cases, such as in wireless nanosensing networks, it is not always important which specific nanomachine sent this message/update but rather the type of the machine and its rough location. Hence, there is not always a need to follow a conventional multi-layer encapsulation approach, where each payload gets accompanied by a rich set of unique identifiers (application layer, service layer, network layer, data link layer, and physical layer). Instead, a single group ID may be associated to specific service run by dozens and hundreds of nanomachines in this location. On one side, this simplifies the protocol stack and reduces the overheads drastically~\cite{samman2010adaptive}. On the other, it requires a deep rethinking of the underlying protocols to enable them to work effectively without distinguishing individual machines by their unique IDs (e.g., MAC address, IP address, DNS name) at every layer.

More examples of cross-layer design could be found in application-specific scenarios. For instance, in the context of wireless chip-scale networks, MAC protocols could benefit from application-level information such as message criticality, which can then be used to drop packets early to reduce temporary congestion without affecting the program's output~\cite{fernando2019replica} or to prioritize transmissions that optimize application metrics (execution time) rather than network metrics (communication latency)~\cite{jog2021one}. Further, the wireless channel within package can be used as a medium for collective computation through the overlap of concurrent transmissions, to implement thread synchronization primitives~\cite{abadal2016wisync} or majority gates~\cite{guirado2023whype} useful for certain parallel computing architectures. This eliminates the need for link or network layer protocols.

\subsection{Prototyping and Testbed Development}
While the vision of electromagnetic nanonetworks is over 15~years old, as of today, no complete \textit{prototypes of nanomachines} exist. While several specific components of the individual nano-radios 
have been experimentally demonstrated (e.g., a plasmonic terahertz signal source was shown in~\cite{barut2022asymmetrically}), these are one-off devices. The robustness and reliability of nanofabrication techniques are the culprits. Today's major semiconductor foundries are designed for CMOS-compliant processes, and as we discussed at length, the miniaturization of nano-transceivers and nano-antennas requires the adoption of less conventional materials and structures, ranging from III-V semiconductors to \gls{2d} materials such as graphene, \gls{hbn} and \gls{mos2}. Major investments are required to overcome this bottleneck.

Moreover, a nano-radio is not enough for many application testing scenarios; a complete nanomachine is needed. For example, on the one hand, in wireless networks within computing packages, the main challenge is the integration of nano-radios within computing architectures, which is technically feasible but requires a non-trivial engineering effort and monetary cost for prototyping. On the other hand, for the \gls{iont}, each individual nano-radio needs to be accompanied by a nano-processor, a nano-memory, nano-sensors and nano-actuators, and an energy nano-system to power them all. Each one of these components is a huge research field by itself~\cite{comp1, comp3, harv1, harv2, harv3, harv4}. 

Even when having nanomachines, additional challenges arise from the environment in which these need to be tested. For example, in the case of biological applications, the nanomachine needs to be embedded in biological tissues, which requires additional material and chemical processing to ensure biocompatibility. For example, in~\cite{jornet2019optogenomic}, experimental results are reported discussing the utilization of microlasers on cerebral organoids.

Beyond prototypes, there is also the need to create \textit{experimental testbeds} to test nanonetwork solutions in repeatable, controlled scenarios.
Unfortunately, most of the existing experimental testbeds only focus on non-electromagnetic nanonetworks 
(e.g., molecular communication~\cite{krishnaswamy2013time,kuscu2015fluorescent,koo2016molecular,deng2017microfluidic,farsad2017novel,grebenstein2018biological,unterweger2018experimental,grebenstein2019molecular,guo2020vertical,lee2020vessel,koo2020deep,kuscu2021fabrication,qiu2023review}, ultrasonic communication~\cite{santagati2014sonar,santagati2016experimental,guan2016distributed,bos2018enabling} and, recently, galvanic coupling~\cite{vizziello2024experimental}). It is relevant to note that many of these testbeds are in fact not employing nanomachines, but utilizing macroscale devices to test the physics of, for example, nanoscale applications. Some of these papers specifically employ phantoms (bought or created from scratch) to mimic the behavior of body tissues in the case of possible medical applications~\cite{phant1, phant2, phant3, phant4, phant5}. In a relevant work towards in-body nanonetworks, the authors of~\cite{vasisht2018body, tao2023magnetic} test their solution for in-body communication and localization based on backscattering at microwave frequencies using chicken and pork tissues, as well as phantoms made of Polyethylene powder and Agarose for muscles, and vegetable oils and gelatin for fat tissues.
In the specific perspective of terahertz testbeds, existing macroscale communication testbeds, such as the TeraNova testbed~\cite{sen2020teranova}, can be utilized to study the propagation of true terahertz signals on-chip or across chips, in a reproducible manner. Similarly, a \gls{vna} with appropriate terahertz probes can also be utilized to characterize chip-to-chip communications~\cite{kim2016statistical}.  


While waiting for the development of prototypes and experimental testbeds, \textit{simulation tools} can be employed.
Existing simulation tools contain both non-electromagnetic and electromagnetic nanonetwork simulators.
On one hand, most of the non-electromagnetic platforms focus on molecular communications~\cite{rizwan2018review,sahin2021evaluation},
e.g., NanoNS~\cite{gul2010nanons} and N3Sim~\cite{llatser2014n3sim} for immobile molecular communications, as well as simulators designed for mobile scenarios~\cite{toth20113,felicetti2012simulation,akkaya2014hla}.
In addition, BNSim~\cite{wei2013efficient} and nanoNS3~\cite{jian2017nanons3} have been developed to deploy bacteria as molecular communication nodes.
On the other hand, multiple simulation tools have been developed for electromagnetic nanonetworks by leveraging TS-OOK modulation for various applications.
For example, Nano-Sim~\cite{piro2013simulating} is the first electromagnetic nanonetwork simulator.
Nano-Sim is suitable for different nanonetwork scenarios by varying the parameters of the TS-OOK modulation.
In addition, Vouivre has been proposed~\cite{boillot2013efficient,boillot2015scalable} to support simulations for wireless communications among numerous nano-robots.
More recently, TeraSim~\cite{hossain2018terasim} has been developed to simulate the THz communications for nanoscale scenarios in addition to the macroscale ones.
Another recent electromagnetic nanonetwork simulation software, BitSimulator~\cite{dhoutaut2018bit}, can simulate nanonetworks with concurrent transmissions among numerous nano-devices to test coding, channel access, routing, and congestion control.
Apart from the communication simulators, a 3D model~\cite{cavalcanti2008medical} has also been developed, which provides a vessel model to expedite the research of in-vessel nanonetworks.

\subsection{Security and Privacy}
Wireless nanonetworking setups have certain distinct features, making the design of efficient security mechanisms especially challenging. On one side, many target use cases deal with sensitive data (e.g., eHealth or even active services, such as target drug delivery)~\cite{yang2020comprehensive,alabdulatif2023internet}. Hence, the desired level of security should be relatively high, corresponding to the risks and losses associated with the system compromise. On the other side, low-complexity and low-power nano-scale devices have performance limitations in implementing \emph{full-scale} state-of-the-art security and privacy solutions (e.g., cryptographic primitives, such as AES-512 symmetric ciphers, RSA-4096 public-key/non-symmetric ciphers for message encryption/authentication). Hence, designing and testing lower-complexity but still sufficiently secure solutions (often grouped under the ``Lightweight Cryptography'' umbrella~\cite{dutta2019lightweight}) is of interest for prospective electromagnetic nanonetworks and the \gls{iont}.

This trend is however also challenged by the slow still sustainable evolution of the devices implementing quantum cryptanalysis making many prospective cryptography primitives (especially, those coming from the lightweight cryptography field) vulnerable to quantum computers~\cite{pandey2023cryptographyc}. \emph{We observe three main directions in this field}. The first one is related to novel methods to protect the transmitted data while staying under 512-bit operation sizes (ideally, as low as just a few bytes to match the envisioned payload in prospective scenarios~\cite{akyildiz2010electromagnetic}). The second direction is related to designing and implementing novel node authentication methods to maintain the integrity of the nanonetwork from malicious nodes. The principal challenge here is that prospective nanonetwork deployments may not have a central node with absolute trust. Hence, instead of a simple authentication by a single trusted node, the distributed network must perform collective authentication procedure by e.g., utilizing computationally-intense shared secret cryptography~\cite{ometov2018multi}.

Third, the cross-network security needs to be maintained over the interface between future electromagnetic nanonetworks and existing/future macro-scale networks (e.g., \glspl{wlan} and cellular). Here, ensuring absolute end-to-end security is a particular challenge to address, as electromagnetic nanonetworks (due to their complexity, capacity, and latency constraints) may be not always capable of handling macro-scale data formats and procedures from the macro-world, including IPSec and \gls{vpn}.

\subsection{Standardization Activities}
While fundamental research, applied research, and prototyping of the key ideas are all progressing fast, a technology transfer is needed to make a real impact on individual human lives and society, at large. Notably, here it becomes different depending on the target use case. For instance, single-vendor use cases, such as wireless NoC, are primarily driven by commercial products. In contrast, larger-scale use cases involving several vendors (e.g., \gls{iont} and wearable networks) also require global standards, so the individual solutions from different vendors may seamlessly co-exist with each other and interact thus forming electromagnetic nanonetworks.

In 2009, an amendment to the IEEE~802.15.3 for \glspl{wpan} was finalized enabling wireless links over $60$\,GHz millimeter wave bands with the data rates of up to $5$\,Gb/s~\cite{ieee_802_15_3c_2009}. The next step was made in 2017 in another amendment to the IEEE~802.15.3, IEEE 802.15.3d--2017 that regulates point-to-point sub-THz connectivity ($252$\,GHz--$325$\,GHz)~\cite{ieee_802_15_3d_2017}. Notably, this standard also supports cm-scale and mm-scale high-rate board-to-board communications enabling the data rates over $100$\,Gb/s using ultra-broadband channels of up to $69$\,GHz~\cite{petrov2020ieee}. The latest revision of the IEEE~802.15.3 was published in February~2024 including both the mmWave and the sub-THz innovative solutions discussed above~\cite{ieee_802_15_3_2023}.

\emph{Still, there is a lot of work to be done in this area in the foreseeable future}. Specifically, the sub-THz interface in~\cite{ieee_802_15_3d_2017} and~\cite{ieee_802_15_3_2023} supports no more than two nodes (pairnet, not a full-scale network). Hence, expanding these solutions to many-node \gls{iont} use cases is needed. Another important challenge is related to supporting mobile nodes, as these IEEE solutions primarily target stationary use cases. Last but not least, an essential direction here is related to enabling support for battery-less devices and energy harvesting~\cite{3gpp_tr_38_848,song2022advances}.


\subsection{Bridging Nano to Macro}
In the original vision of electromagnetic nanonetworks~\cite{akyildiz2010electromagnetic}, nanomachines built in a bottom-up approach were envisioned to interface with each other wirelessly and coordinatedly reach the macroscale through the appropriate interfaces (e.g., nano-controller or nano-to-macro interfaces), and together enable unprecedented applications such as those discussed in Sec.~\ref{subsec:nanothings}. Through the years, the vision grew also to include cases in which nanomachines, or at least some of their components, are integrated into existing micro- and macro-systems (e.g. chip multiprocessors, quantum computers) to provide them with new functionalities or improve their performance, as we discussed in Secs.~\ref{subsec:computing} and~\ref{subsec:quantum}. Here, we argue that modern massive antenna systems for macro-scale communications in the terahertz band and beyond could fall into this category and hence benefit from embedded nanonetworks -- thereby bridging the nanonetworking and macroscale networking paradigms. 

As the next generations of wireless networks (e.g., 6G and beyond) tap into the terahertz spectrum seeking to satisfy their ever-increasing appetite for bandwidth~\cite{akyildiz2022terahertz}, new solutions are required to address the \emph{distance problem}~\cite{akyildiz2018combating}  posed by the high path losses and the sensitivity of terahertz waves to blockage. Ultra-massive MIMO communication schemes in transmission and reception~\cite{akyildiz2016realizing} and \glspl{ris}~\cite{liaskos2018using, pan2021reconfigurable, abadal2020programmable} to add programmability to the channel have emerged as key technologies to enable this transition. In this context, with the extreme miniaturization of the antenna elements (both due to the high frequency and the use of deep subwavelength elements) and the possibility of integrating control units directly at the level of a single antenna element, the opportunity of nanonetwork-enabled antenna systems arises~\cite{liaskos2015design}. 

In the \gls{ris} domain, embedding a nanonetwork of controllers in metasurfaces has been proposed as a scalable and efficient alternative to existing FPGA-attached prototypes~\cite{petrou2022first}. This opens the door to a myriad of possibilities, such as intelligent sensing of waves~\cite{liaskos2019absense}, fast reconfiguration of graphene metasurfaces for multi-wideband operation~\cite{taghvaee2022multiwideband}, simultaneous transmission-reception operation~\cite{ahmed2023survey}, holographic operation~\cite{an2023stacked}, or even powering of the \gls{ris} via harvesting~\cite{zheng2023zero, tyrovolas2023zero, ntontin2024perpetual, ntontin2022wireless}. At the nano-level, studies have been done about the possibility of using wireless communications in the embedded controller nanonetwork~\cite{tasolamprou2019exploration}, for routing packets in the controller nanonetwork in the presence of faults~\cite{kouzapas2020towards}, and to adapt it to the reconfiguration workload imposed by the \gls{ris}~\cite{saeed2021workload}. However, we posit that this nano-to-macro convergence of wireless communications can still be explored to advance further in the realm of nanonetwork-enabled antenna systems.

\section{Conclusion}
\label{sec:conclusion}
The concept of nanonetwork has gone through a natural evolution since its inception, going from the interconnection of nano-sized machines to a broader paradigm with a more relaxed definition, but still having the size and amount of integration as primary factor. In the electromagnetic side of this paradigm, the number of potential communication technologies has increased (including magnetoelectric antennas in the microwave band) and its maturity has increased (thanks to intense efforts in nanophotonics, graphene, and terahertz technology). The application scope of electromagnetic nanonetworks has been broadened as well, adding wireless communications within computing packages to the existing range of IoNT applications in the biomedical, agricultural, logistics, or environmental monitoring domains. Furthermore, quantum computing has recently gained momentum and presents important nanocommunication problems with extreme resource constraints and performance requirements, hence being a suitable candidate to become an application scenario for electromagnetic nanonetworks. Overall, we have seen that the existing communication technologies for nanonetworks are very diverse, and so are the requirements of the different application scenarios. This, together with the wide array of challenges still to be solved, both general/cross-cutting and specific to the different applications, suggests that the field of electromagnetic nanonetworks will contine growing in the years to come.



\section*{Acknowledgment}
S. A. acknowledges support from the EU's Horizon Europe program through the European Research Council (ERC) under grant agreement 101042080 (WINC) and through the European Innovation Council (EIC) PATHFINDER scheme, grant agreement No 101099697 (QUADRATURE). 
C. H. acknowledges the in-part support from the National Natural Science Foundation of China (NSFC) – European Research Council (ERC) Research Program under Project No. 62311530342, and from Alexander von Humboldt Foundation.
The EU partially supported the activity of L. G. under the Italian National Recovery and Resilience Plan (NRRP) of NextGenerationEU, partnership on “Telecommunications of the Future” (PE0000001 - program “RESTART”). I. F. A.  is supported by Rannis (Icelandic Research Fund) for the Grant of Excellence program at the University of Iceland. J. M. J. acknowledges the support of the US National Science Foundation through Award CNS-1955004, Award CNS-2011411, Award CNS-2225590, and CBET-2039189.

\ifCLASSOPTIONcaptionsoff
  \newpage
\fi




\begin{thebibliography}{100}
\providecommand{\url}[1]{#1}
\csname url@samestyle\endcsname
\providecommand{\newblock}{\relax}
\providecommand{\bibinfo}[2]{#2}
\providecommand{\BIBentrySTDinterwordspacing}{\spaceskip=0pt\relax}
\providecommand{\BIBentryALTinterwordstretchfactor}{4}
\providecommand{\BIBentryALTinterwordspacing}{\spaceskip=\fontdimen2\font plus
\BIBentryALTinterwordstretchfactor\fontdimen3\font minus \fontdimen4\font\relax}
\providecommand{\BIBforeignlanguage}[2]{{%
\expandafter\ifx\csname l@#1\endcsname\relax
\typeout{** WARNING: IEEEtran.bst: No hyphenation pattern has been}%
\typeout{** loaded for the language `#1'. Using the pattern for}%
\typeout{** the default language instead.}%
\else
\language=\csname l@#1\endcsname
\fi
#2}}
\providecommand{\BIBdecl}{\relax}
\BIBdecl

\bibitem{akyildiz2008nanonetworks}
I.~F. Akyildiz, F.~Brunetti, and C.~Bl{\'a}zquez, ``Nanonetworks: A new communication paradigm,'' \emph{Computer Networks}, vol.~52, no.~12, pp. 2260--2279, 2008.

\bibitem{kuhn2008managing}
K.~Kuhn, C.~Kenyon, A.~Kornfeld, M.~Liu, A.~Maheshwari, W.-k. Shih, S.~Sivakumar, G.~Taylor, P.~VanDerVoorn, and K.~Zawadzki, ``Managing process variation in intel's 45nm cmos technology.'' \emph{Intel Technology Journal}, vol.~12, no.~2, 2008.

\bibitem{anker2008biosensing}
J.~N. Anker, W.~P. Hall, O.~Lyandres, N.~C. Shah, J.~Zhao, and R.~P. Van~Duyne, ``Biosensing with plasmonic nanosensors,'' \emph{Nature materials}, vol.~7, no.~6, pp. 442--453, 2008.

\bibitem{wang2008energy}
Z.~L. Wang, ``Energy harvesting for self-powered nanosystems,'' \emph{Nano Research}, vol.~1, pp. 1--8, 2008.

\bibitem{jornet2010graphene}
J.~M. Jornet and I.~F. Akyildiz, ``Graphene-based nano-antennas for electromagnetic nanocommunications in the terahertz band,'' in \emph{Proceedings of the Fourth European Conference on Antennas and Propagation}.\hskip 1em plus 0.5em minus 0.4em\relax IEEE, 2010, pp. 1--5.

\bibitem{yang2020comprehensive}
K.~Yang, D.~Bi, Y.~Deng, R.~Zhang, M.~M.~U. Rahman, N.~A. Ali, M.~A. Imran, J.~M. Jornet, Q.~H. Abbasi, and A.~Alomainy, ``A comprehensive survey on hybrid communication in context of molecular communication and terahertz communication for body-centric nanonetworks,'' \emph{IEEE Transactions on Molecular, Biological and Multi-Scale Communications}, vol.~6, no.~2, pp. 107--133, 2020.

\bibitem{akyildiz2015internet}
I.~F. Akyildiz, M.~Pierobon, S.~Balasubramaniam, and Y.~Koucheryavy, ``The internet of bio-nano things,'' \emph{IEEE Communications Magazine}, vol.~53, no.~3, pp. 32--40, 2015.

\bibitem{akyildiz2010electromagnetic}
I.~F. Akyildiz and J.~M. Jornet, ``Electromagnetic wireless nanosensor networks,'' \emph{Nano Communication Networks}, vol.~1, no.~1, pp. 3--19, 2010.

\bibitem{cao2023future}
W.~Cao, H.~Bu, M.~Vinet, M.~Cao, S.~Takagi, S.~Hwang, T.~Ghani, and K.~Banerjee, ``The future transistors,'' \emph{Nature}, vol. 620, no. 7974, pp. 501--515, 2023.

\bibitem{yu2020mmwave}
Q.~Yu, S.~Rami, J.~Waldemer, Y.~Ma, V.~Neeli, J.~Garrett, G.~Liu, J.~Koo, M.~Marulanda, S.~Morarka \emph{et~al.}, ``mmwave and sub-thz technology development in intel 22nm finfet (22ffl) process,'' in \emph{2020 IEEE International Electron Devices Meeting (IEDM)}.\hskip 1em plus 0.5em minus 0.4em\relax IEEE, 2020, pp. 17--4.

\bibitem{pasricha2020survey}
S.~Pasricha and M.~Nikdast, ``A survey of silicon photonics for energy-efficient manycore computing,'' \emph{IEEE Design \& Test}, vol.~37, no.~4, pp. 60--81, 2020.

\bibitem{ntontin2024perpetual}
K.~Ntontin, A.-A.~A. Boulogeorgos, S.~Abadal, A.~Mesodiakaki, S.~Chatzinotas, and B.~Ottersten, ``Perpetual reconfigurable intelligent surfaces through in-band energy harvesting: Architectures, protocols, and challenges,'' \emph{IEEE Vehicular Technology Magazine}, 2024.

\bibitem{ferrari2015science}
A.~C. Ferrari, F.~Bonaccorso, V.~Fal'Ko, K.~S. Novoselov, S.~Roche, P.~B{\o}ggild, S.~Borini, F.~H. Koppens, V.~Palermo, N.~Pugno \emph{et~al.}, ``Science and technology roadmap for graphene, related two-dimensional crystals, and hybrid systems,'' \emph{Nanoscale}, vol.~7, no.~11, pp. 4598--4810, 2015.

\bibitem{briggs2019roadmap}
N.~Briggs, S.~Subramanian, Z.~Lin, X.~Li, X.~Zhang, K.~Zhang, K.~Xiao, D.~Geohegan, R.~Wallace, L.-Q. Chen \emph{et~al.}, ``A roadmap for electronic grade 2d materials,'' \emph{2D Materials}, vol.~6, no.~2, p. 022001, 2019.

\bibitem{dash2022active}
S.~Dash, C.~Psomas, I.~Krikidis, I.~F. Akyildiz, and A.~Pitsillides, ``Active control of thz waves in wireless environments using graphene-based ris,'' \emph{IEEE Transactions on Antennas and Propagation}, vol.~70, no.~10, pp. 8785--8797, 2022.

\bibitem{abadal2022graphene}
S.~Abadal \emph{et~al.}, ``Graphene-based wireless agile interconnects for massive heterogeneous multi-chip processors,'' \emph{IEEE Wireless Communications}, vol.~30, no.~4, pp. 162--169, 2023.

\bibitem{alarcon2023scalable}
E.~Alarc{\'o}n, S.~Abadal, F.~Sebastiano, M.~Babaie, E.~Charbon, P.~H. Bol{\'\i}var, M.~Palesi, E.~Blokhina, D.~Leipold, B.~Staszewski \emph{et~al.}, ``Scalable multi-chip quantum architectures enabled by cryogenic hybrid wireless/quantum-coherent network-in-package,'' in \emph{2023 IEEE International Symposium on Circuits and Systems (ISCAS)}, 2023.

\bibitem{akyildiz2010internet}
I.~F. Akyildiz and J.~M. Jornet, ``The internet of nano-things,'' \emph{IEEE Wireless Communications}, vol.~17, no.~6, pp. 58--63, 2010.

\bibitem{jornet2012internet}
J.~M. Jornet and I.~F. Akyildiz, ``The internet of multimedia nano-things,'' \emph{Nano Communication Networks}, vol.~3, no.~4, pp. 242--251, 2012.

\bibitem{balasubramaniam2012realizing}
S.~Balasubramaniam and J.~Kangasharju, ``Realizing the internet of nano things: challenges, solutions, and applications,'' \emph{computer}, vol.~46, no.~2, pp. 62--68, 2012.

\bibitem{rikhtegar2013brief}
N.~Rikhtegar and M.~Keshtgary, ``A brief survey on molecular and electromagnetic communications in nano-networks,'' \emph{International Journal of Computer Applications}, vol.~79, no.~3, 2013.

\bibitem{dressler2015connecting}
F.~Dressler and S.~Fischer, ``Connecting in-body nano communication with body area networks: Challenges and opportunities of the internet of nano things,'' \emph{Nano Communication Networks}, vol.~6, no.~2, pp. 29--38, 2015.

\bibitem{chude2017molecular}
U.~A. Chude-Okonkwo, R.~Malekian, B.~T. Maharaj, and A.~V. Vasilakos, ``Molecular communication and nanonetwork for targeted drug delivery: A survey,'' \emph{IEEE Communications Surveys \& Tutorials}, vol.~19, no.~4, pp. 3046--3096, 2017.

\bibitem{rizwan2018review}
A.~Rizwan, A.~Zoha, R.~Zhang, W.~Ahmad, K.~Arshad, N.~A. Ali, A.~Alomainy, M.~A. Imran, and Q.~H. Abbasi, ``A review on the role of nano-communication in future healthcare systems: A big data analytics perspective,'' \emph{IEEE Access}, vol.~6, pp. 41\,903--41\,920, 2018.

\bibitem{akyildiz2019moving}
I.~F. Akyildiz, M.~Pierobon, and S.~Balasubramaniam, ``Moving forward with molecular communication: From theory to human health applications [point of view],'' \emph{Proceedings of the IEEE}, vol. 107, no.~5, pp. 858--865, 2019.

\bibitem{marzo2019nanonetworks}
J.~L. Marzo, J.~M. Jornet, and M.~Pierobon, ``Nanonetworks in biomedical applications,'' \emph{Current drug targets}, vol.~20, no.~8, pp. 800--807, 2019.

\bibitem{kabir2021electromagnetic}
M.~H. Kabir, S.~R. Islam, A.~P. Shrestha, F.~Ali, M.~A. Badsha, M.~J. Piran, and D.-T. Do, ``Electromagnetic nanocommunication networks: Principles, applications, and challenges,'' \emph{IEEE Access}, vol.~9, pp. 166\,147--166\,165, 2021.

\bibitem{bi2021survey}
D.~Bi, A.~Almpanis, A.~Noel, Y.~Deng, and R.~Schober, ``A survey of molecular communication in cell biology: Establishing a new hierarchy for interdisciplinary applications,'' \emph{IEEE Communications Surveys \& Tutorials}, vol.~23, no.~3, pp. 1494--1545, 2021.

\bibitem{lemic2021survey}
F.~Lemic, S.~Abadal, W.~Tavernier, P.~Stroobant, D.~Colle, E.~Alarc{\'o}n, J.~Marquez-Barja, and J.~Famaey, ``Survey on terahertz nanocommunication and networking: A top-down perspective,'' \emph{IEEE Journal on Selected Areas in Communications}, vol.~39, no.~6, pp. 1506--1543, 2021.

\bibitem{koshy2022new}
A.~M. Koshy, I.~Jeerapan, and C.~Manjunatha, ``New insights on molecular communication in nano communication networks and their applications,'' \emph{ECS Transactions}, vol. 107, no.~1, p. 9295, 2022.

\bibitem{yin2022biomedical}
X.-X. Yin, A.~Baghai-Wadji, and Y.~Zhang, ``A biomedical perspective in terahertz nano-communications: A review,'' \emph{IEEE Sensors Journal}, vol.~22, no.~10, pp. 9215--9227, 2022.

\bibitem{buniyamin2022nanotechnology}
I.~Buniyamin, R.~M. Akhir, N.~A. Asli, Z.~Khusaimi, M.~F. Malek, and M.~R. Mahmood, ``Nanotechnology applications in biomedical systems,'' \emph{Current Nanomaterials}, vol.~7, no.~3, pp. 167--180, 2022.

\bibitem{jornet2023nanonetworking}
J.~M. Jornet and A.~Sangwan, ``Nanonetworking in the terahertz band and beyond,'' \emph{IEEE Nanotechnology Magazine}, 2023.

\bibitem{csenturk2023internet}
{\c{S}}.~{\c{S}}ent{\"u}rk, b.~K{\"o}k, and F.~{\c{S}}ent{\"u}rk, ``Internet of nano and bio-nano things: A review,'' \emph{Semantic Intelligence: Select Proceedings of ISIC 2022}, pp. 265--276, 2023.

\bibitem{alabdulatif2023internet}
A.~Alabdulatif, N.~N. Thilakarathne, Z.~K. Lawal, K.~E. Fahim, and R.~Y. Zakari, ``Internet of nano-things (iont): A comprehensive review from architecture to security and privacy challenges,'' \emph{Sensors}, vol.~23, no.~5, p. 2807, 2023.

\bibitem{akyildiz2011nanonetworks}
I.~F. Akyildiz, J.~M. Jornet, and M.~Pierobon, ``Nanonetworks: A new frontier in communications,'' \emph{Communications of the ACM}, vol.~54, no.~11, pp. 84--89, 2011.

\bibitem{jornet2012joint}
J.~M. Jornet and I.~F. Akyildiz, ``Joint energy harvesting and communication analysis for perpetual wireless nanosensor networks in the terahertz band,'' \emph{IEEE Transactions on Nanotechnology}, vol.~11, no.~3, pp. 570--580, 2012.

\bibitem{dorfmuller2010plasmonic}
J.~Dorfmuller, R.~Vogelgesang, W.~Khunsin, C.~Rockstuhl, C.~Etrich, and K.~Kern, ``Plasmonic nanowire antennas: experiment, simulation, and theory,'' \emph{Nano letters}, vol.~10, no.~9, pp. 3596--3603, 2010.

\bibitem{agio2013optical}
M.~Agio and A.~Al{\`u}, \emph{Optical antennas}.\hskip 1em plus 0.5em minus 0.4em\relax Cambridge University Press, 2013.

\bibitem{petrov2017polymer}
A.~Petrov, V.~Bessonov, K.~Abrashitova, N.~Kokareva, K.~Safronov, A.~Barannikov, P.~Ershov, N.~Klimova, I.~Lyatun, V.~Yunkin \emph{et~al.}, ``Polymer x-ray refractive nano-lenses fabricated by additive technology,'' \emph{Optics express}, vol.~25, no.~13, pp. 14\,173--14\,181, 2017.

\bibitem{nafari2017modeling}
M.~Nafari and J.~M. Jornet, ``Modeling and performance analysis of metallic plasmonic nano-antennas for wireless optical communication in nanonetworks,'' \emph{IEEE Access}, vol.~5, pp. 6389--6398, 2017.

\bibitem{feng2014single}
L.~Feng, Z.~J. Wong, R.-M. Ma, Y.~Wang, and X.~Zhang, ``Single-mode laser by parity-time symmetry breaking,'' \emph{Science}, vol. 346, no. 6212, pp. 972--975, 2014.

\bibitem{wong2021epitaxially}
W.~W. Wong, Z.~Su, N.~Wang, C.~Jagadish, and H.~H. Tan, ``Epitaxially grown inp micro-ring lasers,'' \emph{Nano Letters}, vol.~21, no.~13, pp. 5681--5688, 2021.

\bibitem{zhang2020tunable}
Z.~Zhang, X.~Qiao, B.~Midya, K.~Liu, J.~Sun, T.~Wu, W.~Liu, R.~Agarwal, J.~M. Jornet, S.~Longhi \emph{et~al.}, ``Tunable topological charge vortex microlaser,'' \emph{Science}, vol. 368, no. 6492, pp. 760--763, 2020.

\bibitem{haffner2015all}
C.~Haffner, W.~Heni, Y.~Fedoryshyn, J.~Niegemann, A.~Melikyan, D.~L. Elder, B.~Baeuerle, Y.~Salamin, A.~Josten, U.~Koch \emph{et~al.}, ``All-plasmonic mach--zehnder modulator enabling optical high-speed communication at the microscale,'' \emph{Nature Photonics}, vol.~9, no.~8, pp. 525--528, 2015.

\bibitem{thomaschewski2022plasmonic}
M.~Thomaschewski, V.~A. Zenin, S.~Fiedler, C.~Wolff, and S.~I. Bozhevolnyi, ``Plasmonic lithium niobate mach--zehnder modulators,'' \emph{Nano Letters}, vol.~22, no.~16, pp. 6471--6475, 2022.

\bibitem{liu2011graphene}
M.~Liu, X.~Yin, E.~Ulin-Avila, B.~Geng, T.~Zentgraf, L.~Ju, F.~Wang, and X.~Zhang, ``A graphene-based broadband optical modulator,'' \emph{Nature}, vol. 474, no. 7349, pp. 64--67, 2011.

\bibitem{li2017single}
B.~Li, S.~Zu, J.~Zhou, Q.~Jiang, B.~Du, H.~Shan, Y.~Luo, Z.~Liu, X.~Zhu, and Z.~Fang, ``Single-nanoparticle plasmonic electro-optic modulator based on mos2 monolayers,'' \emph{ACS nano}, vol.~11, no.~10, pp. 9720--9727, 2017.

\bibitem{chen2012infrared}
H.~Chen, N.~Xi, B.~Song, L.~Chen, J.~Zhao, K.~W.~C. Lai, and R.~Yang, ``Infrared camera using a single nano-photodetector,'' \emph{IEEE Sensors Journal}, vol.~13, no.~3, pp. 949--958, 2012.

\bibitem{nozaki2016photonic}
K.~Nozaki, S.~Matsuo, T.~Fujii, K.~Takeda, M.~Ono, A.~Shakoor, E.~Kuramochi, and M.~Notomi, ``Photonic-crystal nano-photodetector with ultrasmall capacitance for on-chip light-to-voltage conversion without an amplifier,'' \emph{Optica}, vol.~3, no.~5, pp. 483--492, 2016.

\bibitem{thomson2016roadmap}
D.~Thomson, A.~Zilkie, J.~E. Bowers, T.~Komljenovic, G.~T. Reed, L.~Vivien, D.~Marris-Morini, E.~Cassan, L.~Virot, J.-M. F{\'e}d{\'e}li \emph{et~al.}, ``Roadmap on silicon photonics,'' \emph{Journal of Optics}, vol.~18, no.~7, p. 073003, 2016.

\bibitem{sangwan2021beamforming}
A.~Sangwan and J.~M. Jornet, ``Beamforming optical antenna arrays for nano-bio sensing and actuation applications,'' \emph{Nano Communication Networks}, vol.~29, p. 100363, 2021.

\bibitem{qiao2021higher}
X.~Qiao, B.~Midya, Z.~Gao, Z.~Zhang, H.~Zhao, T.~Wu, J.~Yim, R.~Agarwal, N.~M. Litchinitser, and L.~Feng, ``Higher-dimensional supersymmetric microlaser arrays,'' \emph{Science}, vol. 372, no. 6540, pp. 403--408, 2021.

\bibitem{vakil2011transformation}
A.~Vakil and N.~Engheta, ``Transformation optics using graphene,'' \emph{Science}, vol. 332, no. 6035, pp. 1291--1294, 2011.

\bibitem{burke2006quantitative}
P.~J. Burke, S.~Li, and Z.~Yu, ``Quantitative theory of nanowire and nanotube antenna performance,'' \emph{IEEE transactions on nanotechnology}, vol.~5, no.~4, pp. 314--334, 2006.

\bibitem{llatser2012graphene}
I.~Llatser, C.~Kremers, A.~Cabellos-Aparicio, J.~M. Jornet, E.~Alarc{\'o}n, and D.~N. Chigrin, ``Graphene-based nano-patch antenna for terahertz radiation,'' \emph{Photonics and Nanostructures-Fundamentals and Applications}, vol.~10, no.~4, pp. 353--358, 2012.

\bibitem{perruisseau2013graphene}
J.~Perruisseau-Carrier, M.~Tamagnone, J.~S. Gomez-Diaz, and E.~Carrasco, ``Graphene antennas: Can integration and reconfigurability compensate for the loss?'' in \emph{Proc. of European Microwave Conference}, 2013, pp. 369--372.

\bibitem{tamagnone2012reconfigurable}
M.~Tamagnone, J.~Gomez-Diaz, J.~R. Mosig, and J.~Perruisseau-Carrier, ``Reconfigurable terahertz plasmonic antenna concept using a graphene stack,'' \emph{Applied Physics Letters}, vol. 101, no.~21, 2012.

\bibitem{zakrajsek2017design}
L.~Zakrajsek, E.~Einarsson, N.~Thawdar, M.~Medley, and J.~M. Jornet, ``Design of graphene-based plasmonic nano-antenna arrays in the presence of mutual coupling,'' in \emph{2017 11th European Conference on Antennas and Propagation (EUCAP)}.\hskip 1em plus 0.5em minus 0.4em\relax IEEE, 2017, pp. 1381--1385.

\bibitem{singh2020design}
A.~Singh, M.~Andrello, N.~Thawdar, and J.~M. Jornet, ``Design and operation of a graphene-based plasmonic nano-antenna array for communication in the terahertz band,'' \emph{IEEE Journal on Selected Areas in Communications}, vol.~38, no.~9, pp. 2104--2117, 2020.

\bibitem{akyildiz2016realizing}
I.~F. Akyildiz and J.~M. Jornet, ``Realizing ultra-massive {MIMO} (1024$\times$ 1024) communication in the (0.06--10) terahertz band,'' \emph{Nano Communication Networks}, vol.~8, pp. 46--54, 2016.

\bibitem{crabb2021hydrodynamic}
J.~Crabb, X.~Cantos-Roman, J.~M. Jornet, and G.~R. Aizin, ``Hydrodynamic theory of the dyakonov-shur instability in graphene transistors,'' \emph{Physical Review B}, vol. 104, no.~15, p. 155440, 2021.

\bibitem{crabb2022plasma}
J.~Crabb, X.~C. Roman, J.~Jornet, and G.~Aizin, ``Plasma instability in graphene field-effect transistors with a shifted gate,'' \emph{Applied Physics Letters}, vol. 121, no.~14, 2022.

\bibitem{crabb2022amplitude}
J.~Crabb, X.~Cantos-Roman, G.~R. Aizin, and J.~M. Jornet, ``Amplitude and frequency modulation with an on-chip graphene-based plasmonic terahertz nanogenerator,'' \emph{IEEE Transactions on Nanotechnology}, vol.~21, pp. 539--546, 2022.

\bibitem{singh2016graphene}
P.~K. Singh, G.~Aizin, N.~Thawdar, M.~Medley, and J.~M. Jornet, ``Graphene-based plasmonic phase modulator for terahertz-band communication,'' in \emph{2016 10th European Conference on Antennas and Propagation (EuCAP)}.\hskip 1em plus 0.5em minus 0.4em\relax IEEE, 2016, pp. 1--5.

\bibitem{crabb2023chip}
J.~Crabb, X.~Cantos-Roman, G.~Aizin, and J.~M. Jornet, ``On-chip integration of a plasmonic fet source and a nano-patch antenna for efficient terahertz wave radiation,'' \emph{Nanomaterials}, vol.~13, no.~24, p. 3114, 2023.

\bibitem{Park2012}
J.-D. Park, S.~Kang, S.~Thyagarajan, E.~Alon, and A.~Niknejad, ``{A 260 GHz fully integrated CMOS transceiver for wireless chip-to-chip communication},'' in \emph{Proceedings of the VLSIC '12}, 2012, pp. 48--49.

\bibitem{Yu2014}
X.~Yu, J.~Baylon, P.~Wettin, D.~Heo, P.~{Pratim Pande}, and S.~Mirabbasi, ``{Architecture and Design of Multi-Channel Millimeter-Wave Wireless Network-on-Chip},'' \emph{IEEE Design {\&} Test}, vol.~31, no.~6, pp. 19--28, 2014.

\bibitem{Thyagarajan2015}
S.~V. Thyagarajan, S.~Kang, and A.~M. Niknejad, ``{A 240 GHz Fully Integrated Wideband QPSK Receiver in 65 nm CMOS},'' \emph{IEEE Journal of Solid-State Circuits}, vol.~50, no.~10, pp. 2268--2280, 2015.

\bibitem{Fritsche2017}
D.~Fritsche, P.~St{\"{a}}rke, C.~Carta, and F.~Ellinger, ``{A Low-Power SiGe BiCMOS 190-GHz Transceiver Chipset With Demonstrated Data Rates up to 50 Gbit/s Using On-Chip Antennas},'' \emph{IEEE Transactions on Microwave Theory and Techniques}, vol.~65, no.~9, pp. 3312--3323, 2017.

\bibitem{Tokgoz2018}
K.~K. Tokgoz, S.~Maki, J.~Pang, N.~Nagashima, I.~Abdo, S.~Kawai, T.~Fujimura, Y.~Kawano, T.~Suzuki, T.~Iwai, K.~Okada, and A.~Matsuzawa, ``{A 120Gb/s 16QAM CMOS millimeter-wave wireless transceiver},'' \emph{Proceedings of the ISSCC '18}, pp. 168--170, 2018.

\bibitem{Byeon2020}
C.~W. Byeon, K.~C. Eun, and C.~S. Park, ``{A 2.65-pJ/Bit 12.5-Gb/s 60-GHz OOK CMOS Transmitter and Receiver for Proximity Communications},'' \emph{IEEE Transactions on Microwave Theory and Techniques}, vol.~68, no.~7, pp. 2902--2910, 2020.

\bibitem{Yi2021}
C.~Yi, D.~Kim, S.~Solanki, J.~H. Kwon, M.~Kim, S.~Jeon, Y.~C. Ko, and I.~Lee, ``{Design and Performance Analysis of THz Wireless Communication Systems for Chip-to-Chip and Personal Area Networks Applications},'' \emph{IEEE Journal on Selected Areas in Communications}, vol.~39, no.~6, pp. 1785--1796, 2021.

\bibitem{callender2022fully}
S.~Callender, A.~Agrawal, A.~Whitcombe, R.~Bhat, M.~Rahman, C.~C. Lee, P.~Sagazio, G.~C. Dogiamis, B.~R. Carlton, C.~Hull \emph{et~al.}, ``A fully integrated 160-gb/s d-band transmitter achieving 1.1-pj/b efficiency in 22-nm finfet,'' \emph{IEEE Journal of Solid-State Circuits}, vol.~57, no.~12, pp. 3582--3598, 2022.

\bibitem{agrawal2023128}
A.~Agrawal, A.~Whitcombe, W.~Shin, R.~Bhat, S.~Kundu, P.~Sagazio, H.~Chandrakumar, T.~W. Brown, B.~R. Carlton, C.~Hull \emph{et~al.}, ``A 128-gb/s $ d $-band receiver with integrated pll and adc achieving 1.95-pj/b efficiency in 22-nm finfet,'' \emph{IEEE Journal of Solid-State Circuits}, 2023.

\bibitem{saeed2018graphene}
M.~Saeed, A.~Hamed, Z.~Wang, M.~Shaygan, D.~Neumaier, and R.~Negra, ``Graphene integrated circuits: new prospects towards receiver realisation,'' \emph{Nanoscale}, vol.~10, no.~1, pp. 93--99, 2018.

\bibitem{Saeed2021}
M.~Saeed, P.~Palacios, M.-D. Wei, E.~Baskent, C.-Y. Fan, B.~Uzlu, T.~Wang, A.~Hemmetter, Z.~Wang, D.~Neumaier, M.~C. Lemme, and R.~Negra, ``{Graphene-Based Microwave Circuits: A Review},'' \emph{Advanced Materials}, 2021.

\bibitem{hamed2020graphene}
A.~Hamed, M.~Saeed, and R.~Negra, ``Graphene-based frequency-conversion mixers for high-frequency applications,'' \emph{IEEE Transactions on Microwave Theory and Techniques}, vol.~68, no.~6, pp. 2090--2096, 2020.

\bibitem{you2020laser}
R.~You, Y.-Q. Liu, Y.-L. Hao, D.-D. Han, Y.-L. Zhang, and Z.~You, ``Laser fabrication of graphene-based flexible electronics,'' \emph{Advanced Materials}, vol.~32, no.~15, p. 1901981, 2020.

\bibitem{rudrapati2020graphene}
R.~Rudrapati, \emph{Graphene: Fabrication methods, properties, and applications in modern industries}.\hskip 1em plus 0.5em minus 0.4em\relax IntechOpen London, UK, 2020, vol.~1.

\bibitem{wittmann2023assessment}
S.~Wittmann, S.~Pindl, S.~Sawallich, M.~Nagel, A.~Michalski, H.~Pandey, A.~Esteki, S.~Kataria, and M.~C. Lemme, ``Assessment of wafer-level transfer techniques of graphene with respect to semiconductor industry requirements,'' \emph{Advanced Materials Technologies}, vol.~8, no.~8, p. 2201587, 2023.

\bibitem{akyildiz2022terahertz}
I.~F. Akyildiz, C.~Han, Z.~Hu, S.~Nie, and J.~M. Jornet, ``Terahertz band communication: An old problem revisited and research directions for the next decade,'' \emph{IEEE Transactions on Communications}, vol.~70, no.~6, pp. 4250--4285, 2022.

\bibitem{kokkoniemi2016wideband}
J.~Kokkoniemi, V.~Petrov, D.~Moltchanov, J.~Lehtomaeki, Y.~Koucheryavy, and M.~Juntti, ``Wideband terahertz band reflection and diffuse scattering measurements for beyond {5G} indoor wireless networks,'' in \emph{European Wireless 2016; 22th European Wireless Conference}, 2016, pp. 1--6.

\bibitem{abadal2019wave}
S.~Abadal, C.~Han, and J.~M. Jornet, ``Wave propagation and channel modeling in chip-scale wireless communications: A survey from millimeter-wave to terahertz and optics,'' \emph{IEEE access}, vol.~8, pp. 278--293, 2019.

\bibitem{elayan2017terahertz}
H.~Elayan, R.~M. Shubair, J.~M. Jornet, and P.~Johari, ``Terahertz channel model and link budget analysis for intrabody nanoscale communication,'' \emph{IEEE transactions on nanobioscience}, vol.~16, no.~6, pp. 491--503, 2017.

\bibitem{elayan2017multi}
H.~Elayan, R.~M. Shubair, J.~M. Jornet, and R.~Mittra, ``Multi-layer intrabody terahertz wave propagation model for nanobiosensing applications,'' \emph{Nano communication networks}, vol.~14, pp. 9--15, 2017.

\bibitem{750kokkoniemi2016frequency}
J.~Kokkoniemi, J.~Lehtomäki, V.~Petrov, D.~Moltchanov, and M.~Juntti, ``Frequency domain penetration loss in the terahertz band,'' in \emph{2016 Global Symposium on Millimeter Waves (GSMM) \& ESA Workshop on Millimetre-Wave Technology and Applications}, 2016, pp. 1--4.

\bibitem{nan2017acoustically}
T.~Nan, H.~Lin, Y.~Gao, A.~Matyushov, G.~Yu, H.~Chen, N.~Sun, S.~Wei, Z.~Wang, M.~Li \emph{et~al.}, ``Acoustically actuated ultra-compact nems magnetoelectric antennas,'' \emph{Nature communications}, vol.~8, no.~1, p. 296, 2017.

\bibitem{zaeimbashi2019nanoneurorfid}
M.~Zaeimbashi, H.~Lin, C.~Dong, X.~Liang, M.~Nasrollahpour, H.~Chen, N.~Sun, A.~Matyushov, Y.~He, X.~Wang \emph{et~al.}, ``Nanoneurorfid: A wireless implantable device based on magnetoelectric antennas,'' \emph{IEEE Journal of Electromagnetics, RF and Microwaves in Medicine and Biology}, vol.~3, no.~3, pp. 206--215, 2019.

\bibitem{will2022tutorial}
A.~Will-Cole, A.~E. Hassanien, S.~D. Calisgan, M.-G. Jeong, X.~Liang, S.~Kang, V.~Rajaram, I.~Martos-Repath, H.~Chen, A.~Risso \emph{et~al.}, ``Tutorial: Piezoelectric and magnetoelectric n/mems—materials, devices, and applications,'' \emph{Journal of Applied Physics}, vol. 131, no.~24, 2022.

\bibitem{Vizziello1}
M.~Swaminathan, A.~Vizziello, D.~Duong, P.~Savazzi, and K.~R. Chowdhury, ``Beamforming in the body: Energy-efficient and collision-free communication for implants,'' in \emph{IEEE INFOCOM 2017 - IEEE Conference on Computer Communications}, 2017, pp. 1--9.

\bibitem{Vizziello2}
\BIBentryALTinterwordspacing
A.~Vizziello, P.~Savazzi, and G.~Magenes, ``Electromyography data transmission via galvanic coupling intra-body communication link,'' in \emph{Proceedings of the Eight Annual ACM International Conference on Nanoscale Computing and Communication}, ser. NANOCOM '21.\hskip 1em plus 0.5em minus 0.4em\relax New York, NY, USA: Association for Computing Machinery, 2021. [Online]. Available: \url{https://doi.org/10.1145/3477206.3477450}
\BIBentrySTDinterwordspacing

\bibitem{GC6}
A.~Vizziello, P.~Savazzi, G.~Magenes, and P.~Gamba, ``Phy design and implementation of a galvanic coupling testbed for intra-body communication links,'' \emph{IEEE Access}, vol.~8, pp. 184\,585--184\,597, 2020.

\bibitem{MDPI2020}
N.~H. Sebastián, N.~V. Villaseñor, F.-J. Renero-Carrillo, D.~D. Alonso, and W.~C. Arriaga, ``{Design of a Fully Integrated Inductive Coupling System: A Discrete Approach Towards Sensing Ventricular Pressure },'' \emph{MDPI Sensors}, vol.~20, no.~5, 2020.

\bibitem{WCMC}
D.~Naranjo-Hernandez, A.~Callejo-Leblic, Z.~Lucev, M.~Seyedi, and Y.~Gao, ``{Past results, present trends, and future challenges in intrabody communication},'' \emph{Hindawi Wireless Communications and Mobile Computing}, 2018.

\bibitem{Galluccio1}
G.~E. Santagati, T.~Melodia, L.~Galluccio, and S.~Palazzo, ``Medium access control and rate adaptation for ultrasonic intrabody sensor networks,'' \emph{IEEE/ACM Transactions on Networking}, vol.~23, no.~4, pp. 1121--1134, 2015.

\bibitem{Galluccio2}
E.~C. Sciacca and L.~Galluccio, ``Impulse response analysis of an ultrasonic human body channel,'' \emph{Computer Networks}, vol. 171, p. 107149, 2020.

\bibitem{harvest1}
``{Autonomous in-vivo brain-machine-interface in 28nm-cmos technology with ultrasound-based power-harvester and communication-link (Brain28nm Project)},'' 2017.

\bibitem{harvest2}
A.~Ballo, A.~Grasso, and M.~Privitera, ``{A High Efficiency and High Power Density Active AC/DC Converter for Battery-Less US-powered IMDs in a 28-nm CMOS Technology},'' \emph{IEEE Access}, 2024.

\bibitem{harvest3}
------, ``{Active and passive rectification methods for US-powered IMDs: A comparison in a 28-nm bulk CMOS technology},'' \emph{Analog Integrated Circuits and Signal Processing}, vol. 117, no.~1, 2023.

\bibitem{15}
R.~Das, F.~Moradi, and H.~Heidari, ``Biointegrated and wirelessly powered implantable brain devices: A review,'' \emph{IEEE Transactions on Biomedical Circuits and Systems}, vol.~14, no.~2, pp. 343--358, 2020.

\bibitem{environmentsurvey}
S.~Hooshmand, P.~Kassanos, M.~Keshavarz, P.~Duru, C.~I. Kayalan, I.~Kale, and M.~K. Bayazit, ``Wearable nano-based gas sensors for environmental monitoring and encountered challenges in optimization,'' \emph{MDPI Sensors}, vol.~23, no.~20, 2023.

\bibitem{environmentsurvey1}
N.~Ally and B.~Gumbi, ``A review on metal nanoparticles as nano-sensors for environmental detection of emerging contaminants,'' \emph{Elsevier Materials Today}, 2023.

\bibitem{wedage2023climate}
L.~T. Wedage, B.~Butler, S.~Balasubramaniam, Y.~Koucheryavy, J.~M. Jornet, and M.~C. Vuran, ``Climate change sensing through terahertz communication infrastructure: A disruptive application of 6g networks,'' \emph{IEEE Network}, 2023.

\bibitem{manufacturing}
``{The Revolutionary Impact of Nanotechnology on Supply Chain Management},'' \url{https://omnialog.centric.ae/en/resources/transportation-operations/the-potential-role-of-nanotechnology-in-enhancing-product-quality-and-safety-in-supply-chains}, 2023, [Online; accessed 24-February-2024].

\bibitem{agrisurvey}
V.~Goyal, D.~Rani, S.~Mehrotra, C.~Deng, and Y.~Wang, ``Unlocking the potential of nano-enabled precision agriculture for efficient and sustainable farming,'' \emph{MDPI Plants}, vol.~12, no.~21, 2023.

\bibitem{agrisurvey1}
N.~Yadava, V.~K. Gargc, A.~K. Chhillard, and J.~S. Ranab, ``Recent advances in nanotechnology for the improvement of conventional agricultural systems: A review,'' \emph{Elsevier Plant Nano Biology}, vol.~4, 2023.

\bibitem{smartcities}
R.~Sanchez-Corcuera, A.~Nunez-Marcos, J.~Sesma-Solance, A.~Bilbao-Jayo, R.~Mulero, U.~Zulaika, G.~Azkune, and A.~Almeida, ``{Smart cities survey: Technologies, application domains and challenges for the cities of the future},'' \emph{International Journal of Distributed Sensor Networks}, vol.~15, no.~6, 2019.

\bibitem{vizziello2023intra}
A.~Vizziello, M.~Magarini, P.~Savazzi, and L.~Galluccio, ``Intra-body communications for nervous system applications: Current technologies and future directions,'' \emph{Computer Networks}, vol. 227, p. 109718, 2023.

\bibitem{deisseroth2011optogenetics}
K.~Deisseroth, ``Optogenetics,'' \emph{Nature methods}, vol.~8, no.~1, pp. 26--29, 2011.

\bibitem{93}
K.~L. Montgomery, A.~J. Yeh, J.~S. Ho, V.~Tsao, S.~M. Iyer, L.~Grosenick, E.~A. Ferenczi, Y.~Tanabe, K.~Deisseroth, S.~L. Delp, and A.~S.~Y. Poon, ``Wirelessly powered, fully internal optogenetics for brain, spinal and peripheral circuits in mice,'' \emph{Nature Methods}, vol.~12, pp. 969--974, 2015.

\bibitem{Park2015}
S.~I. Park, D.~S. Brenner, G.~Shin, C.~D. Morgan, B.~A. Copits, H.~U. Chung, M.~Y. Pullen, K.~N. Noh, S.~Davidson, S.~J. Oh, J.~Yoon, K.-I. Jang, V.~K. Samineni, M.~E. Norman, J.~G. Grajales-Reyes, S.~K. Vogt, S.~S. Sundaram, K.~Wilson, J.~S. Ha, R.~Xu, T.~Pan, T.~il~Kim, Y.~Huang, M.~C. Montana, J.~P. Golden, M.~R. Bruchas, R.~W. Gereau, and J.~A. Rogers, ``Soft, stretchable, fully implantable miniaturized optoelectronic systems for wireless optogenetics,'' \emph{Nature biotechnology}, vol.~33, pp. 1280 -- 1286, 2015.

\bibitem{balasubramaniam2018wireless}
S.~Balasubramaniam, S.~A. Wirdatmadja, M.~T. Barros, Y.~Koucheryavy, M.~Stachowiak, and J.~M. Jornet, ``Wireless communications for optogenetics-based brain stimulation: Present technology and future challenges,'' \emph{IEEE Communications Magazine}, vol.~56, no.~7, pp. 218--224, 2018.

\bibitem{wirdatmadja2017wireless}
S.~A. Wirdatmadja, M.~T. Barros, Y.~Koucheryavy, J.~M. Jornet, and S.~Balasubramaniam, ``Wireless optogenetic nanonetworks for brain stimulation: Device model and charging protocols,'' \emph{IEEE transactions on nanobioscience}, vol.~16, no.~8, pp. 859--872, 2017.

\bibitem{jornet2019optogenomic}
J.~M. Jornet, Y.~Bae, C.~R. Handelmann, B.~Decker, A.~Balcerak, A.~Sangwan, P.~Miao, A.~Desai, L.~Feng, E.~K. Stachowiak \emph{et~al.}, ``Optogenomic interfaces: Bridging biological networks with the electronic digital world,'' \emph{Proceedings of the IEEE}, vol. 107, no.~7, pp. 1387--1401, 2019.

\bibitem{miccio2015red}
L.~Miccio, P.~Memmolo, F.~Merola, P.~A. Netti, and P.~Ferraro, ``Red blood cell as an adaptive optofluidic microlens,'' \emph{Nature communications}, vol.~6, no.~1, p. 6502, 2015.

\bibitem{johari2017nanoscale}
P.~Johari and J.~M. Jornet, ``Nanoscale optical wireless channel model for intra-body communications: Geometrical, time, and frequency domain analyses,'' \emph{IEEE Transactions on Communications}, vol.~66, no.~4, pp. 1579--1593, 2017.

\bibitem{wirdatmadja2019analysis}
S.~Wirdatmadja, P.~Johari, A.~Desai, Y.~Bae, E.~K. Stachowiak, M.~K. Stachowiak, J.~M. Jornet, and S.~Balasubramaniam, ``Analysis of light propagation on physiological properties of neurons for nanoscale optogenetics,'' \emph{IEEE Transactions on Neural Systems and Rehabilitation Engineering}, vol.~27, no.~2, pp. 108--117, 2019.

\bibitem{sangwan2022joint}
A.~Sangwan and J.~M. Jornet, ``Joint communication and bio-sensing with plasmonic nano-systems to prevent the spread of infectious diseases in the internet of nano-bio things,'' \emph{IEEE Journal on Selected Areas in Communications}, vol.~40, no.~11, pp. 3271--3284, 2022.

\bibitem{elayan2023terahertz}
H.~Elayan, A.~W. Eckford, and R.~Adve, ``Terahertz induced protein interactions in a random medium,'' \emph{IEEE Transactions on Molecular, Biological and Multi-Scale Communications}, 2023.

\bibitem{elayan2017photothermal}
H.~Elayan, P.~Johari, R.~M. Shubair, and J.~M. Jornet, ``Photothermal modeling and analysis of intrabody terahertz nanoscale communication,'' \emph{IEEE transactions on nanobioscience}, vol.~16, no.~8, pp. 755--763, 2017.

\bibitem{yu2022magnetoelectric}
Z.~Yu, F.~T. Alrashdan, W.~Wang, M.~Parker, X.~Chen, F.~Y. Chen, J.~Woods, Z.~Chen, J.~T. Robinson, and K.~Yang, ``Magnetoelectric backscatter communication for millimeter-sized wireless biomedical implants,'' in \emph{Proceedings of the 28th Annual International Conference on Mobile Computing And Networking}, 2022, pp. 432--445.

\bibitem{gong2024}
Z.~Gong, Z.~An, D.~Dai, J.~Tong, S.~Long, and L.~Yang, ``Enabling cross-medium wireless networks with miniature mechanical antennas,'' in \emph{Proceedings of the 30th Annual International Conference on Mobile Computing And Networking}, 2024.

\bibitem{canovas2020understanding}
S.~Canovas-Carrasco, R.~Asorey-Cacheda, A.-J. Garcia-Sanchez, J.~Garcia-Haro, K.~Wojcik, and P.~Kulakowski, ``Understanding the applicability of terahertz flow-guided nano-networks for medical applications,'' \emph{IEEE Access}, vol.~8, pp. 214\,224--214\,239, 2020.

\bibitem{garcia2023dynamic}
A.-J. Garcia-Sanchez, R.~Asorey-Cacheda, J.~Garcia-Haro, and J.-L. Gomez-Tornero, ``Dynamic multihop routing in terahertz flow-guided nanosensor networks: A reinforcement learning approach,'' \emph{IEEE Sensors Journal}, vol.~23, no.~4, pp. 3408--3422, 2023.

\bibitem{gomez2023optimizing}
J.~T. G{\'o}mez, J.~Simonjan, J.~M. Jornet, and F.~Dressler, ``Optimizing terahertz communication between nanosensors in the human cardiovascular system and external gateways,'' \emph{IEEE Communications Letters}, 2023.

\bibitem{Nychis2012}
G.~Nychis, C.~Fallin, T.~Moscibroda, O.~Mutlu, and S.~Seshan, ``{On-chip networks from a networking perspective: congestion and scalability in many-core interconnects},'' in \emph{Proceedings of the SIGCOMM}, 2012, pp. 407--18.

\bibitem{Shao2019}
Y.~S. Shao, J.~Clemons, and {R. Venkatesan \emph{et al.}}, ``{Simba: Scaling Deep-Learning Inference with Multi-Chip-Module-Based Architecture},'' in \emph{Proceedings of the MICRO-52}, 2019.

\bibitem{Shamim2017}
S.~Shamim, N.~Mansoor, R.~S. Narde, V.~Kothandapani, A.~Ganguly, and J.~Venkataraman, ``{A Wireless Interconnection Framework for Seamless Inter and Intra-chip Communication in Multichip Systems},'' \emph{IEEE Transactions on Computers}, vol.~66, no.~3, pp. 389--402, 2017.

\bibitem{nafari2017chip}
M.~Nafari, L.~Feng, and J.~M. Jornet, ``On-chip wireless optical channel modeling for massive multi-core computing architectures,'' in \emph{2017 IEEE Wireless Communications and Networking Conference (WCNC)}.\hskip 1em plus 0.5em minus 0.4em\relax IEEE, 2017, pp. 1--6.

\bibitem{Petrov2017}
V.~Petrov, D.~Moltchanov, M.~Komar, A.~Antonov, P.~Kustarev, S.~Rakheja, and Y.~Koucheryavy, ``{Terahertz Band Intra-Chip Communications: Can Wireless Links Scale Modern x86 CPUs?}'' \emph{IEEE Access}, vol.~5, no.~c, pp. 6095--6109, 2017.

\bibitem{Timoneda2018ADAPT}
X.~Timoneda, S.~Abadal, A.~Franques, D.~Manessis, J.~Zhou, J.~Torrellas, E.~Alarc{\'{o}}n, and A.~Cabellos-Aparicio, ``{Engineer the Channel and Adapt to it: Enabling Wireless Intra-Chip Communication},'' \emph{IEEE Transactions on Communications}, vol.~68, no.~5, pp. 3247--3258, 2020.

\bibitem{imani2021smart}
M.~F.~Imani, S.~Abadal, and P.~Del~Hougne, ``Metasurface-programmable wireless network-on-chip,'' \emph{Advanced Science}, vol.~9, no.~26, p. 2201458, 2022.

\bibitem{rodriguez2023collective}
F.~Rodr{\'\i}guez-Gal{\'a}n, A.~Bandara, E.~P. De~Santana, P.~H. Bol{\'\i}var, E.~Alarc{\'o}n, and S.~Abadal, ``Collective communication patterns using time-reversal terahertz links at the chip scale,'' in \emph{GLOBECOM 2023-2023 IEEE Global Communications Conference}.\hskip 1em plus 0.5em minus 0.4em\relax IEEE, 2023, pp. 5098--5103.

\bibitem{jog2021one}
S.~Jog, Z.~Liu, A.~Franques, V.~Fernando, S.~Abadal, J.~Torrellas, and H.~Hassanieh, ``One protocol to rule them all: Wireless network-on-chip using deep reinforcement learning,'' in \emph{18th USENIX Symposium on Networked Systems Design and Implementation (NSDI 21)}, 2021, pp. 973--989.

\bibitem{fernando2019replica}
V.~Fernando, A.~Franques, S.~Abadal, S.~Misailovic, and J.~Torrellas, ``Replica: A wireless manycore for communication-intensive and approximate data,'' in \emph{Proceedings of the Twenty-Fourth International Conference on Architectural Support for Programming Languages and Operating Systems}, 2019, pp. 849--863.

\bibitem{Wang2023}
J.~Wang, M.~I. Ibrahim, I.~B. Harris, N.~M. Monroe, M.~I.~W. Khan, X.~Yi, D.~R. Englund, and R.~Han, ``{THz Cryo-CMOS Backscatter Transceiver: A Contactless 4 Kelvin-300 Kelvin Data Interface},'' in \emph{Proc. of ISSCC '23}.\hskip 1em plus 0.5em minus 0.4em\relax IEEE, 2023, pp. 14--16.

\bibitem{Low_2019}
G.~H. Low and I.~L. Chuang, ``Hamiltonian simulation by qubitization,'' \emph{Quantum}, vol.~3, p. 163, jul 2019.

\bibitem{Peruzzo_2014}
A.~Peruzzo, J.~McClean, P.~Shadbolt, M.-H. Yung, X.-Q. Zhou, P.~J. Love, A.~Aspuru-Guzik, and J.~L. O’brien, ``A variational eigenvalue solver on a photonic quantum processor,'' \emph{Nature Communications}, vol.~5, no.~1, jul 2014.

\bibitem{Woerner_2019}
S.~Woerner \emph{et~al.}, ``Quantum risk analysis,'' \emph{npj Quantum Information}, vol.~5, no.~1, 2019.

\bibitem{fi15030094}
R.~Ur~Rasool, H.~F. Ahmad, W.~Rafique, A.~Qayyum, J.~Qadir, and Z.~Anwar, ``Quantum computing for healthcare: A review,'' \emph{Future Internet}, vol.~15, no.~3, 2023.

\bibitem{choi2023ibm}
C.~Q. Choi, ``Ibm's quantum leap: The company will take quantum tech past the 1,000-qubit mark in 2023,'' \emph{IEEE Spectrum}, vol.~60, no.~1, pp. 46--47, 2023.

\bibitem{rodrigo2021double}
S.~Rodrigo, S.~Abadal, E.~Alarcon, M.~Bandic, H.~Van~Someren, and C.~G. Almud{\'e}ver, ``On double full-stack communication-enabled architectures for multicore quantum computers,'' \emph{IEEE micro}, vol.~41, no.~5, pp. 48--56, 2021.

\bibitem{smith2022scaling}
K.~N. Smith, G.~S. Ravi, J.~M. Baker, and F.~T. Chong, ``Scaling superconducting quantum computers with chiplet architectures,'' in \emph{Proceedings of the IEEE MICRO-55}, 2022, pp. 1092--1109.

\bibitem{jnane2022multicore}
H.~Jnane, B.~Undseth, Z.~Cai, S.~C. Benjamin, and B.~Koczor, ``Multicore quantum computing,'' \emph{Physical Review Applied}, vol.~18, no.~4, p. 044064, 2022.

\bibitem{gold2021entanglement}
A.~Gold, J.~Paquette, A.~Stockklauser, M.~J. Reagor, M.~S. Alam, A.~Bestwick, N.~Didier, A.~Nersisyan, F.~Oruc, A.~Razavi \emph{et~al.}, ``Entanglement across separate silicon dies in a modular superconducting qubit device,'' 2021.

\bibitem{escofet2023interconnect}
P.~Escofet, S.~B. Rached, S.~Rodrigo, C.~G. Almudever, E.~Alarc{\'o}n, and S.~Abadal, ``Interconnect fabrics for multi-core quantum processors: A context analysis,'' in \emph{Proceedings of the 16th International Workshop on Network on Chip Architectures}, 2023, pp. 34--39.

\bibitem{rached2024spatio}
S.~B. Rached, C.~G. Almudever, E.~Alarcon, and S.~Abadal, ``Spatio-temporal characterization of qubit routing in connectivity-constrained quantum processors,'' \emph{Proceedings of ISCAS '24}, 2024.

\bibitem{rodrigo2022characterizing}
S.~Rodrigo \emph{et~al.}, ``Characterizing the spatio-temporal qubit traffic of a quantum intranet aiming at modular quantum computer architectures,'' in \emph{Proceedings of the ACM NANOCOM}, 2022.

\bibitem{nakamura_1999_coherent}
Y.~Nakamura, Y.~A. Pashkin, and J.~Tsai, ``Coherent control of macroscopic quantum states in a single-cooper-pair box,'' \emph{Nature}, vol. 398, pp. 786--788, 04 1999.

\bibitem{pieter_2007_linear}
P.~Kok, W.~J. Munro, K.~Nemoto, T.~C. Ralph, J.~P. Dowling, and G.~J. Milburn, ``Linear optical quantum computing with photonic qubits,'' \emph{Rev. Mod. Phys.}, vol.~79, pp. 135--174, Jan 2007.

\bibitem{PhysRevLettQDS}
A.~Imamog, D.~D. Awschalom, G.~Burkard, D.~P. DiVincenzo, D.~Loss, M.~Sherwin, A.~Small \emph{et~al.}, ``Quantum information processing using quantum dot spins and cavity qed,'' \emph{Phys. Rev. Lett.}, vol.~83, pp. 4204--4207, Nov 1999.

\bibitem{PhysRevLettTI}
J.~I. Cirac and P.~Zoller, ``Quantum computations with cold trapped ions,'' \emph{Phys. Rev. Lett.}, vol.~74, pp. 4091--4094, May 1995.

\bibitem{kaushal2019shuttlingbased}
V.~Kaushal, B.~Lekitsch, A.~Stahl, J.~Hilder, D.~Pijn, C.~Schmiegelow, A.~Bermudez, M.~M{\"u}ller, F.~Schmidt-Kaler, and U.~Poschinger, ``Shuttling-based trapped-ion quantum information processing,'' \emph{AVS Quantum Science}, vol.~2, no.~1, 2020.

\bibitem{llewellyn2020chip}
D.~Llewellyn, Y.~Ding, I.~I. Faruque, S.~Paesani, D.~Bacco, R.~Santagati, Y.-J. Qian, Y.~Li, Y.-F. Xiao, M.~Huber \emph{et~al.}, ``Chip-to-chip quantum teleportation and multi-photon entanglement in silicon,'' \emph{Nature Physics}, vol.~16, no.~2, pp. 148--153, 2020.

\bibitem{c877b567b53e44cda1a0346b58a90d0f}
C.~Zhou, P.~Lu, M.~Praquin, T.-C. Chien, R.~Kaufman, X.~Cao, M.~Xia, R.~S. Mong, W.~Pfaff, D.~Pekker \emph{et~al.}, ``\BIBforeignlanguage{English (US)}{Realizing all-to-all couplings among detachable quantum modules using a microwave quantum state router},'' \emph{\BIBforeignlanguage{English (US)}{npj Quantum Information}}, vol.~9, no.~1, 2023.

\bibitem{marinelli2023dynamically}
B.~Marinelli, J.~Luo, H.~Ren, B.~M. Niedzielski, D.~K. Kim, R.~Das, M.~Schwartz, D.~I. Santiago, and I.~Siddiqi, ``Dynamically reconfigurable photon exchange in a superconducting quantum processor,'' \emph{arXiv preprint arXiv:2303.03507}, 2023.

\bibitem{Krantz_2019}
P.~Krantz, M.~Kjaergaard, F.~Yan, T.~P. Orlando, S.~Gustavsson, and W.~D. Oliver, ``A quantum engineer{\textquotesingle}s guide to superconducting qubits,'' \emph{Applied Physics Reviews}, vol.~6, no.~2, jun 2019.

\bibitem{potovcnik2021millikelvin}
A.~Poto{\v{c}}nik, S.~Brebels, J.~Verjauw, R.~Acharya, A.~Grill, D.~Wan, M.~Mongillo, R.~Li, T.~Ivanov, S.~Van~Winckel \emph{et~al.}, ``Millikelvin temperature cryo-cmos multiplexer for scalable quantum device characterisation,'' \emph{Quantum Science and Technology}, vol.~7, no.~1, p. 015004, 2021.

\bibitem{lecocq2021control}
F.~Lecocq, F.~Quinlan, K.~Cicak, J.~Aumentado, S.~Diddams, and J.~Teufel, ``Control and readout of a superconducting qubit using a photonic link,'' \emph{Nature}, vol. 591, no. 7851, pp. 575--579, 2021.

\bibitem{magnard2020microwave}
P.~Magnard, S.~Storz, P.~Kurpiers, J.~Sch{\"a}r, F.~Marxer, J.~L{\"u}tolf, T.~Walter, J.-C. Besse, M.~Gabureac, K.~Reuer \emph{et~al.}, ``Microwave quantum link between superconducting circuits housed in spatially separated cryogenic systems,'' \emph{Physical Review Letters}, vol. 125, no.~26, p. 260502, 2020.

\bibitem{calo2022reconfigurable}
G.~Cal{\`o}, L.~Gabriele, G.~Bellanca, J.~Nanni, M.~Barbiroli, F.~Fuschini, V.~Tralli, D.~Bertozzi, G.~Serafino, and V.~Petruzzelli, ``Reconfigurable optical wireless switches for on-chip interconnection,'' \emph{IEEE Journal of Quantum Electronics}, 2022.

\bibitem{DiTomaso2015}
D.~DiTomaso, A.~Kodi, D.~Matolak, S.~Kaya, S.~Laha, and W.~Rayess, ``{A-WiNoC: Adaptive Wireless Network-on-Chip Architecture for Chip Multiprocessors},'' \emph{IEEE Transactions on Parallel and Distributed Systems}, vol.~26, no.~12, pp. 3289--3302, 2015.

\bibitem{pellegrini2013antennas}
A.~Pellegrini, A.~Brizzi, L.~Zhang, K.~Ali, Y.~Hao, X.~Wu, C.~Constantinou, Y.~Nechayev, P.~Hall, N.~Chahat \emph{et~al.}, ``Antennas and propagation for body-centric wireless communications at millimeter-wave frequencies: A review [wireless corner],'' \emph{IEEE Antennas and Propagation Magazine}, vol.~55, no.~4, pp. 262--287, 2013.

\bibitem{extra_chang2008}
M.~F. Chang, J.~Cong, A.~Kaplan, M.~Naik, G.~Reinman, E.~Socher, and S.-W. Tam, ``{CMP} network-on-chip overlaid with multi-band rf-interconnect,'' in \emph{2008 IEEE 14th International Symposium on High Performance Computer Architecture}, 2008, pp. 191--202.

\bibitem{extra_Mineo2016}
A.~Mineo, M.~Palesi, G.~Ascia, and V.~Catania, ``Runtime tunable transmitting power technique in mm-wave winoc architectures,'' \emph{IEEE Transactions on Very Large Scale Integration (VLSI) Systems}, vol.~24, no.~4, pp. 1535--1545, 2016.

\bibitem{Baniya2019}
P.~Baniya and K.~L. Melde, ``{Switched-Beam Endfire Planar Array With Integrated 2-D Butler Matrix for 60 GHz Chip-to-Chip Space-Surface Wave Communications},'' \emph{IEEE Antennas and Wireless Propagation Letters}, vol.~18, no.~2, pp. 236--240, 2019.

\bibitem{Duraisamy2016}
K.~Duraisamy, H.~Lu, P.~P. Pande, and A.~Kalyanaraman, ``{High-Performance and Energy-Efficient Network-on-Chip Architectures for Graph Analytics},'' \emph{ACM Trans. Embed. Comput. Syst}, vol.~15, no.~26, pp. 1--26, 2016.

\bibitem{gade2019}
S.~H. Gade, M.~{Meraj Ahmed}, S.~Deb, and A.~Ganguly, ``{Energy efficient chip-To-chip wireless interconnection for heterogeneous architectures},'' \emph{ACM Transactions on Design Automation of Electronic Systems}, vol.~24, no.~5, 2019.

\bibitem{37}
N.~B. Asan, J.~Velander, Y.~Redzwan, R.~Augustine, E.~Hassan, D.~Noreland, T.~Voigt, and T.~J. Blokhuis, ``{Reliability of the fat tissue channel for intra-body microwave communication},'' in \emph{Proceedings of IEEE Conference on Antenna Measurements Applications}, 2017.

\bibitem{38}
N.~B. Asan, C.~P. Penichet, S.~R.~M. Shah, D.~Noreland, E.~Hassan, A.~Rydberg, T.~J. Blokhuis, T.~Voigt, and R.~Augustine, ``{Data packet transmission through fat tissue for wireless intrabody networks},'' \emph{IEEE Journal of Electromagnetics, RF and Microwaves in Medicine and Biology}, vol.~12, 2017.

\bibitem{ganguly2022interconnects}
A.~Ganguly, S.~Abadal, I.~Thakkar, N.~E. Jerger, M.~Riedel, M.~Babaie, R.~Balasubramonian, A.~Sebastian, S.~Pasricha, and B.~Taskin, ``Interconnects for dna, quantum, in-memory, and optical computing: Insights from a panel discussion,'' \emph{IEEE micro}, vol.~42, no.~3, 2022.

\bibitem{jornet2014femtosecond}
J.~M. Jornet and I.~F. Akyildiz, ``Femtosecond-long pulse-based modulation for terahertz band communication in nanonetworks,'' \emph{IEEE Transactions on Communications}, vol.~62, no.~5, pp. 1742--1754, 2014.

\bibitem{han2017sync}
C.~Han, I.~F. Akyildiz, and W.~H. Gerstacker, ``Timing acquisition and error analysis for pulse-based terahertz band wireless systems,'' \emph{IEEE Transactions on Vehicular Technology}, vol.~66, no.~11, pp. 10\,102--10\,113, 2017.

\bibitem{hossain2019stochastic}
Z.~Hossain, C.~N. Mollica, J.~F. Federici, and J.~M. Jornet, ``Stochastic interference modeling and experimental validation for pulse-based terahertz communication,'' \emph{IEEE Transactions on Wireless Communications}, vol.~18, no.~8, pp. 4103--4115, 2019.

\bibitem{jornet2011low}
J.~M. Jornet and I.~F. Akyildiz, ``Low-weight channel coding for interference mitigation in electromagnetic nanonetworks in the terahertz band,'' in \emph{2011 IEEE international conference on communications (ICC)}.\hskip 1em plus 0.5em minus 0.4em\relax IEEE, 2011, pp. 1--6.

\bibitem{jornet2012phlame}
J.~M. Jornet, J.~{Capdevila Pujol}, and J.~{Solé Pareta}, ``{PHLAME: A} physical layer aware mac protocol for electromagnetic nanonetworks in the terahertz band,'' \emph{Nano Communication Networks}, vol.~3, no.~1, pp. 74--81, 2012.

\bibitem{harv1}
X.~Bu, B.~Zhou, J.~Li, C.~Gao, and J.~Guo, ``Orange peel-like triboelectric nanogenerators with multiscale micro-nano structure for energy harvesting and touch sensing applications,'' \emph{Elsevier Nano Energy}, vol. 122, April 2024.

\bibitem{harv2}
A.~Ali, S.~Ali, H.~Shaukat, E.~Khalid, L.~Behram, H.~Rani, W.~A. Altabey, S.~A. Kouritem, and M.~Noori, ``Advancements in piezoelectric wind energy harvesting: A review,'' \emph{Elsevier Results in Engineering}, vol.~21, March 2024.

\bibitem{harv3}
S.~Ajori and F.~Sadeghi, ``Design of high-frequency carbon nanotube–carbon nanotorus oscillators for energy harvesting: A molecular dynamics study,'' \emph{Langmuir 2024}, vol.~40, February 2024.

\bibitem{harv4}
G.~de~Marzo, V.~M. Mastronardi, M.~T. Todaro, L.~Blasi, V.~Antonaci, L.~Algieri, M.~Scaraggi, and M.~D. Vittorio, ``Sustainable electronic biomaterials for body-compliant devices: Challenges and perspectives for wearable bio-mechanical sensors and body energy harvesters,'' \emph{Elsevier nano Energy}, vol. 123, May 2024.

\bibitem{pierobon2014routing}
M.~Pierobon, J.~M. Jornet, N.~Akkari, S.~Almasri, and I.~F. Akyildiz, ``A routing framework for energy harvesting wireless nanosensor networks in the terahertz band,'' \emph{Wireless networks}, vol.~20, pp. 1169--1183, 2014.

\bibitem{tsioliaridou2017packet}
A.~Tsioliaridou, C.~Liaskos, E.~Dedu, and S.~Ioannidis, ``Packet routing in 3d nanonetworks: A lightweight, linear-path scheme,'' \emph{Nano communication networks}, vol.~12, pp. 63--71, 2017.

\bibitem{tsioliaridou2015corona}
A.~Tsioliaridou, C.~Liaskos, S.~Ioannidis, and A.~Pitsillides, ``Corona: A coordinate and routing system for nanonetworks,'' in \emph{Proceedings of the second annual international conference on nanoscale computing and communication}, 2015, pp. 1--6.

\bibitem{samman2010adaptive}
F.~A. Samman, T.~Hollstein, and M.~Glesner, ``Adaptive and deadlock-free tree-based multicast routing for networks-on-chip,'' \emph{IEEE Transactions on Very Large Scale Integration (VLSI) Systems}, vol.~18, no.~7, pp. 1067--1080, 2010.

\bibitem{abadal2016wisync}
S.~Abadal, A.~Cabellos-Aparicio, E.~Alarcon, and J.~Torrellas, ``Wisync: An architecture for fast synchronization through on-chip wireless communication,'' \emph{ACM SIGPLAN Notices}, vol.~51, no.~4, pp. 3--17, 2016.

\bibitem{guirado2023whype}
R.~Guirado, A.~Rahimi, G.~Karunaratne, E.~Alarc{\'o}n, A.~Sebastian, and S.~Abadal, ``Whype: A scale-out architecture with wireless over-the-air majority for scalable in-memory hyperdimensional computing,'' \emph{IEEE Journal on Emerging and Selected Topics in Circuits and Systems}, vol.~13, no.~1, pp. 137--149, 2023.

\bibitem{barut2022asymmetrically}
B.~Barut, X.~Cantos-Roman, J.~Crabb, C.-P. Kwan, R.~Dixit, N.~Arabchigavkani, S.~Yin, J.~Nathawat, K.~He, M.~D. Randle \emph{et~al.}, ``Asymmetrically engineered nanoscale transistors for on-demand sourcing of terahertz plasmons,'' \emph{Nano Letters}, vol.~22, no.~7, pp. 2674--2681, 2022.

\bibitem{comp1}
C.~Yang, Z.~Wei, H.~Gao, C.~H. Heng, and Y.~Zheng, ``Towards ultra-low power transceivers for pico-iot,'' \emph{IEEE Nanotechnology Magazine}, vol.~18, no.~1, February 2024.

\bibitem{comp3}
P.~K. Bulasara and S.~R. Sahoo, ``A robust hybrid model with low energy consumption for biosensor nano-networks,'' \emph{Elsevier Journal of King Saud University - Computer and Information Sciences}, vol.~36, no.~1, January 2024.

\bibitem{krishnaswamy2013time}
B.~Krishnaswamy, C.~M. Austin, J.~P. Bardill, D.~Russakow, G.~L. Holst, B.~K. Hammer, C.~R. Forest, and R.~Sivakumar, ``Time-elapse communication: Bacterial communication on a microfluidic chip,'' \emph{IEEE Transactions on Communications}, vol.~61, no.~12, pp. 5139--5151, 2013.

\bibitem{kuscu2015fluorescent}
M.~Kuscu, A.~Kiraz, and O.~B. Akan, ``Fluorescent molecules as transceiver nanoantennas: The first practical and high-rate information transfer over a nanoscale communication channel based on fret,'' \emph{Scientific reports}, vol.~5, no.~1, p. 7831, 2015.

\bibitem{koo2016molecular}
B.-H. Koo, C.~Lee, H.~B. Yilmaz, N.~Farsad, A.~Eckford, and C.-B. Chae, ``Molecular {MIMO}: From theory to prototype,'' \emph{IEEE Journal on Selected Areas in Communications}, vol.~34, no.~3, pp. 600--614, 2016.

\bibitem{deng2017microfluidic}
Y.~Deng, M.~Pierobon, and A.~Nallanathan, ``A microfluidic feed forward loop pulse generator for molecular communication,'' in \emph{Proc. of IEEE Global Communications Conference}, 2017, pp. 1--7.

\bibitem{farsad2017novel}
N.~Farsad, D.~Pan, and A.~Goldsmith, ``A novel experimental platform for in-vessel multi-chemical molecular communications,'' in \emph{Proc. of IEEE Global Communications Conference}, 2017, pp. 1--6.

\bibitem{grebenstein2018biological}
L.~Grebenstein, J.~Kirchner, R.~S. Peixoto, W.~Zimmermann, W.~Wicke, A.~Ahmadzadeh, V.~Jamali, G.~Fischer, R.~Weigel, A.~Burkovski \emph{et~al.}, ``Biological optical-to-chemical signal conversion interface: A small-scale modulator for molecular communications,'' in \emph{Proc. of ACM International Conference on Nanoscale Computing and Communication}, 2018, pp. 1--6.

\bibitem{unterweger2018experimental}
H.~Unterweger, J.~Kirchner, W.~Wicke, A.~Ahmadzadeh, D.~Ahmed, V.~Jamali, C.~Alexiou, G.~Fischer, and R.~Schober, ``Experimental molecular communication testbed based on magnetic nanoparticles in duct flow,'' in \emph{Proc. of IEEE International Workshop on Signal Processing Advances in Wireless Communications (SPAWC)}, 2018, pp. 1--5.

\bibitem{grebenstein2019molecular}
L.~Grebenstein, J.~Kirchner, W.~Wicke, A.~Ahmadzadeh, V.~Jamali, G.~Fischer, R.~Weigel, A.~Burkovski, and R.~Schober, ``A molecular communication testbed based on proton pumping bacteria: Methods and data,'' \emph{IEEE Transactions on Molecular, Biological and Multi-Scale Communications}, vol.~5, no.~1, pp. 56--62, 2019.

\bibitem{guo2020vertical}
W.~Guo, I.~Atthanayake, and P.~Thomas, ``Vertical underwater molecular communications via buoyancy: Gaussian velocity distribution of signal,'' in \emph{Proc. of IEEE International Conference on Communications (ICC)}, 2020, pp. 1--6.

\bibitem{lee2020vessel}
C.~Lee, B.-H. Koo, and C.-B. Chae, ``In-vessel molecular mimo communications,'' in \emph{Proc. of IEEE Wireless Communications and Networking Conference Workshops (WCNCW)}, 2020, pp. 1--2.

\bibitem{koo2020deep}
B.-H. Koo, H.~J. Kim, J.-Y. Kwon, and C.-B. Chae, ``Deep learning-based human implantable nano molecular communications,'' in \emph{Proc. of IEEE International Conference on Communications (ICC)}, 2020, pp. 1--7.

\bibitem{kuscu2021fabrication}
M.~Kuscu, H.~Ramezani, E.~Dinc, S.~Akhavan, and O.~B. Akan, ``Fabrication and microfluidic analysis of graphene-based molecular communication receiver for {Internet of Nano Things (IoNT)},'' \emph{Scientific reports}, vol.~11, no.~1, p. 19600, 2021.

\bibitem{qiu2023review}
S.~Qiu, Z.~Wei, Y.~Huang, M.~Abbaszadeh, J.~Charmet, B.~Li, and W.~Guo, ``Review of physical layer security in molecular internet of nano-things,'' \emph{IEEE Transactions on NanoBioscience}, 2023.

\bibitem{santagati2014sonar}
G.~E. Santagati and T.~Melodia, ``Sonar inside your body: Prototyping ultrasonic intra-body sensor networks,'' in \emph{Proc. of IEEE Conference on Computer Communications}, 2014, pp. 2679--2687.

\bibitem{santagati2016experimental}
------, ``Experimental evaluation of impulsive ultrasonic intra-body communications for implantable biomedical devices,'' \emph{IEEE Transactions on Mobile Computing}, vol.~16, no.~2, pp. 367--380, 2016.

\bibitem{guan2016distributed}
Z.~Guan, G.~E. Santagati, and T.~Melodia, ``Distributed algorithms for joint channel access and rate control in ultrasonic intra-body networks,'' \emph{IEEE/ACM Transactions on Networking}, vol.~24, no.~5, pp. 3109--3122, 2016.

\bibitem{bos2018enabling}
T.~Bos, W.~Jiang, J.~Dahooge, M.~Verhelst, and W.~Dehaene, ``Enabling ultrasound in-body communication: Fir channel models and qam experiments,'' \emph{IEEE transactions on biomedical circuits and systems}, vol.~13, no.~1, pp. 135--144, 2018.

\bibitem{vizziello2024experimental}
A.~Vizziello, P.~Savazzi, R.~R. Guerra, and F.~Dell’Acqua, ``Experimental channel characterization of human body communication based on measured impulse response,'' \emph{IEEE Transactions on Communications}, 2024.

\bibitem{phant1}
V.~Mattsson, M.~D. Perez, L.~Joseph, and R.~Augustine, ``Machine learning algorithm to extract properties of ate phantoms from microwave measurements,'' \emph{International Journal of Microwave and Wireless Technologies}, January 2024.

\bibitem{phant2}
A.~Hamdi, A.~Nahali, and R.~Brahem, ``Optimal and efficient sensor design for 5g-based internet-of-body healthcare monitoring network,'' \emph{Springer Journal of Network and Systems Management}, vol.~32, January 2024.

\bibitem{phant3}
M.~Särestöniemi, D.~Singh, R.~Dessai, C.~Heredia, S.~Myllymäki, and T.~Myllylä, ``Realistic 3d phantoms for validation of microwave sensing in health monitoring applications,'' \emph{MDPI Sensors}, March 2024.

\bibitem{phant4}
B.~Li, Y.~Wang, J.~Zhao, and J.~Shi, ``Ultra-wideband antennas for wireless capsule endoscope system: A review,'' \emph{IEEE Open Access Journal of Antennas and Propagation}, vol.~5, no.~2, April 2024.

\bibitem{phant5}
G.~E. Santagati, N.~Dave, and T.~Melodia, ``Design and performance evaluation of an implantable ultrasonic networking platform for the internet of medical things,'' \emph{IEEE/ACM Transactions on Networking}, vol.~28, no.~2, February 2020.

\bibitem{vasisht2018body}
D.~Vasisht, G.~Zhang, O.~Abari, H.-M. Lu, J.~Flanz, and D.~Katabi, ``In-body backscatter communication and localization,'' in \emph{Proceedings of the 2018 Conference of the ACM Special Interest Group on Data Communication}, 2018, pp. 132--146.

\bibitem{tao2023magnetic}
B.~Tao, E.~Sie, J.~Shenoy, and D.~Vasisht, ``Magnetic backscatter for in-body communication and localization,'' in \emph{Proceedings of the 29th Annual International Conference on Mobile Computing and Networking}, 2023, pp. 1--15.

\bibitem{sen2020teranova}
P.~Sen, D.~A. Pados, S.~N. Batalama, E.~Einarsson, J.~P. Bird, and J.~M. Jornet, ``The teranova platform: An integrated testbed for ultra-broadband wireless communications at true terahertz frequencies,'' \emph{Computer Networks}, vol. 179, p. 107370, 2020.

\bibitem{kim2016statistical}
S.~Kim and A.~Zaji{\'c}, ``Statistical modeling and simulation of short-range device-to-device communication channels at sub-thz frequencies,'' \emph{IEEE Transactions on Wireless Communications}, vol.~15, no.~9, pp. 6423--6433, 2016.

\bibitem{sahin2021evaluation}
E.~Sahin, O.~Dagdeviren, and M.~A. Akkas, ``An evaluation of internet of nano-things simulators,'' in \emph{Proc. of IEEE International Conference on Computer Science and Engineering (UBMK)}, 2021, pp. 670--675.

\bibitem{gul2010nanons}
E.~Gul, B.~Atakan, and O.~B. Akan, ``{NanoNS}: A nanoscale network simulator framework for molecular communications,'' \emph{Nano Communication Networks}, vol.~1, no.~2, pp. 138--156, 2010.

\bibitem{llatser2014n3sim}
I.~Llatser, D.~Demiray, A.~Cabellos-Aparicio, D.~T. Altilar, and E.~Alarc{\'o}n, ``{N3Sim}: Simulation framework for diffusion-based molecular communication nanonetworks,'' \emph{Simulation Modelling Practice and Theory}, vol.~42, pp. 210--222, 2014.

\bibitem{toth20113}
{\'A}.~Toth, D.~B{\'a}nky, and V.~Grolmusz, ``3-{D} brownian motion simulator for high-sensitivity nanobiotechnological applications,'' \emph{IEEE transactions on nanobioscience}, vol.~10, no.~4, pp. 248--249, 2011.

\bibitem{felicetti2012simulation}
L.~Felicetti, M.~Femminella, and G.~Reali, ``A simulation tool for nanoscale biological networks,'' \emph{Nano Communication Networks}, vol.~3, no.~1, pp. 2--18, 2012.

\bibitem{akkaya2014hla}
A.~Akkaya, G.~Genc, and T.~Tugcu, ``{HLA} based architecture for molecular communication simulation,'' \emph{Simulation Modelling Practice and Theory}, vol.~42, pp. 163--177, 2014.

\bibitem{wei2013efficient}
G.~Wei, P.~Bogdan, and R.~Marculescu, ``Efficient modeling and simulation of bacteria-based nanonetworks with{ BNSim},'' \emph{IEEE Journal on Selected Areas in Communications}, vol.~31, no.~12, pp. 868--878, 2013.

\bibitem{jian2017nanons3}
Y.~Jian, B.~Krishnaswamy, C.~M. Austin, A.~O. Bicen, A.~Einolghozati, J.~E. Perdomo, S.~C. Patel, F.~Fekri, I.~F. Akyildiz, C.~R. Forest \emph{et~al.}, ``{nanoNS3}: A network simulator for bacterial nanonetworks based on molecular communication,'' \emph{Nano communication networks}, vol.~12, pp. 1--11, 2017.

\bibitem{piro2013simulating}
G.~Piro, L.~A. Grieco, G.~Boggia, and P.~Camarda, ``Simulating wireless nano sensor networks in the ns-3 platform,'' in \emph{Proc. of IEEE International Conference on Advanced Information Networking and Applications Workshops}, 2013, pp. 67--74.

\bibitem{boillot2013efficient}
N.~Boillot, D.~Dhoutaut, and J.~Bourgeois, ``Efficient simulation environment of wireless radio communications in mems modular robots,'' in \emph{Proc. of IEEE International Conference on Green Computing and Communications and IEEE Internet of Things and IEEE Cyber, Physical and Social Computing}, 2013, pp. 638--645.

\bibitem{boillot2015scalable}
------, ``Scalable simulation of wireless electro-magnetic nanonetworks,'' in \emph{Proc. of IEEE International Conference on Embedded and Ubiquitous Computing}, 2015, pp. 83--89.

\bibitem{hossain2018terasim}
Z.~Hossain, Q.~Xia, and J.~M. Jornet, ``{TeraSim}: An ns-3 extension to simulate terahertz-band communication networks,'' \emph{Nano Communication Networks}, vol.~17, pp. 36--44, 2018.

\bibitem{dhoutaut2018bit}
D.~Dhoutaut, T.~Arrabal, and E.~Dedu, ``Bit simulator, an electromagnetic nanonetworks simulator,'' in \emph{Proc. of ACM International Conference on Nanoscale Computing and Communication}, 2018.

\bibitem{cavalcanti2008medical}
A.~Cavalcanti, B.~Shirinzadeh, and L.~C. Kretly, ``Medical nanorobotics for diabetes control,'' \emph{Nanomedicine: Nanotechnology, Biology and Medicine}, vol.~4, no.~2, pp. 127--138, 2008.

\bibitem{dutta2019lightweight}
I.~K. Dutta, B.~Ghosh, and M.~Bayoumi, ``Lightweight cryptography for internet of insecure things: {A} survey,'' in \emph{Proc. of the IEEE 9th Annual Computing and Communication Workshop and Conference (CCWC)}, 2019, pp. 0475--0481.

\bibitem{pandey2023cryptographyc}
A.~K. Pandey, A.~Banati, B.~Rajendran, S.~D. Sudarsan, and K.~K.~S. Pandian, ``Cryptographic challenges and security in post quantum cryptography migration: {A} prospective approach,'' in \emph{Proc. of the IEEE International Conference on Public Key Infrastructure and its Applications (PKIA)}, 2023, pp. 1--8.

\bibitem{ometov2018multi}
A.~Ometov, S.~Bezzateev, N.~Mäkitalo, S.~Andreev, T.~Mikkonen, and Y.~Koucheryavy, ``Multi-factor authentication: {A} survey,'' \emph{Cryptography}, vol.~2, no.~1, 2018.

\bibitem{ieee_802_15_3c_2009}
IEEE, ``Ieee standard for {I}nformation {T}echnology -- {L}ocal and metropolitan area networks -- {S}pecific requirements-- {P}art 15.3: {A}mendment 2: {M}illimeter-wave-based alternative physical layer extension,'' 802.15.3c--2009, 2009.

\bibitem{ieee_802_15_3d_2017}
------, ``Ieee standard for high data rate wireless multi-media networks -- {A}mendment 2: 100 gb/s wireless switched point-to-point physical layer,'' 802.15.3d--2017, 2017.

\bibitem{petrov2020ieee}
V.~Petrov, T.~Kurner, and I.~Hosako, ``Ieee 802.15.3d: First standardization efforts for sub-terahertz band communications toward 6g,'' \emph{IEEE Communications Magazine}, vol.~58, no.~11, pp. 28--33, 2020.

\bibitem{ieee_802_15_3_2023}
IEEE, ``Ieee standard for wireless multimedia networks,'' 802.15.3--2023, 2024.

\bibitem{3gpp_tr_38_848}
3GPP, ``{Study on Ambient IoT (Internet of Things) in RAN},'' TR 38.848, V18.0.0, 2023.

\bibitem{song2022advances}
C.~Song, Y.~Ding, A.~Eid, J.~G.~D. Hester, X.~He, R.~Bahr, A.~Georgiadis, G.~Goussetis, and M.~M. Tentzeris, ``Advances in wirelessly powered backscatter communications: {F}rom antenna/{RF} circuitry design to printed flexible electronics,'' \emph{Proceedings of the IEEE}, vol. 110, no.~1, pp. 171--192, 2022.

\bibitem{akyildiz2018combating}
I.~F. Akyildiz, C.~Han, and S.~Nie, ``Combating the distance problem in the millimeter wave and terahertz frequency bands,'' \emph{IEEE Communications Magazine}, vol.~56, no.~6, pp. 102--108, 2018.

\bibitem{liaskos2018using}
C.~Liaskos, A.~Tsioliaridou, A.~Pitsillides, S.~Ioannidis, and I.~Akyildiz, ``Using any surface to realize a new paradigm for wireless communications,'' \emph{Communications of the ACM}, vol.~61, no.~11, pp. 30--33, 2018.

\bibitem{pan2021reconfigurable}
C.~Pan, H.~Ren, K.~Wang, J.~F. Kolb, M.~Elkashlan, M.~Chen, M.~Di~Renzo, Y.~Hao, J.~Wang, A.~L. Swindlehurst \emph{et~al.}, ``Reconfigurable intelligent surfaces for 6g systems: Principles, applications, and research directions,'' \emph{IEEE Communications Magazine}, vol.~59, no.~6, pp. 14--20, 2021.

\bibitem{abadal2020programmable}
S.~Abadal, T.-J. Cui, T.~Low, and J.~Georgiou, ``Programmable metamaterials for software-defined electromagnetic control: Circuits, systems, and architectures,'' \emph{IEEE Journal on Emerging and Selected Topics in Circuits and Systems}, vol.~10, no.~1, pp. 6--19, 2020.

\bibitem{liaskos2015design}
C.~Liaskos, A.~Tsioliaridou, A.~Pitsillides, I.~F. Akyildiz, N.~V. Kantartzis, A.~X. Lalas, X.~Dimitropoulos, S.~Ioannidis, M.~Kafesaki, and C.~Soukoulis, ``Design and development of software defined metamaterials for nanonetworks,'' \emph{IEEE Circuits and Systems Magazine}, vol.~15, no.~4, pp. 12--25, 2015.

\bibitem{petrou2022first}
L.~Petrou, K.~M. Kossifos, M.~A. Antoniades, and J.~Georgiou, ``The first family of application-specific integrated circuits for programmable and reconfigurable metasurfaces,'' \emph{Scientific reports}, vol.~12, no.~1, p. 5826, 2022.

\bibitem{liaskos2019absense}
C.~Liaskos, G.~Pyrialakos, A.~Pitilakis, S.~Abadal, A.~Tsioliaridou, A.~Tasolamprou, O.~Tsilipakos, N.~Kantartzis, S.~Ioannidis, E.~Alarcon \emph{et~al.}, ``Absense: Sensing electromagnetic waves on metasurfaces via ambient compilation of full absorption,'' in \emph{Proceedings of the Sixth Annual ACM International Conference on Nanoscale Computing and Communication}, 2019, pp. 1--6.

\bibitem{taghvaee2022multiwideband}
H.~Taghvaee, A.~Pitilakis, O.~Tsilipakos, A.~C. Tasolamprou, N.~V. Kantartzis, M.~Kafesaki, A.~Cabellos-Aparicio, E.~Alarcon, and S.~Abadal, ``Multiwideband terahertz communications via tunable graphene-based metasurfaces in 6g networks: Graphene enables ultimate multiwideband thz wavefront control,'' \emph{IEEE Vehicular Technology Magazine}, vol.~17, no.~2, pp. 16--25, 2022.

\bibitem{ahmed2023survey}
M.~Ahmed, A.~Wahid, S.~S. Laique, W.~U. Khan, A.~Ihsan, F.~Xu, S.~Chatzinotas, and Z.~Han, ``A survey on star-ris: Use cases, recent advances, and future research challenges,'' \emph{IEEE Internet of Things Journal}, 2023.

\bibitem{an2023stacked}
J.~An, C.~Xu, D.~W.~K. Ng, G.~C. Alexandropoulos, C.~Huang, C.~Yuen, and L.~Hanzo, ``Stacked intelligent metasurfaces for efficient holographic mimo communications in 6g,'' \emph{IEEE Journal on Selected Areas in Communications}, 2023.

\bibitem{zheng2023zero}
Y.~Zheng, S.~A. Tegos, Y.~Xiao, P.~D. Diamantoulakis, Z.~Ma, and G.~K. Karagiannidis, ``Zero-energy device networks with wireless-powered riss,'' \emph{IEEE Transactions on Vehicular Technology}, 2023.

\bibitem{tyrovolas2023zero}
D.~Tyrovolas, S.~A. Tegos, V.~K. Papanikolaou, Y.~Xiao, P.-V. Mekikis, P.~D. Diamantoulakis, S.~Ioannidis, C.~K. Liaskos, and G.~K. Karagiannidis, ``Zero-energy reconfigurable intelligent surfaces (zeris),'' \emph{IEEE Transactions on Wireless Communications}, 2023.

\bibitem{ntontin2022wireless}
K.~Ntontin, A.-A.~A. Boulogeorgos, E.~Bj{\"o}rnson, W.~A. Martins, S.~Kisseleff, S.~Abadal, E.~Alarc{\'o}n, A.~Papazafeiropoulos, F.~I. Lazarakis, and S.~Chatzinotas, ``Wireless energy harvesting for autonomous reconfigurable intelligent surfaces,'' \emph{IEEE Transactions on Green Communications and Networking}, vol.~7, no.~1, pp. 114--129, 2022.

\bibitem{tasolamprou2019exploration}
A.~C. Tasolamprou, A.~Pitilakis, S.~Abadal, O.~Tsilipakos, X.~Timoneda, H.~Taghvaee, M.~S. Mirmoosa, F.~Liu, C.~Liaskos, A.~Tsioliaridou \emph{et~al.}, ``Exploration of intercell wireless millimeter-wave communication in the landscape of intelligent metasurfaces,'' \emph{IEEE access}, vol.~7, pp. 122\,931--122\,948, 2019.

\bibitem{kouzapas2020towards}
D.~Kouzapas, C.~Skitsas, T.~Saeed, V.~Soteriou, M.~Lestas, A.~Philippou, S.~Abadal, C.~Liaskos, L.~Petrou, J.~Georgiou \emph{et~al.}, ``Towards fault adaptive routing in metasurface controller networks,'' \emph{Journal of Systems Architecture}, vol. 106, p. 101703, 2020.

\bibitem{saeed2021workload}
T.~Saeed, S.~Abadal, C.~Liaskos, A.~Pitsillides, H.~Taghvaee, A.~Cabellos-Aparicio, V.~Soteriou, E.~Alarcon, I.~F. Akyildiz, and M.~Lestas, ``Workload characterization and traffic analysis for reconfigurable intelligent surfaces within 6g wireless systems,'' \emph{IEEE Transactions on Mobile Computing}, vol.~22, no.~5, pp. 3079--3094, 2021.

\end{thebibliography}



\begin{IEEEbiography}
[{\includegraphics[width=1in,height=1.25in,clip,keepaspectratio]{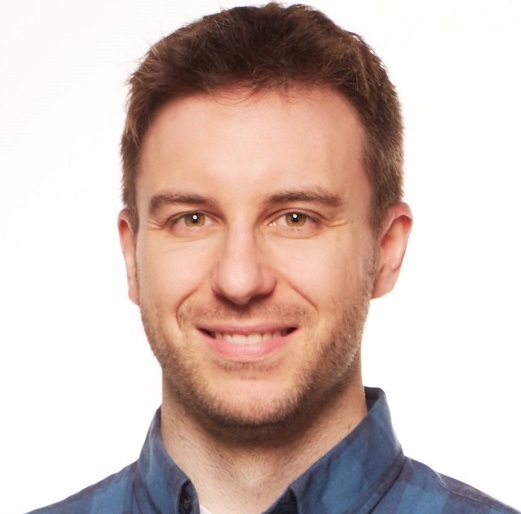}}]{Sergi Abadal} received the PhD in Computer Architecture from the Department of Computer Architecture, Universitat Polit\`{e}cnica de Catalunya (UPC), Barcelona, Spain, in July 2016, where he also had received the M.Sc. and B.Sc. in Telecommunication Engineering in 2011 and 2010, respectively. He has held several visiting positions in Georgia Tech in 2009, University of Illinois Urbana-Champaign in 2015 and 2016, and the Foundation of Research and Technology – Hellas in 2018. Currently, he is principal investigator of an ERC Starting Grant and research director of the NaNoNetworking Center of Catalunya (N3Cat) at UPC. He is author of over 140 publications in top-tier conferences and journals. In terms of service, he served in over 50 workshops and conferences including 7 as general chair and 5 as program chair. He is also active in editorial work with different positions including Associate Editor-in-Chief of Digital Media of the IEEE Journal on Emerging and Selected Topics in Circuits and Systems (since 2024) and Associate Editor of the IEEE Transactions on Mobile Computing (since 2023), IEEE Transactions on Molecular, Biological, and Multi-Scale Communications (since 2023), Elsevier’s Nano Communication Networks Journal (since 2018, Editor of the Year 2019). Thanks to his efforts, he received the ACM NanoCom Outstanding Milestone Award in 2022, NEC Labs Fellowship Award in 2022, and the Young Investigator Award of the Nano Communication Networks Journal in 2019. His current research interests are in the areas of chip-scale wireless communications, including channel modeling and protocol design, the application of these techniques for the creation next-generation wireless networks for on-chip or within metamaterials, and their implications at the system design level.
\end{IEEEbiography}

\begin{IEEEbiography}[{\includegraphics[width=1in,height=1.25in,clip,keepaspectratio]{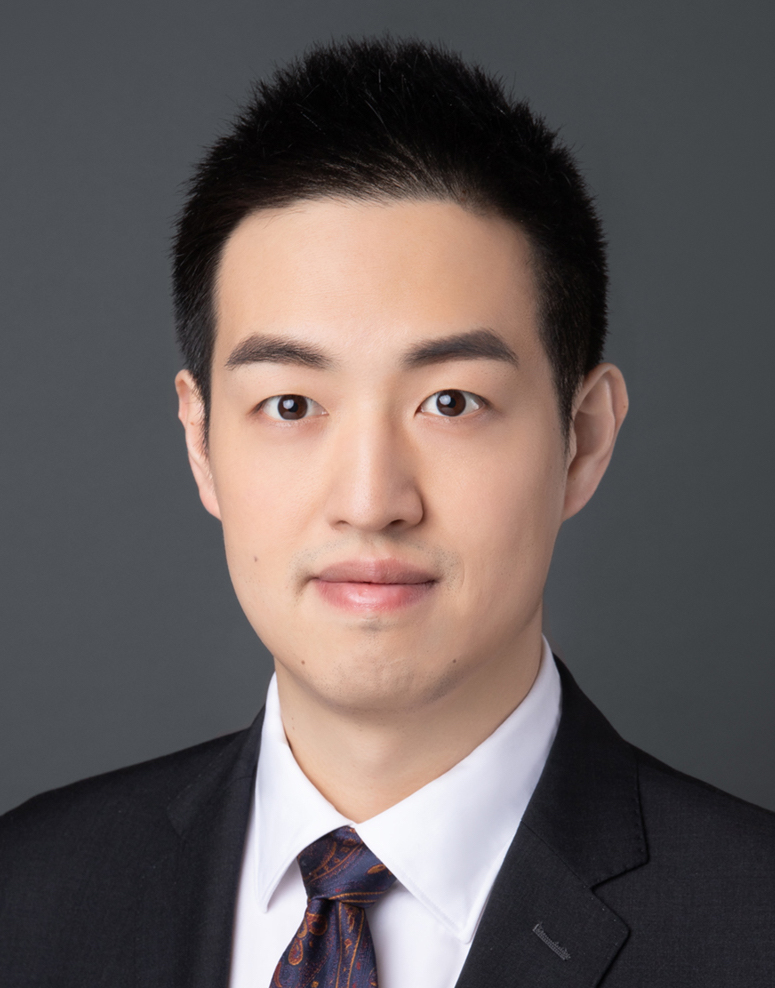}}]{Chong Han} received Ph.D. degree in Electrical and Computer Engineering from Georgia Institute of Technology, USA in 2016. He is currently a John Wu \& Jane Sun Endowed Associate Professor with University of Michigan-Shanghai Jiao Tong University (UM-SJTU) Joint Institute, Shanghai Jiao Tong University, China, and director of the Terahertz Wireless Communications (TWC) Laboratory. Since 2021, he is also affiliated with Department of Electronic Engineering and Cooperative Medianet Innovation Center (CMIC), Shanghai Jiao Tong University. He is a co-founder and vice-chair of IEEE ComSoc Special Interest Group (SIG) on Terahertz Communications, since 2021. He is the recipient of 2024 Friedrich Wilhelm Bessel Research Award from Alexander von Humboldt Foundation in Germany, 2023 IEEE ComSoc Asia-Pacific Outstanding Young Researcher Award, among other wards. He is a (guest) editor with IEEE Trans. Wireless Communications, IEEE JSAC, IEEE JSTSP, etc. He is a TPC chair to organize multiple IEEE and ACM conferences and workshops, including GC’2023 SAC THz communications. His research interests include terahertz and millimeter-wave communications. He is a senior member of IEEE.
\end{IEEEbiography}


\begin{IEEEbiography}[{\includegraphics[width=1in,height=1.25in,clip,keepaspectratio]{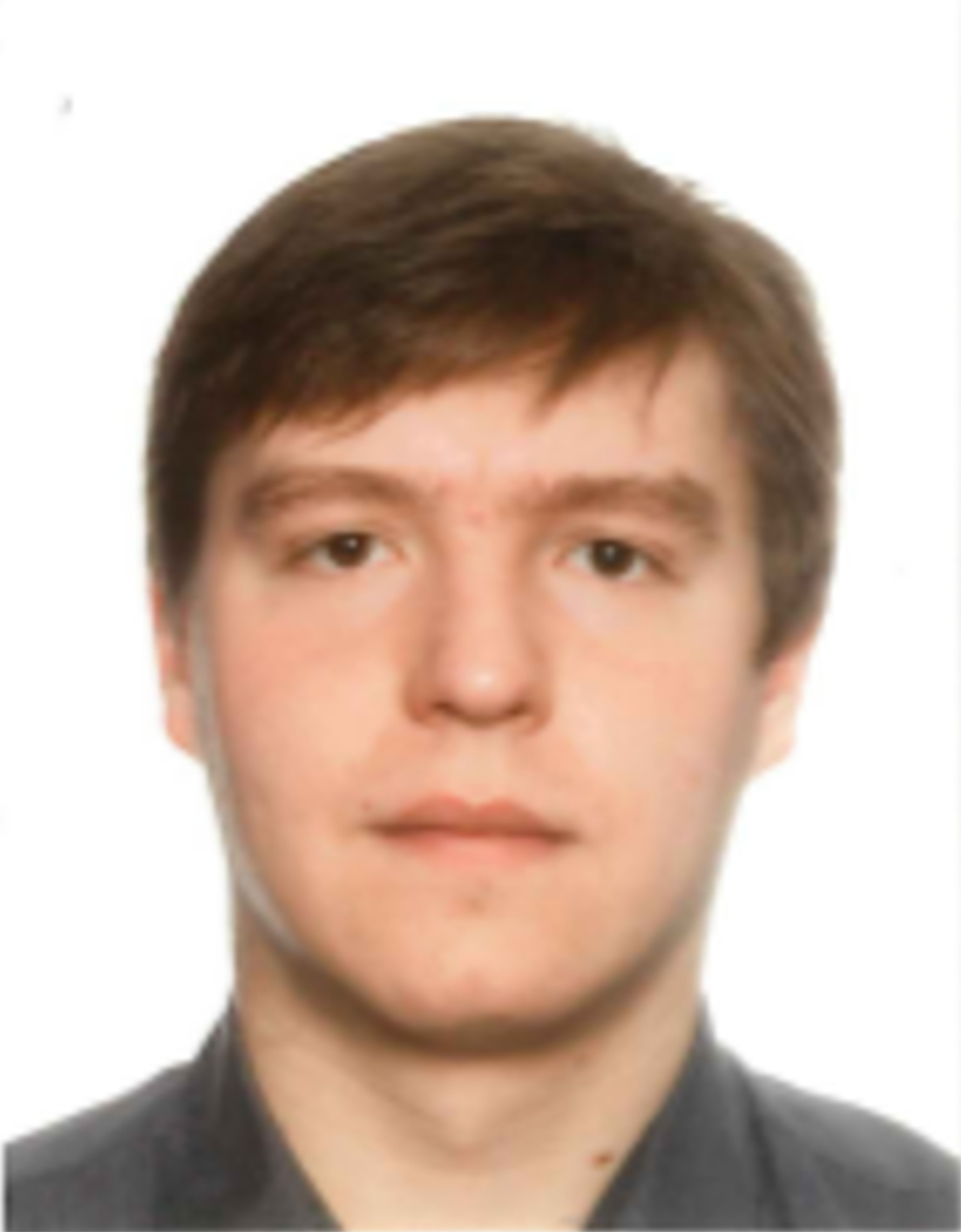}}] {Vitaly Petrov } is an Assistant Professor with the Division of Communication Systems, KTH~Royal Institute of Technology, Stockholm, Sweden. Prior to joining KTH in 2024, he was a Principal Research Scientist at Northeastern University, Boston, MA, USA (2022 -- 2024) and a Senior Standardization Specialist and a 3GPP RAN1 delegate with Nokia Bell Labs and later Nokia Standards (2020 -- 2022). Vitaly obtained his M.Sc. degree in Information Systems Security from SUAI University, St.~Petersburg, Russia, in 2011, his M.Sc. degree in IT and Communications Engineering from Tampere University of Technology, Tampere, Finland, in 2014, and his Ph.D. degree in Communications Engineering from Tampere University, Finland, in 2020. Vitaly has also been a visiting researcher with the University of Texas at Austin, Georgia Institute of Technology, and King’s College London. His current research interests include short-range and long-range terahertz band communications and networking. He is a recipient of the Best Student Paper Award at IEEE~VTC-Fall 2015, the Best Student Poster Award at IEEE~WCNC~2017, and the Best Student Journal Paper Award from IEEE~Finland in 2019.
\end{IEEEbiography}

\begin{IEEEbiography}[{\includegraphics[width=1in,height=1.25in,clip,keepaspectratio]{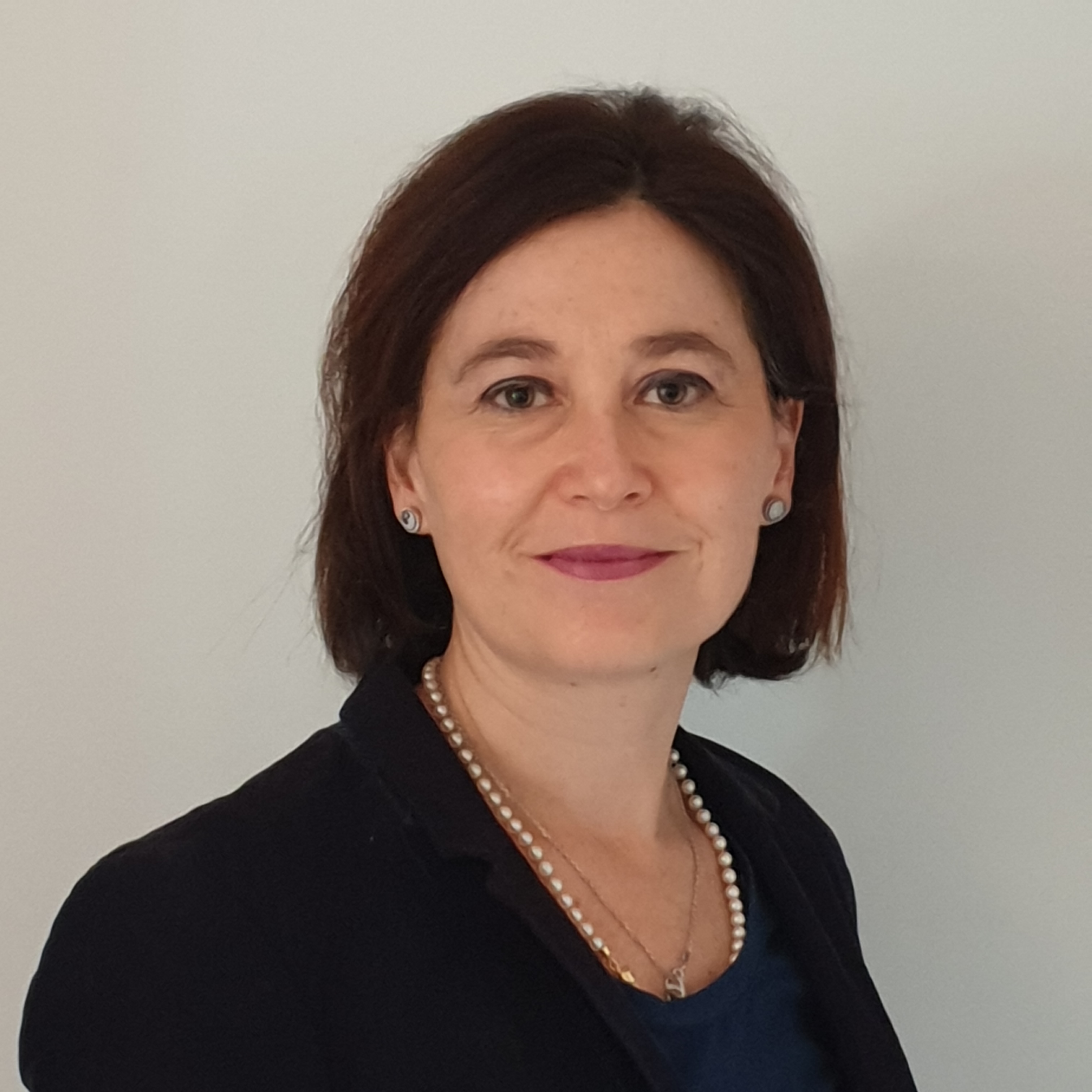}}]{Laura Galluccio}  received her laurea degree in Electrical Engineering from University of Catania, Catania, Italy, in 2001. In March 2005 she got her Ph.D. in Electrical, Computer and Telecommunications Engineering at the same. Since 2002 she is also at the Italian National Consortium of Telecommunications (CNIT), where she worked as a Research Fellow within the VICOM (Virtual Immersive Communications) and the SATNEX Projects. Since November 2010 to October 2019 she has been Assistant Professor at University of Catania. From November 2019 she is Associate Professor at the same university. Her research interests include microfluidic and ultrasonic networks, ad hoc and sensor networks, protocols and algorithms for wireless networks, and network performance analysis.
From May to July 2005 she has been a Visiting Scholar at the COMET Group, Columbia University, NY under the guidance of Prof. Andrew T. Campbell.
In September 2015 she has been Visiting Professor at Central Supelec, Gif-sur-Yvette, Paris. She is senior member of the IEEE.
In 2022 Dr. Galluccio was the recipient of the ACM Nanocom Outstanding Milestone Award for her seminal contribution to neural system interfacing for applications in the Internet of Bio-NanoThings and microfluidic nano networks research. In 2023 Dr. Galluccio has been inserted by the ACM Sigmobile N2Women association in the list of the 10 best female researchers denoted as "Stars in Computer Networking and Communications". 
Dr. Galluccio has been Leading Guest Editor of various special issues in prestigious journals  and magazines such as Elsevier Ad Hoc Networks Journal, IEEE Wireless Communications Magazine, and JSAC. She serves and has served in the Editorial Board of IEEE Transactions on Wireless Communications, IEEE Communication Letters, Elsevier Computer Networks, Wireless Communications and Mobile Computing (WCMC), Wiley and Elsevier Ad Hoc Networks Journal and has been involved in the Organizing Committee and TPCs of numerous conferences such as ACM Nanocom, IEEE Infocom, IEEE ICC, IEEE Globecom, etc.
Dr. Galluccio has been instructors for multiple tutorials in international conferences in the field of intrabody communications and also been Keynote Speaker for the IEEE International Conference on Wireless Communications Signal Processing and Networking (WispNet), March 2021.

\end{IEEEbiography}

\begin{IEEEbiography}[{\includegraphics[width=1in,height=1.25in,clip,keepaspectratio]{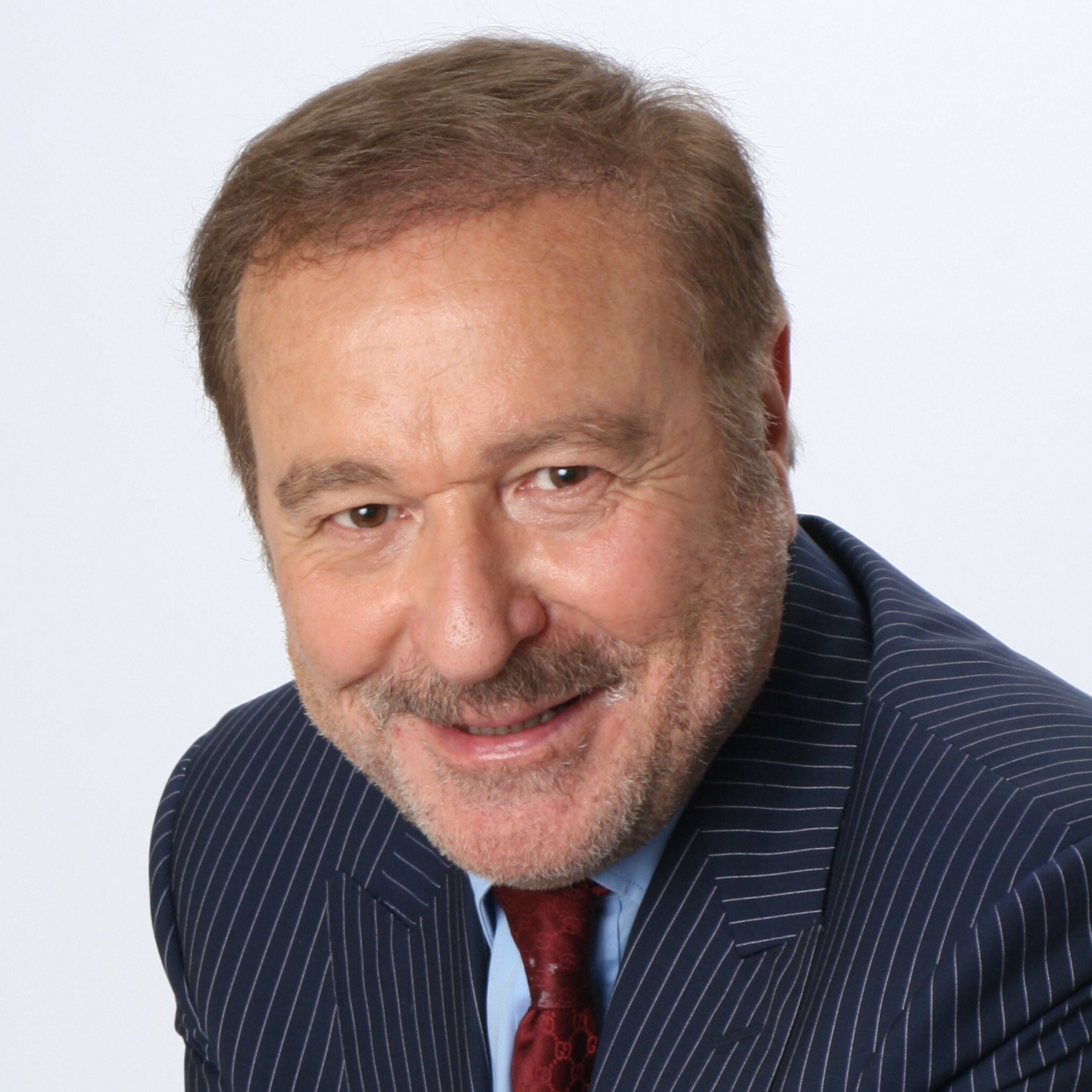}}]{Ian F. Akyildiz} (Life Fellow, IEEE) received the B.S., M.S., and Ph.D. degrees in electrical and computer engineering from the University of Erlangenrnberg, Germany, in 1978, 1981, and 1984, respectively. He has been the Founder and the President of Truva Inc., a consulting company based in Alpharetta, GA, USA, since 1989. He has been an Adjunct Professor with the University of Iceland, since 2020; the University of Helsinki, since 2021, and the University of Cyprus, since 2017. His current research interests include terahertz band communication, 6G/7G wireless systems, reconfigurable intelligent surfaces, hologram communication, extended reality wireless communications, the Internet of Space Things/CUBESATs, the Internet of Bio-Nano Things, molecular communications, and underwater and underground communications. He has been an Advisory Board Member of the Technology Innovation Institute (TII), Abu Dhabi, United Arab Emirates, since June 2020. He has also been the Founder and the Editor-in-Chief of the newly established of International Telecommunication Union Journal on Future and Evolving Technologies (ITU J-FET), since August 2020. He has served as the Ken Byers Chair Professor of Telecommunications, the Past Chair of the Telecom Group at the ECE, and the Director of the Broadband Wireless Networking Laboratory, Georgia Institute of Technology, from 1985 to 2020. He had many international affiliations during his career and established research centers in Spain, South Africa, Finland, Saudi Arabia, Germany, Russia, and India. He has been an ACM Fellow, since 1997. He received numerous awards from IEEE, ACM, and other professional organizations, including Humboldt Award from Germany and TUBITAK Award from Turkey.
\end{IEEEbiography}

\begin{IEEEbiography}[{\includegraphics[width=1in,height=1.25in,clip,keepaspectratio]{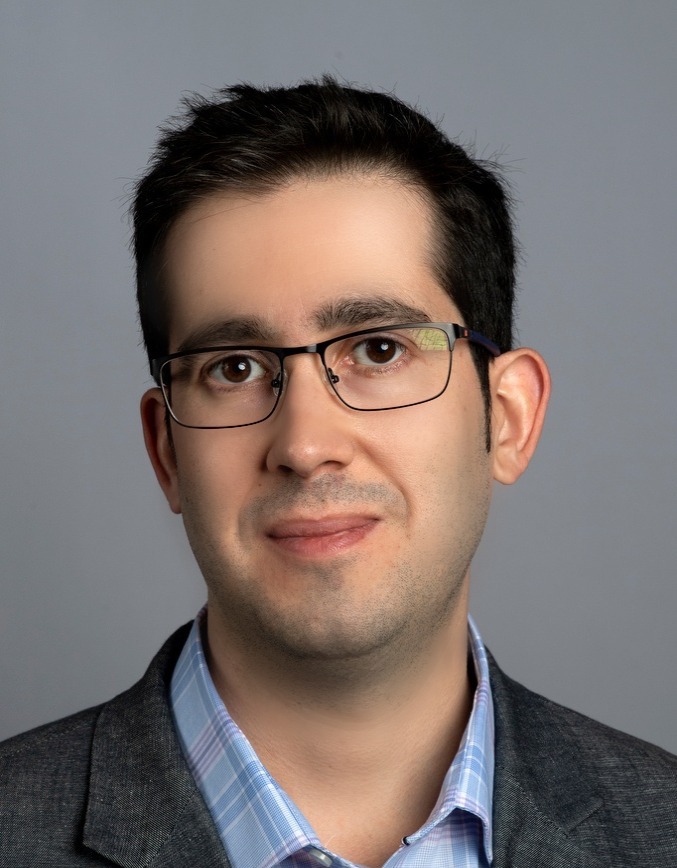}}]{Josep M. Jornet} (M'13--SM'20--F'24) is a Professor in the Department of Electrical and Computer Engineering, the director of the Ultrabroadband Nanonetworking (UN) Laboratory, and the Associate Director of the Institute for the Wireless Internet of Things at Northeastern University (NU). He received a Degree in Telecommunication Engineering and a Master of Science in Information and Communication Technologies from the Universitat Politècnica de Catalunya, Spain, in 2008. He received his Ph.D. degree in Electrical and Computer Engineering from the Georgia Institute of Technology, Atlanta, GA, in August 2013. Between August 2013 and August 2019, he was in the Department of Electrical Engineering at the University at Buffalo (UB), The State University of New York (SUNY). He is a leading expert in terahertz communications, in addition to wireless nano-bio-communication networks and the Internet of Nano-Things. In these areas, he has co-authored more than 250 peer-reviewed scientific publications, including one book, and has been granted five US patents. His work has received over 17,000 citations (h-index of 60 as of April 2024). He is serving as the lead PI on multiple grants from U.S. federal agencies including the National Science Foundation, the Air Force Office of Scientific Research, and the Air Force Research Laboratory as well as industry. He is the recipient of multiple awards, including the 2017 IEEE ComSoc Young Professional Best Innovation Award, the 2017 ACM NanoCom Outstanding Milestone Award, the NSF CAREER Award in 2019, the 2022 IEEE ComSoc RCC Early Achievement Award, and the 2022 IEEE Wireless Communications Technical Committee Outstanding Young Researcher Award, among others, as well as four best paper awards. He is a Fellow of the IEEE and an IEEE ComSoc Distinguished Lecturer (Class of 2022-2023, Extended to 2024). He is also the Editor-in-Chief of the Elsevier Nano Communication Networks journal and Editor for IEEE Transactions on Communications and Nature Scientific Reports.
\end{IEEEbiography}




\end{document}